\documentclass[12pt]{article}
\usepackage{amsmath,amssymb}
\usepackage[english]{babel}
\usepackage[cp866]{inputenc}

\usepackage{hyperref}

\numberwithin{equation}{section}

\newcommand{\bq}{\begin{eqnarray}}
\newcommand{\eq}{\end{eqnarray}}
\newcommand{\bbq}{\begin{equation*}}
\newcommand{\eeq}{\end{equation*}}

\newcommand{\ra}{\rightarrow}
\newcommand{\la}{\Lambda_Q}
\newcommand{\mph}{\mu_{\Phi}}
\newcommand{\ov}{\overline}
\newcommand{\lym}{\Lambda_{YM}}

\newcommand{\bo}{{\rm b_o}}
\newcommand{\nd}{{\ov N}_c}

\newcommand{\sq}{\textsf{Q}}
\newcommand{\dq}{\textsf{q}}
\newcommand{\odq}{\ov{\textsf{q}}}
\newcommand{\oq}{\ov{\textsf{Q}}}

\newcommand{\bd}{{\rm\ov b}_{\rm o}}

\newcommand{\ma}{m^{\rm pole}_{\sq,2}}

\newcommand{\qq}{{\ov Q}Q}
\newcommand{\qql}{\langle\qq\rangle_L}
\newcommand{\qqs}{\langle\qq\rangle_S}
\newcommand{\mgl}{\mu_{{\rm gl},\,L}}
\newcommand{\mgs}{\mu_{{\rm gl},\,S}}
\newcommand{\mql}{m^{\rm pole}_{Q,\,L}}
\newcommand{\mqs}{m^{\rm pole}_{Q,\,S}}
\newcommand{\mg}{\mu_{\rm gl}}
\newcommand{\mo}{\mu_{\Phi,\rm o}}

\newcommand{\qop}{\mu^{\rm pole}_{q,1}}

\newcommand{\qtp}{\mu^{\rm pole}_{q,2}}
\newcommand{\mug}{{\ov\mu}_{\rm gl}}
\newcommand{\muo}{{\ov\mu}_{{\rm gl},\,1}}
\newcommand{\mut}{{\ov\mu}_{{\rm gl},\,2}}

\newcommand{\mgt}{\mu_{\rm gl, 2}}
\newcommand{\mgo}{\mu_{\rm gl, 1}}

\newcommand{\oqt}{\ov{\textsf{Q}}_2}

\newcommand{\sqt}{\textsf{Q}^2}
\newcommand{\Qo}{({\ov Q}Q)_1}
\newcommand{\Qt}{({\ov Q}Q)_2}
\newcommand{\cw}{{\mathcal W}}
\newcommand{\qo}{({\ov q}q)_1}
\newcommand{\qt}{{(\ov q}q)_2}

\begin{document}
\begin{center}{\bf \large Mass spectra in $\mathbf{{\cal N}=1\,\, SQCD}$ \\
with additional colorless but flavored fields} \end{center}
\vspace{1cm}
\begin{center} \bf Victor L. Chernyak $^{\,a,\,b}$ \end{center}
\begin{center}(e-mail: v.l.chernyak@inp.nsk.su) \end{center}
\begin{center} a)\, Budker Institute of Nuclear Physics, 630090 Novosibirsk, Russia \end{center}
\begin{center} b)\, Novosibirsk State University, 630090 Novosibirsk, Russia \end{center}
\vspace{1cm}
\begin{center}{\bf Abstract} \end{center}
\vspace{1cm}

Considered is the ${\cal N}=1$ SQCD-like theory with $SU(N_c)$ colors and $0< N_F<2N_c$ flavors of equal mass $0< m_Q\ll\la$ quarks $\,Q^i_a,{\ov Q}^{\,a}_j$. Besides, it includes $N^2_F$ additional colorless but flavored fields $\Phi_{i}^{j}$, with the large mass parameter $\mph\gg\la$, interacting with quarks through the Yukawa coupling in the superpotential. The mass spectra of this $\Phi$-theory are first directly calculated at $0<N_F<N_c$ where the quarks are weakly coupled, in all different vacua with the unbroken or spontaneously broken flavor symmetry $U(N_F)\ra U(n_1)\times U(n_2)$.

Further, the mass spectra of this direct $\Phi$-theory and its Seiberg's dual variant, the $d\Phi$-theory, are calculated at $3N_c/2<N_F<2N_c$ and various values of $\mph/\la\gg 1$ (in strong coupling regimes),  using the dynamical scenario introduced by the author in his previous article \cite{ch3}. This scenario assumes that quarks can be in two different standard phases only\,: either this is the HQ (heavy quark) phase with $\langle Q\rangle=0$ where they are confined, or they are higgsed with some components $\langle Q^i_a\rangle\neq 0$, at appropriate values of the lagrangian parameters. Within the used dynamical scenario, it is shown that mass spectra of the direct $\Phi$ and dual $d\Phi$ - theories are parametrically different.

Besides it is shown in the direct $\Phi$-theory that a qualitatively new phenomenon takes place: under appropriate conditions, the seemingly heavy and dynamically irrelevant fields $\Phi$ `return back' and there appear two additional generations of light $\Phi$-particles with small masses $\mu(\Phi)\ll\la$.

Also considered is the $X$-theory which is the ${\cal N}=2$ SQCD with $SU(N_c)$ colors and $0< N_F<2N_c$ flavors of light quarks, broken down to ${\cal N}=1$ by the large mass parameter of the adjoint scalar superfield $X$, \, $\mu_X\gg\Lambda_2$. The tight interrelations between these $X$ and $\Phi$ theories are described, in particular, the conditions under which they are equivalent.\\

\newpage

\tableofcontents
\numberwithin{equation}{section}

\section{Introduction}

\hspace*{3mm} Our main purpose in this article is to calculate the mass spectra in the two ${\cal N}=1$ SQCD-like theories outlined in the abstract: the direct $\Phi$-theory and Seiberg's dual variant \cite{S1,S2}, the ${d\Phi}$-theory. In the next section 2 the definitions of these direct $\Phi$- and dual $d\Phi$-theories and their most general properties are presented in some details.

In section 3 we calculate the mass spectra of the direct $\Phi$-theory at $0< N_F< N_c$. As we show, for these values of $N_F$ all quarks interactions have logarithmically small couplings and so all calculations do not require any additional dynamical assumptions and are, in a sense, standard and straightforward.

Starting from section 4 we consider both the direct and dual theories with $N_c<N_F<2N_c$. In section 4 exact results are given for multiplicities of vacua and the nontrivial parametric behavior of quark and gluino condensates in different vacua and at different values of $\mph/\la\gg 1$, where $\mph$ is the large mass parameter of the fields $\Phi$ and $\la$ is the scale factor of the gauge coupling. These results for the quark (and gluino) condensates constitute a base for further calculations of mass spectra in sections 6-11.

In section 5 we discuss a new nontrivial phenomenon of the appearance (at the appropriate conditions) of additional generations of {\it light} colorless $\Phi$-particles in the direct theory. We show that, due to their Yukawa interactions with light quarks, the seemingly heavy and dynamically irrelevant fields $\Phi$ (fions) with the large original mass parameter $\mph(\mu\sim\la)\gg\la$ can 'return back', and there appear two additional generations of light $\Phi$-particles with small masses $\mu^{\rm pole}(\Phi)\ll\la$.

In sections 6-11 we deal with calculations of mass spectra in the direct and dual theories at
$3N_c/2<N_F<2N_c$ where both theories are, in general, in the strong coupling regimes with the gauge coupling of the direct theory $a=(N_c g^2/8\pi^2)\sim 1$. At present, unfortunately, no way is known to obtain {\it direct solutions} (i.e. without any additional assumptions) of ${\cal N}=1$ SQCD-like theories in strong coupling regimes. Therefore, to calculate mass spectra of ${\cal N}=1$ theories in such cases one has to introduce and use some {\it assumptions} about the dynamics of these theories in the strong coupling regions. In other words, one has to rely on a definite dynamical scenario. 

The definite dynamical scenario used in this paper was introduced in \cite{ch3}, and the mass spectra were calculated in the standard direct ${\cal N}=1$ SQCD with the superpotential $W={\rm Tr}\,(\,{\ov Q}m_Q Q)$ and in Seiberg's dual variant \cite{S1, S2} within this scenario. It was shown that the mass spectra of the standard direct theory and its Seiberg's dual variant are {\it parametrically different}.

We recall that this dynamical scenario introduced in \cite{ch3} assumes that in considered ${\cal N}=1$ SQCD-like theories $\Phi$ and $d\Phi$ the quarks can be in two {\it standard} phases only. These are: a) the HQ (heavy quark) phase where they are not higgsed but confined, $\langle Q^i\rangle=0\,;$\,\, b) the Higgs phase where they are not confined but higgsed, with some components $\langle Q^i\rangle\neq 0$. Moreover, the 'standard phases' imply that these two phases are realized in a standard way, even in the strong coupling region $a\sim 1$. This means that, unlike e.g. ${\cal N}=2$ SQCD with its very special properties, in these ${\cal N}=1$ SQCD-like theories without adjoint colored scalar superfields, {\it there appear no additional non-standard parametrically lighter particles (e.g. parametrically lighter magnetic monopoles or dyons) in the spectrum in the strong coupling region $a\sim 1$, in comparison with that in the weak coupling one}.

In sections 6-11 below we calculate the mass spectra of the $\Phi$ and ${d\Phi}$ theories within this dynamical scenario (mainly at the left end of the conformal window, $0<\bd/N_F=(2N_F-3N_c)/N_F\ll 1$)
and show that, similarly to the standard direct ${\cal N}=1$ SQCD with the superpotential $W={\rm Tr}\,(\,{\ov Q}m_Q Q)$ considered in \cite{ch3}, the mass spectra of the direct $\Phi$-theory and its Seiberg's dual variant, the $d\Phi$-theory, are parametrically different, so that these two theories are not equivalent.

We would like to emphasize however that, by itself, this does not mean that Seiberg's proposal
\cite{S1,S2} about the equivalence of the direct and dual theories, although not proven and remains a hypothesis up to now, is not correct. Still, it may be right but maybe not. The reason is, clearly, that the results about parametric differences of mass spectra of the direct and dual theories obtained in \cite{ch3} and in this paper are based on definite additional dynamical assumptions. In other words, on using the dynamical scenario introduced in \cite{ch3}. This dynamical scenario from \cite{ch3} satisfies all those tests which were used as checks of Seiberg's hypothesis. Moreover, it looks self-consistent and is not in contradiction with any known at present strictly proven results. Therefore, it has to be considered at present as possibly right. However, because it is not proven, it really may be right but may be not. Therefore, what is still missing at present in this story is a proof that, for instance, either the Seiberg hypothesis is right or that the dynamical scenario introduced in \cite{ch3} is right. Nevertheless, the results obtained within this scenario in \cite{ch3} and in this paper demonstrate that the checks on which the Seiberg hypothesis about the equivalence of the direct and dual theories is based (i.e. the 't Hooft triangles for the effectively massless particles and some correspondences in the superconformal regime), although necessary, {\it may well be insufficient}.\\

Finally, we consider in section 12 the $X$-theory which is ${\cal N}=2$ SQCD broken down to ${\cal N}=1$ by the large mass parameter $\mu_X$ of the adjoint scalar superfield. The tight interrelations between these $X$ and $\Phi$ - theories are described, in particular, the conditions under which they are equivalent.\\

The direct $\Phi$ and dual $d\Phi$ - theories considered in this paper have much in common with the standard ${\cal N}=1$ SQCD (and its dual variant) considered in \cite{ch3}. It is implied that the reader is familiar with the paper \cite{ch3} and with the calculation methods used therein. These methods (by the way, sufficiently standard, the non-standard is only the dynamical scenario itself) are heavily used in this paper. For this reason, some technical details are omitted in the text below, and we refer to \cite{ch3} where all additional technical details of similar calculations can be found. But besides, for the convenience of the reader, we recall below in section 2.1 assumptions of the dynamical scenario introduced in \cite{ch3}.

\section{Definitions and some generalities}

\subsection{\quad Direct $\mathbf \Phi$ - theory}

\hspace*{4mm} The field content of this direct ${\cal N}=1\,\,\,\Phi$ - theory includes $SU(N_c)$ gluons and $0< N_F<2N_c$ flavors of quarks ${\ov Q}_j, Q^i$. Besides, there are $N^2_F$ colorless but flavored fields $\Phi^{j}_{i}$ (fions) with the large mass parameter $\mph\gg\la$.

The Lagrangian at scales $\mu\gg\la$ (or at $\mu\gg\mu_H$ if $\mu_H\gg\la$, where $\mu_H$ is the next largest physical mass below $\mu^{\rm pole}_1(\Phi)\gg\la$, see the Appendix A\,;\, $\nd=N_F-N_c$\,, the exponents with gluons in the Kahler term K are implied here and everywhere below) looks as
\bq
K=\frac{1}{f^2}{\rm Tr}\,\Bigl (\Phi^\dagger \Phi\Bigr )+z(\la,\mu){\rm Tr}\Bigl (\,Q^\dagger Q+(Q\ra {\ov Q})\,\Bigr )\,,\quad \cw=-\frac{2\pi}{\alpha(\mu,\la)} S+\cw_{\Phi}+\cw_Q\,, \label{(2.1)}
\eq
\bbq
\cw_{\Phi}=\frac{\mph}{2}\Biggl [{\rm Tr}\,(\Phi^2)-\frac{1}{\nd}\Bigl ({\rm Tr}\,\Phi\Bigr )^2\Biggr ],\quad\cw_Q={\rm Tr}\,{\ov Q}(m_Q-\Phi) Q,\quad
z_Q(\la,\mu)\sim \Bigl (\ln\frac{\mu}{\la}\Bigr )^{N_c/\bo}\gg 1\,.
\eeq
Here\,: $\mph$ and $m_Q$  are the mass parameters, $S=-W^{a}_{\beta}W^{a,\,\beta}/32\pi^2$ where $W^a_{\beta}$ is the gauge field strength, $a=1...N_c^2-1,\, \beta=1,2$,\, $\alpha(\mu,\la)=g^2(\mu,\la)/4\pi$ is the gauge coupling with its scale factor $\la$, $f$ is the Yukawa coupling, $a_f=N_cf^2/8\pi^2< 1$,\, $\bo=3N_c-N_F$. {\it This normalization of fields is used everywhere below in the main text}. Besides, the perturbative NSVZ $\beta$-function for massless SUSY theories \cite{NSVZa,NSVZb} is used in this paper.

Therefore, finally, the $\Phi$-theory we deal with has the parameters\,: $N_c\,,\,0<N_F<2N_c\,,\,\mph$,\,
$\la,\, m_Q,\, f$, with the {\it strong hierarchies} $\mph\gg\la\gg m_Q$. Everywhere below in the text the mass parameter $\mph$ will be varied while $m_Q$ and $\la$ will stay intact.\\

The Konishi anomalies \cite{Konishi} from \eqref{(2.1)} for the $i$-th flavor look as (${\it i}=1\, ...\, N_F$)
\bbq
\langle\Phi_{i}\rangle\langle\frac{\partial \cw_{\Phi}}{\partial \Phi_{i}}\rangle=0\,,\quad
\langle m_{Q,i}^{\rm tot}\rangle\langle {\ov Q}_i Q^i\rangle=\langle S\rangle\,,\quad \langle m_{Q,\,i}^{\rm tot}\rangle=m_Q-\langle\Phi_{i}\rangle\,,
\eeq
\bq
\langle\Phi^{i}_{j}\rangle=\frac{1}{\mph}\Biggl ( \langle{\ov Q}_j Q^i\rangle
-\delta_{j}^{i}\frac{1}{N_c}{\rm Tr}\,\langle\qq\rangle\Biggr )\,,\quad \langle{\ov Q}_j Q^i\rangle
=\delta_{j}^{i}\langle{\ov Q}_i Q^i\rangle\,, \label{(2.2)}
\eq
and, in cases with $\mu_H<\la$,\, $\langle m_{Q,i}^{\rm tot}\rangle$ is the value of the quark total running mass at $\mu=\la$.

At all scales $\mu$ until the field $\Phi$ remains too heavy and non-dynamical, i.e. until its perturbative running mass $\mu_{\Phi}^{\rm pert}(\mu)>\mu$, it can be integrated out and the Lagrangian takes the form

\bbq
K=z_Q(\la,\mu){\rm Tr}\Bigl (Q^\dagger Q+Q\ra {\ov Q}\Bigr ),\quad \cw =-\frac{2\pi}{\alpha(\mu,\la)}S+\cw_Q\,,
\eeq
\bq
\cw_Q=m_Q{\rm Tr}({\ov Q} Q)-\frac{1}{2\mph}\Biggl ({\rm Tr}\,({\ov Q}Q)^2-\frac{1}{N_c}\Bigl({\rm Tr}\,{\ov Q} Q \Bigr)^2 \Biggr ).\label{(2.3)}
\eq

The Konishi anomalies from \eqref{(2.3)} for the i-th flavor look as
\bbq
\langle S\rangle=\langle\frac{\lambda\lambda}{32\pi^2}\rangle=\langle {\ov Q}_i\frac{\partial \cw_Q}{\partial {\ov Q}_i}\rangle=m_Q\langle {\ov Q}_i Q^i \rangle-\frac{1}{\mph}\Biggl (\sum_j\langle{\ov Q}_i Q^j\rangle\langle{\ov Q}_j Q^i\rangle-\frac{1}{N_c}\langle{\ov Q}_i Q^i\rangle\langle {\rm Tr}\,{\ov Q} Q \rangle\Biggr )=
\eeq
\bq
=\langle {\ov Q}_i Q^i \rangle\Biggl [\,m_Q-\frac{1}{\mph}\Biggl ( \langle {\ov Q}_i Q^i \rangle-
\frac{1}{N_c}\langle {\rm Tr}\,{\ov Q} Q \rangle\Biggr )  \Biggr ]\,,\quad i=1\,...\, N_F\,,\label{(2.4)}
\eq
\bbq
\langle {\ov Q}_i\frac{\partial \cw_Q}{\partial {\ov Q}_i}-{\ov Q}_j\frac{\partial \cw_Q}{\partial
{\ov Q}_j}\rangle=\langle{\ov Q}_i Q^i -{\ov Q}_j Q^j\rangle \Biggl [\,m_Q-\frac{1}{\mph}\Biggl ( \langle {\ov Q}_i Q^i+{\ov Q}_j Q^j\rangle-\frac{1}{N_c}\langle {\rm Tr}\,{\ov Q} Q \rangle\Biggr )  \Biggr ]=0\,.
\eeq

It is most easily seen from \eqref{(2.4)} that there are only two types of vacua\,: a) the vacua with the unbroken flavor symmetry, $\langle{\ov Q}_j Q^i\rangle=\delta^i_j\langle{\ov Q}Q\rangle$,\, b) the vacua with the spontaneously broken flavor symmetry, and the breaking is of the type $U(N_F)\ra U(n_1)\times U(n_2)$ only: $\langle{\ov Q}_j Q^i\rangle=\delta^i_j\langle {\ov Q}_1 Q^1\rangle\equiv\delta^i_j\langle\Qo\rangle,\, i,j=1,...n_1,\,\, \langle{\ov Q}_j Q^i\rangle=\langle {\ov Q}_2 Q^2\rangle\equiv\delta^i_j\langle\Qt\rangle,\, i,j=n_1+1,...N_F$. In these vacua one obtains from \eqref{(2.4)}
\bbq
\langle {\Qo+\Qt-\frac{1}{N_c}{\rm Tr}\, {\ov Q} Q}\rangle_{\rm br}=m_Q\mph,\quad
\langle S\rangle_{\rm br}=\frac{1}{\mph}\langle\Qo\rangle_{\rm br}\langle\Qt\rangle_{\rm br},\quad
\langle\Qo\rangle_{\rm br}\neq\langle\Qt\rangle_{\rm br}\,,
\eeq
\bq
\langle m^{\rm tot}_{Q,1}\rangle_{\rm br}=m_Q-\langle\Phi_1\rangle_{\rm br}=\frac{\langle\Qt\rangle_{\rm br}}{\mph},\quad \langle m^{\rm tot}_{Q,2}\rangle_{\rm br}=m_Q-\langle\Phi_2\rangle_{\rm br}=\frac{\langle\Qo\rangle_{\rm br}}{\mph}\,.\label{(2.5)}
\eq

We now recall details of the dynamical scenario introduced in \cite{ch3} and used in this paper in sections 6-11 for calculations of mass spectra in the conformal window $3N_c/2<N_F<2N_c$ in the strong coupling regime, both in the direct and dual theories.\\

1) Recall first that NSVZ $\beta$-function \cite{NSVZa}\cite{NSVZb} predicts {\it exact values} of quark anomalous dimensions, $\gamma_Q$ of the direct quark or $\gamma_q$ of the dual one, in the conformal regime at $3N_c/2<N_F<3N_c\,$. In the direct theory
\bq
\frac{d a^{-1}(\mu)}{d\ln\mu}=\hat\beta(a)=\frac{\bo-N_F\gamma_Q}{1-a}=0\,\,\ra\,\, \gamma_Q=\frac{\bo}
{N_F}\,,\,\, \bo=3N_c-N_F>0\,,\,\, a(\mu)=\frac{N_c g^2(\mu)}{8\pi^2}\,.\label{(2.6)}
\eq

Therefore, the renormalization factor of the quark Kahler term is also known exactly in the conformal regime: $z_Q(\la,\mu)=(\mu/\la)^{\gamma_Q}\ll 1$ at $\mu\ll\la$\,, while $a(\mu\ll\la)\ra a_{*}={\rm const},\,\,a_{*}=O(1)$ in general. In the direct theory, when the fion field $\Phi$ is effectively massless and participates actively in the conformal regime, its anomalous dimension and renormalization factor are also known exactly from the conformal symmetry: $\gamma_{\Phi}=-2\gamma_Q,\,\, z_{\Phi}(\la,\mu)=(\mu/\la)^{\gamma_{\Phi}}\gg 1$ at $\mu\ll\la$. In the dual theory, correspondingly: $\gamma_q=\bd/N_F\,,\,\,\gamma_M=-2\gamma_q\,,\,\, \bd=3\nd-N_F=2N_F-3N_c>0$, and the dual gauge coupling ${\ov a}(\mu\ll\la)\ra {\ov a}_{*}={\rm const}$. But at the left end of the conformal window there appears {\it additional small parameter}\,: $0<\bd/N_F=(2N_F-3N_c)/N_F\ll 1,\,\, \gamma_q=\bd/N_F\approx {\ov a}_{*}\ll 1$. The {\it explicit parametric dependence of various particle masses on this small parameter} is widely used in the text. It allows to trace {\it the parametric differences of mass spectra of direct and dual theories}.  \\

2) At some lower scales $\mu_i\ll\la$ the conformal regime is broken explicitly by nonzero particles masses. These may be e.g. the quark pole masses $m^{\rm pole}_{Q,i},\,\,i=1\, {\rm or}\,\, 2,$ or gluon masses $\mu^{\rm pole}_{{\rm gl},\,i}$ due to higgsed quarks. And {\it this is a first place where we need to use the additional assumption of the dynamical scenario from} \cite{ch3}. This states that (at least parametrically, i.e. up to non-parametric factors $O(1)$, this is sufficient for our purposes) the scales of these masses are given by the standard expressions: $m^{\rm pole}_{Q,i}\sim m/z_Q(\la,m^{\rm pole}_{Q,i})\,,\,\,(\mu^{\rm pole}_{\rm gl,i})^2\sim z_Q(\la,\mu^{\rm pole}_{\rm gl,i})\langle ({\ov Q}Q)_i\rangle$. If quarks $Q^i$ are in the strong coupling regime, $a_{*}(\mu= m^{\rm pole}_{Q,i}\ll\la)=N_c (g^*)^2/2\pi=O(1)$, and are in the HQ (heavy quark) phase, i.e. not higgsed but confined, then the value $m^{\rm pole}_{Q,i}$ determines the typical mass scale of hadrons made from these quarks.\\

3) It is additionally assumed that, unlike the very specific ${\cal N}=2$ SQCD, in considered ${\cal N}=1$ SQCD-like theories without colored adjoint scalar fields, the dynamics is really standard, i.e. no additional {\it parametrically lighter} solitons (e.g. magnetic monopoles or dyons) are formed at those scales where the conformal regime is broken explicitly by the quark masses or gluon masses originating from higgsed quarks.
\footnote{\,
Note that appearance of such {\it additional light solitons} will change the 't Hooft triangles at lower energies. \label{(f1)}
}
I.e., in this respect, the dynamics is {\it qualitatively similar} to those in the weak coupling regime.\\

4) Finally, to deal with the ${\cal N}=1$ SYM theory, originating after decoupling of heavy quarks at lower energies, we use the effective superpotential proposed by Veneziano-Yankielowicz \cite{VY}.\\

The use of the values of quark condensates $\langle\Qo\rangle$ and $\langle\Qt\rangle$ in various vacua calculated in section 3, the known RG evolution in the superconformal regime and described above assumptions of the dynamical scenario is sufficient to calculate parametrically (i.e. up to non-parametric factors $O(1)\,$) mass spectra of direct and dual theories in the conformal window $3N_c/2<N_F<2N_c$. This is done in sections 6-11.\\

5) Moreover, we explicitly calculate in these sections the {\it parametric dependencies} of particle masses on the {\it additional small parameter}, $0<\bd/N_F=(2N_F-3N_c)/N_F\ll 1$, appearing at the left end of the conformal window, see section 2.2 below and section 4 in \cite{ch3} for more details. This allows to trace explicitly the {\it parametric differences} in mass spectra of the direct and Seiberg's dual theories. \\

\subsection{\quad Dual $\mathbf {d\Phi}$ - theory}

In parallel with the direct $\Phi$ - theory with $N_c<N_F<2N_c$\,,\, we consider also the Seiberg dual variant \cite{S1,S2} (the $d\Phi$ - theory), with the dual Lagrangian at $\mu=\la$
\bq
K=\frac{1}{f^2}{\rm Tr}\,\Phi^\dagger \Phi+ {\rm Tr}\Bigl ( q^\dagger q + (q\ra\ov q)\, \Bigr )+{\rm Tr}\, \frac{M^{\dagger}M}{\mu_2^2}\,,\quad
\cw=\, -\,\frac{2\pi}{\ov \alpha(\mu=\la)}\, {\ov s}+{\ov \cw}_M+\cw_q\,,\label{(2.7)}
\eq
\bbq
{\ov \cw}_M=\frac{\mph}{2}\Biggl [{\rm Tr}\,(\Phi^2)-\frac{1}{\nd}\Bigl ({\rm Tr}\,\Phi\Bigr )^2\Biggr ]+ {\rm Tr}\, M(m_Q-\Phi),\quad \cw_q= -\,\frac{1}{\mu_1}\,\rm {Tr} \Bigl ({\ov q}\,M\, q \Bigr )\,.
\eeq
Here\,:\, the number of dual colors is $\nd=N_F-N_c,\, \bd=3\nd-N_F$, and $M^i_j\ra ({\ov Q}_j Q^i)$ are the $N_F^2$  elementary mion fields, ${\ov a}(\mu)=\nd{\ov \alpha}(\mu)/2\pi=\nd{\ov g}^2(\mu)/8\pi^2$ is the dual running gauge coupling (with its scale parameter $\Lambda_q$),\,\,${\ov s}=-{\rm \ov w}^{b}_{\beta}{\rm \ov w}^{b,\,\beta}/32\pi^2$,\,\, ${\rm \ov w}^b_{\beta}$ is the dual gluon field strength. The gluino condensates of the direct and dual theories are matched in all vacua, $\langle{-\,\ov s}\rangle=\langle S\rangle=\lym^3$, as well as $\langle M_j^i(\mu=\la)\rangle=\langle{\ov Q}_j Q_i (\mu=\la)\rangle$, and the scale parameter $\Lambda_q$ of the dual gauge coupling is taken as $|\Lambda_q|=\la$, see the Appendix in \cite{ch3} for more details. At $3/2<N_F/N_c<2$ this dual theory can be taken as UV free at $\mu\gg\la$, and this requires that its Yukawa coupling at $\mu=\la,\, f(\mu=\la)=\mu_2/\mu_1$, cannot be larger than its gauge coupling ${\ov g}(\mu=\la)$, i.e. $\mu_2/\mu_1\lesssim 1$. The same requirement to the value of the Yukawa coupling follows from the conformal behavior of this theory at $3/2<N_F/N_c<2$ and $\mu<\la$, i.e. $f(\mu=\la)=\mu_2/\mu_1\approx f_*=O(1)$ at $\bd/N_F=O(1)$. We consider below this dual theory at $\mu\leq\la$ only, where it claims to be equivalent to the direct $\Phi$ - theory. As was explained in \cite{ch3}, one has to take $\mu_1\sim\la$ at $\bd/N_F=(3\nd-N_F)/N_F=O(1)$ in \eqref{(2.7)} to match the gluino condensates in the direct and dual theories. Therefore, $\mu_2\sim\mu_1\sim\la$ in this case also. But to match the gluino condensates in the direct and dual theories at the left end of the conformal window, i.e. at $0<\bd/N_F\ll 1$, one has to take $(\mu_2/\mu_1)^2\approx f^2_*=O(\bd/N_F)\ll 1$ and $\mu_1\sim Z_q\la\ll\la,\, Z_q\sim\exp\{-\nd/7\bd\}\ll 1$ (with the exponential accuracy, i.e. powers of the small parameter $0<\bd/N_F\ll 1$ are not traced here and only the powers of $Z_q$ are traced, this is sufficient for our purposes, so that at $\bd/N_F=O(1)$ one has to put $Z_q\ra 1$, see \cite{ch3} for more details).\\

Really, the fields $\Phi$ remain always too heavy and dynamically irrelevant in this $d\Phi$ - theory at $3 N_c/2<N_F<2 N_c$, so that they can be integrated out once and forever and, finally, we write the Lagrangian of the dual theory at $\mu=\la$ in the form
\bbq
K= {\rm Tr}\Bigl ( q^\dagger q +(q\ra\ov q) \Bigr )+{\rm Tr}\,\frac{M^{\dagger}M}{Z^2_q\la^2}\,,\quad
\cw=\, -\,\frac{2\pi}{\ov \alpha(\mu=\la)}\, {\ov s}+\cw_M+\cw_q\,,
\eeq
\bq
\cw_M=m_Q{\rm Tr}\,M -\frac{1}{2\mph}\Biggl [{\rm Tr}\, (M^2)- \frac{1}{N_c}({\rm Tr}\, M)^2 \Biggr ]\,,\quad
\cw_q= -\,\frac{1}{Z_q\la}\,\rm {Tr} \Bigl ({\ov q}\,M\, q \Bigr )\,.\label{(2.8)}
\eq

The Konishi anomalies for the $i$-th flavor look here as (${\it i}=1\, ...\, N_F$)
\bq
\langle M_i\rangle\langle N_i\rangle=Z_q\la\langle S\rangle\,,\quad \frac{\langle N_i\rangle}{Z_q\la}=m_Q-\frac{1}{\mph}\Bigl (\langle M_i-\frac{1}{N_c}{\rm Tr}\,M \rangle\Bigr )=\langle m_{Q,i}^{\rm tot}\rangle\,,\label{(2.9)}
\eq
\bbq
\langle N_i\rangle\equiv\langle{\ov q}^i q_i(\mu=\la)\rangle\,,\quad {\rm no\,\, summation\,\, over\,\, i}\,.
\eeq

In vacua with the broken flavor symmetry these can be rewritten as
\bbq
\langle M_1+M_2-\frac{1}{N_c}{\rm Tr}\, M\rangle_{\rm br}=m_Q\mph,\quad
\langle S\rangle_{\rm br}=\frac{1}{\mph}\langle M_1\rangle_{\rm br}\langle M_2\rangle_{\rm br},\quad
\langle M_1\rangle_{\rm br}\neq\langle M_2\rangle_{\rm br}\,,
\eeq
\bq
\frac{\langle N_1\rangle_{\rm br}}{Z_q\la}=\frac{\langle S\rangle_{\rm br}}{\langle M_{1}\rangle_{\rm br}}=\frac{\langle M_{2}\rangle_{\rm br}}{\mph}=m_Q-\frac{1}{\mph}\Bigl (\langle M_{1}-\frac{1}{N_c}{\rm Tr}\, M\rangle_{\rm br} \Bigr )=\langle m^{\rm tot}_{Q,1}\rangle_{\rm br}\,,\label{(2.10)}
\eq
\bbq
\frac{\langle N_2\rangle_{\rm br}}{Z_q\la}=\frac{\langle S\rangle_{\rm br}}{\langle M_{2}\rangle_{\rm br}}=\frac{\langle M_{1}\rangle_{\rm br}}{\mph}=m_Q-\frac{1}{\mph}\Bigl (\langle M_{2}-\frac{1}{N_c}{\rm Tr}\, M\rangle_{\rm br} \Bigr )=\langle m^{\rm tot}_{Q,1}\rangle_{\rm br}\,.
\eeq

\section {Vacua, condensates and mass spectra at $\mathbf {0< N_F< N_c}$}

\hspace*{4mm} Clearly, there is no dual theory for this range of $N_F$ values. Moreover (see below in this section), in vacua of the direct theory with the unbroken flavor symmetry all quarks are higgsed in the weak (logarithmically small) coupling regime with the large masses of higgsed gluons, $\mu_{\rm gl}\gg \la$. In vacua with the broken flavor symmetry all quarks: a) are either also higgsed in the weak (logarithmically small) coupling regime with $\mu_{\rm gl,\,\it i}\gg \la,\, i=1,2$\,, at $\la\ll\mph\ll {\tilde\mu}_{\Phi}$,\,\, ${\tilde\mu}_{\Phi}\sim \la (\la/m_Q)^{(\bo-n_1)/n_1}\gg\la$\,;\,\, b) or, in $\rm br1$-vacua with $1\leq n_1\leq [N_F/2]$, the quarks ${\ov Q}_1,\, Q^1$ with flavors $n_1$ are higgsed in the weak (logarithmically small) coupling regime, while the quarks ${\ov Q}_2,\, Q^2$ with flavors $n_2=N_F-n_1$ are in the HQ-phase at $\mph\gg {\tilde\mu}_{\Phi}$, they are weakly confined (i.e. the tension of confining string originated from unbroken $SU(N_c-n_1)$ color group is much smaller than quark masses, $\sqrt\sigma\ll m_{Q,\, perturb}^{\rm pole}$) and also perturbatively logarithmically weakly coupled and non-relativistic inside hadrons (in $\rm br2$-vacua $n_1\leftrightarrow n_2$). Therefore, finally, in all vacua and at all values $\la\ll\mph$, the quarks are parametrically weakly coupled and their dynamics is simple and qualitatively evident.

For this reason, {\it we need no any additional assumptions about the quark dynamics to calculate the mass spectra at}\, $0< N_F< N_c$. In other words, because the HQ- and Higgs-phases of quarks are at logarithmically weak couplings, there is no need to mention about any assumed dynamical scenario at all (it is really needed to calculate the mass spectra in the strong coupling region only).

The calculations methods used below in this section have much in common with those in the standard SQCD with $m_Q/\la\ll 1$ and $0< N_F< N_c$ in section 2 of \cite{ch1}. It is implied that the reader is familiar with \cite{ch1}, so that some technical ins and outs are omitted below (see section 2 in \cite{ch1} for much more details). But really, as mentioned above, because all quarks are parametrically weakly coupled, all calculations in this section 3 are highly standard and, we hope, self-evident.

\subsection{ Unbroken flavor symmetry}

There is $N_{\rm unbrok}=(2N_c-N_F)$ such vacua and all quarks are higgsed in all of them, but the hierarchies in the mass spectrum are parametrically different depending on the value of $\mph$ (see below). In any case, all $N_F^2$ fions are very heavy and dynamically irrelevant in these vacua at scales $\mu<\mu^{\rm pole}_{1}(\Phi)$ (see the Appendix A) and can be integrated out from the beginning.

All quarks are higgsed at the high scale $\mu=\mu_{\rm gl},\, \la\ll\mu_{\rm gl}\ll\mu^{\rm pole}_1(\Phi)$,
\bq
\mu^2_{\rm gl}=N_c g^2(\mu=\mu_{\rm gl})z_Q(\la,\mu_{\rm gl})\langle\Pi\rangle,\quad \langle\Pi\rangle=\langle{\ov Q}_1 Q^1(\mu=\la)\rangle\equiv\langle{\ov Q} Q\rangle, \quad g^2(\mu)=4\pi\alpha(\mu),\label{(3.1)}
\eq
where (in the approximation of leading logs and $C_F=(N_c^2-1)/2N_c\approx N_c/2$)
\bq
\frac{2\pi}{\alpha(\mu_{\rm gl})}\approx \bo\ln{\frac{\mu_{\rm gl}}{\la}}\,\,,\,\, z_Q(\la,\mu_{\rm gl})\sim\Bigl (\frac{\alpha(\la)}{\alpha(\mu_{\rm gl})}\Bigr)^{2C_F/\bo}\sim \Bigl (\ln\frac{\mu_{\rm gl}}{\la}\Bigr )^{N_c/\bo}\gg 1\,,\quad \bo=3N_c-N_F\,.\label{(3.2)}
\eq

Hence, after integrating out all heavy higgsed gluons and their superpartners at $\mu<\mu_{\rm gl}$ one remains with the $SU(N_c-N_F)$ pure Yang-Mills theory with the scale factor $\lym$ of its gauge coupling. Finally, after integrating out remained gluons at $\mu<\lym$ via the Veneziano-Yankielowicz (VY) procedure \cite{VY,TVY} (see section 2 in \cite{ch1} for more details), one obtains the Lagrangian of $N_F^2$ pions
\bq
K=z_Q(\la,\mu_{\rm gl})2\,{\rm Tr}\,\sqrt {\Pi^{\dagger}\Pi}\,,\quad \cw=-\nd S+\cw_{\Pi}\,,\label{(3.3)}
\eq
\bbq
S=\Biggl (\,\frac{\la^{\bo}}{\det \Pi}\,\Biggr )^{\frac{1}{N_c-N_F}}\,,\quad
\cw_{\Pi}=m_Q{\rm Tr}\,\Pi -\frac{1}{2\mph}\Biggl [{\rm Tr}\, ({\Pi}^2)- \frac{1}{N_c}({\rm Tr}\, \Pi)^2  \Biggr ],
\eeq
\bbq
\langle\Pi^i_j\rangle=\delta^i_j\,\langle\Pi\rangle=\delta^i_j\,\langle{\ov Q}_1 Q^1(\mu=\la)\rangle,\quad i,j=1\,...\,N_F\,.
\eeq

It follows from \eqref{(3.3)} that depending on the value of $\mph/\la \gg 1$\, there are two different regimes.\\

{\bf i)}\,\, At $\la\ll\mph \ll \mo=\la(\la/m_Q)^{(2N_c-N_F)/N_c},\,\,\mo\gg\la$, the term $m_Q{\rm Tr}({\ov Q}Q)$ in the superpotential \eqref{(3.3)} gives only a small correction and one obtains
\bq
\langle \Pi \rangle_{\rm o}\sim \la^2\Bigl (\frac{\mph}{\la}\Biggr )^{\frac{N_c-N_F}{2N_c-N_F}}\gg \la^2\,.
\label{(3.4)}
\eq

There are $(2N_c-N_F)$ such vacua, this agrees with \cite{CKM}.
\footnote{\,
To see that there are just $2N_c-N_F$ vacua and not less, one has to separate slightly all quark masses, $m_Q\ra m_Q^{(i)},\,i=1...N_F,\, 0<(\delta m_Q)^{ij}=(m_Q^{(i)}-m_Q^{(j)})\ll {\ov m}_Q$. All quark mass terms give only small power corrections to (3.4), but just these corrections show the $Z_{2N_c-N_F}$ multiplicity of vacua.
}
The masses of heavy gluons and their superpartners are given in \eqref{(3.1)} while from \eqref{(3.3)} the pion masses are
\bq
\mu_{\rm o}(\Pi)\sim \frac{\langle \Pi \rangle_{\rm o}}{z_Q(\la,\mu_{\rm gl})\mph}\sim \frac{\la}{z_Q(\la,
\mu_{\rm gl})}\Bigl (\frac{\la}{\mph}\Biggr )^{\frac{N_c}{2N_c-N_F}}\gg m_Q\,.\label{(3.5)}
\eq
Besides, the scale of the gluino condensate of unbroken $SU(N_c-N_F)$ is
\bq
\lym=\langle S\rangle^{1/3}\sim \Biggl (\frac{\la^{\bo}}{\det \langle\Pi\rangle_{\rm o}}\Biggr )^{\frac{1}{3(N_c-N_F)}}\sim \la\Bigl(\frac{\la}{\mph}\Biggr )^{\frac{N_F}{3(2N_c-N_F)}}\,,\quad \mu_{\rm o}(\Pi)\ll \lym\ll \la\ll \mu_{\rm gl}\,,\label{(3.6)}
\eq
and there is a large number of gluonia with the mass scale $\sim \lym$ (except for the case $N_F=N_c-1$ when the whole gauge group is higgsed, there is no gluonia with masses $\sim \lym$ and the non-perturbative superpotential in \eqref{(3.3)} originates not from the unbroken $SU(N_c-N_F)$ but directly from the instanton contribution \cite{ADS}).\\

{\bf ii)}\,\,\, $(2N_c-N_F)$ vacua split into two groups of vacua with parametrically different mass spectra at $\mph\gg \mo$. There are $N_c$\,\, SQCD vacua with $\langle \Pi \rangle_{\rm SQCD}\sim\la^2(\la/m_Q)^{(N_c-N_F)/N_c}$ differing by $Z_{N_c}$ phases of the quark and gluino condensates (in these, the last term $\sim \Pi^2/\mph$ in the superpotential \eqref{(3.3)} can be neglected), and $(N_c-N_F)$ of nearly degenerate classical vacua with parametrically larger condensates $\langle \Pi \rangle_{\rm cl}\sim m_Q\mph$ (in these, the first non-perturbative quantum term $\sim S$ in the superpotential \eqref{(3.3)} gives only small corrections with $Z_{N_c-N_F}$ phases, but the multiplicity of vacua originates just from these small corrections). The properties of SQCD vacua have been described in detail in chapter 2 of \cite{ch1}, the pion masses are $\mu_{\rm SQCD}(\Pi)\sim m_Q/z_Q(\la,\mu^{SQCD}_{\rm gl})\ll m_Q$ therein, where $z_Q(\la,\mu^{SQCD}_{\rm gl})\gg 1$ is the logarithically large perturbative renormalization factor. In $(N_c-N_F)$ classical vacua the gluon and pion masses are given in \eqref{(3.1)} and \eqref{(3.5)} but now
\bq
\langle \Pi \rangle_{\rm cl}\sim  m_Q\mph \gg \la^2\,,\quad
\mu_{\rm cl}(\Pi)\sim \frac{m_Q}{z_Q(\la,\mu^{\rm cl}_{\rm gl})}\,,\label{(3.7)}
\eq
and in all vacua (except for the case $N_F=N_c-1$ ) there is a large number of gluonia with the mass scale
\bq
\sim \lym^{\rm SQCD}=\langle S \rangle^{1/3}\sim\Biggl (\frac{\la^{\bo}}{\det \langle\Pi\rangle_{\rm SQCD}}\Biggr )^{\frac{1}{3(N_c-N_F)}}\sim \la\Biggl (\frac{m_Q}{\la}\Biggr)^{N_F/3N_c}\quad {\rm in\,\, N_c \,\,\,SQCD \,\, vacua}\,,\label{(3.8)}
\eq
\bq
\sim \lym^{\rm class}\sim\Biggl (\frac{\la^{\bo}}{\det \langle\Pi\rangle_{\rm cl}}\Biggr )^{\frac{1}{3(N_c-N_F)}}\sim \la\Biggl (\frac{\la^2}{m_Q\mph}\Biggr )^{\frac{N_F}{3(N_c-N_F)}}\,\, {\rm in\,\, (N_c-N_F) \,\,classical\,\, vacua}.\label{(3.9)}
\eq

Finally, the change of regimes ${\bf i}\leftrightarrow {\bf ii}$ occurs at
\bq
\Biggl (\frac{\mo}{\la}\Biggr )^{\frac{N_c-N_F}{2N_c-N_F}}\sim \frac{m_Q \mo}{\la^2}\gg 1\quad \ra \quad
\mo \sim \la\Biggl (\frac{\la}{m_Q}\Biggr )^{\frac{2N_c-N_F}{N_c}}\gg\la\,.\label{(3.10)}
\eq
\vspace*{2mm}

\subsection{ Spontaneously broken flavor symmetry\,: $U(N_F)\ra U(n_1)\times U(n_2)$}

\hspace*{4mm} The quark condensates $\langle{\ov Q}_j Q^i\rangle\sim \delta^i_j C_i$ split into two groups in these vacua with the spontaneously broken flavor symmetry\,: there are $1\leq n_1\leq [N_F/2]$ equal values $\langle\Pi_1\rangle=\langle{\ov Q}_1 Q^1\rangle\equiv\langle\Qo\rangle$ and $n_2=(N_F-n_1)\geq n_1$ equal values $\langle\Pi_2\rangle=\langle{\ov Q}_2 Q^2\rangle\equiv\langle\Qt\rangle\neq \langle\Qo\rangle$ (unless stated explicitly, here and everywhere below in the text it is implied that $1-(n_1/N_c),\,\, 1-(n_2/N_c)$ and $(2N_c-N_F)/N_c$ are all $O(1)$\,). And there will be two different phases, depending on the value of $\mph/\la \gg 1$\, (see below).\\

{\,\,\,\bf 3.2.1}\,\,\, At $\la\ll\mph\ll\mo$ all qualitative properties are similar to those for an unbroken symmetry. All quarks are higgsed at high scales $\mu_{\rm gl, 1}\sim \mu_{\rm gl, 2}\gg \la$ and the low energy Lagrangian has the form \eqref{(3.3)}. The term $m_Q{\rm Tr} ({\ov Q}Q)$ in the superpotential in \eqref{(3.3)} gives only small corrections, while \eqref{(2.5)} can be rewritten here in the form
\bbq
\langle\Pi_1+\Pi_2\rangle_{\rm br}=\frac{1}{N_c}{\rm Tr}\,\langle \Pi\rangle_{\rm br}+m_Q\mph\approx \frac{1}{N_c}\langle n_1\Pi_1+n_2\Pi_2\rangle_{\rm br}\,\, \ra \,\, \Bigl (1-\frac{n_1}{N_c}\Bigr )\langle\Pi_1\rangle_{\rm br}\approx -\Bigl (1-\frac{n_2}{N_c}\Bigr )\langle\Pi_2\rangle_{\rm br},
\eeq
\bq
\langle S\rangle_{\rm br}=\Biggl (\frac{\la^{\bo}}{\langle\Pi_1\rangle^{n_1}_{\rm br}\langle\Pi_2\rangle^{n_2}_{\rm br}}\Biggr )^
{\frac{1}{N_c-N_F}}=\frac{\langle\Pi_1\rangle_{\rm br}\langle\Pi_2\rangle_{\rm br}}{\mph}\,,\label{(3.11)}
\eq
\bq
\mu^2_{\rm gl, 1}\sim\mu^2_{\rm gl, 2}\sim  g^2(\mu=\mu_{\rm gl})z_Q(\la,\mu_{\rm gl})
\langle\Pi_{1,2}\rangle_{\rm br},\,\, \langle \Pi_1 \rangle_{\rm br}\sim\langle\Pi_2\rangle_{\rm br}\sim \la^2\Biggl (\frac{\mph}{\la}\Biggr )^{\frac{N_c-N_F}{2N_c-N_F}}\,. \label{(3.12)}
\eq

The pion masses in this regime look as follows, see \eqref{(3.3)}\,:\, a) due to the spontaneous breaking of the flavor symmetry, $U(N_F)\ra U(n_1)\times U(n_2)$, there always will be $2n_1 n_2$ exactly massless Nambu-Goldstone particles and in this case these are the hybrids $\Pi_{12}$ and $\Pi_{21}$; \,b) other $n_1^2+n_2^2$\,\, `normal' pions have masses as in \eqref{(3.5)}.

There are
\bq
N^{\rm tot}_{\rm brok}=\sum_{n_1=1}^{n_1=[N_F/2]}N_{\rm brok}(n_1)=\sum_{n_1=1}^{n_1=[N_F/2]}
(2N_c-N_F){\ov C}^{\, n_1}_{N_F}\,,\quad C^{\, n_1}_{N_F}=\frac{N_F!}{n_1!\, n_2!} \label{(3.13)}
\eq
such vacua (the factor $2N_c-N_F$ originates from $Z_{2N_c-N_F}$ (see the footnote 1)\,, for even $N_F$ the last term with $n_1=N_F/2$ enters \eqref{(3.13)} with the additional factor $1/2$, i.e. ${\ov C}^{\, n_1}_{N_F}$ differ from the standard $C^{\,n_1}_{N_F}$ in \eqref{(3.13)} only by ${\ov C}^{\,n_1={\rm k}}_{N_F=2{\rm k}}=C^{\,n_1={\rm k}}_{N_F=2{\rm k}}/2$ ), so that the total number of vacua
\footnote {\,
By convention, we ignore the continuous multiplicity of vacua due to the spontaneous flavor symmetry breaking. Another way, one can separate slightly all quark masses (see the footnote 1), so that all Nambu-Goldstone bosons will acquire small masses $O(\delta m_Q)\ll {\ov m}_Q$.
}
is
\bq
N_{\rm tot}=\Bigl ( N_{\rm unbrok}=2N_c-N_F\Bigr )+N^{\rm tot}_{\rm brok}\,,\label{(3.14)}
\eq
this agrees with \cite{CKM}.

{\,\,\,\bf 3.2.2}\,\,\, The change of the regime in these vacua with the broken symmetry occurs at $\mo\ll\mph\ll{\tilde\mu}_{\Phi}$\,, see \eqref{(3.10)},\eqref{(3.20)}, when all quarks are still higgsed but there appears a large hierarchy between the values of quark condensates at $\mph\gg\mo$\,, see \eqref{(2.5)}. Instead of $\langle \Pi_1 \rangle\sim \langle \Pi_2 \rangle $, they look now as: \\

{\bf a)}\,\, $\rm{br}1$\, ($\rm{br}1=\rm{breaking}-1$) - vacua
\bq
\hspace*{-4mm}\langle \Pi_1 \rangle_{\rm br1}\approx \Biggl (\rho_1=\frac{N_c}{N_c-n_1}\Biggr ) m_Q\mph\gg \la^2,\,\, \langle \Pi_2 \rangle_{\rm br1}\approx\la^2\Bigl (\frac{\la}{m_Q\rho_1} \Bigr )^{\frac{N_c-n_2}{N_c-n_1}}\Bigl (\frac{\la}{\mph} \Bigr )^{\frac{n_1}{N_c-n_1}}\ll \langle \Pi_1 \rangle_{\rm br1}.\label{(3.15)}
\eq

Unlike the mainly quantum $\langle\Pi\rangle_{\rm o}$ or mainly classical $\langle\Pi\rangle_{\rm cl}$ vacua with unbroken symmetry, these vacua are pseudo-classical\,: the largest value of the condensate $\langle \Pi_1 \rangle_{\rm br1}\sim m_Q\mph$ is classical while the smaller value of $\langle \Pi_2\rangle_{\rm br1}\sim\langle S\rangle_{\rm br1}/m_Q$ is of quantum origin, see \eqref{(2.5)}. There are $N_{\rm br1}(n_1)=(N_c-n_1){\ov C}_{N_F}^{\, n_1}$ such vacua at given values of $n_1$ and $n_2$.\\
{\bf b)}\,\, $\rm{br2}$ - vacua. These are obtained from \eqref{(3.15)} by $n_1\leftrightarrow n_2$ and there are $N_{\rm br2}(n_1)=(N_c-n_2){\ov C}_{N_F}^{\, n_1}$ such vacua. Of course, the total number of vacua, $N_{\rm brok}(n_1)=N_{\rm br1}(n_1)+N_{\rm br2}(n_1)=(2N_c-N_F){\ov C}_{N_F}^{\, n_1}$ remains the same at $\mph\lessgtr\mo$.\\

We consider $\rm br1$ vacua (all results in $\rm br2$ vacua can be obtained by $n_1\leftrightarrow n_2$). In the range $\mo\ll\mph\ll {\tilde\mu}_{\Phi}$ (see below) where all quarks are higgsed finally, the masses of higgsed gluons look now as
\bq
\mu^2_{\rm gl,1}\sim  g^2(\mu=\mu_{\rm gl,1})z_Q(\la,\mu_{\rm gl,1})\langle\Pi_1\rangle\gg
\mu^2_{\rm gl,2 }\,.\label{(3.16)}
\eq
The superpotential in the low energy Lagrangian of pions looks as in \eqref{(3.3)}, but the Kahler term of pions is different. We write it in the form \,:\, $K\sim z_Q(\la,\mu_{\rm gl,1}){\rm Tr}\sqrt{\Pi^{\dagger}
_z\Pi_z}$\,. The $N_F\times N_F$ matrix $\Pi_z$ of pions looks as follows. Its $n_2\times n_2$ part consists of fields $z^{\prime}_Q(\mu_{\rm gl,1},\mu_{\rm gl,2})\Pi_{22}$, where $z^{\prime}_Q\ll 1$ is the perturbative logarithmic renormalization factor of ${\oq}_2,\, {\sq}^2$ quarks with unhiggsed colors which appears due to their additional RG evolution in the range of scales $\mu_{\rm gl,2}<\mu<\mu_{\rm gl,1}$, while at $\mu=\mu_{\rm gl,2}$ they are also higgsed. All other pion fields $\Pi_{11}, \Pi_{12}$ and $\Pi_{21}$ are normal. As a result, the pion masses look as follows. $2n_1n_2$ hybrid pions $\Pi_{12}$ and $\Pi_{21}$ are massless, while the masses of $n_1^2$\, $\Pi_{11}$ and $n_2^2$\, $\Pi_{22}$ are
\bq
\mu(\Pi_{11})\sim \frac{m_Q}{z_Q(\la,\mu_{\rm gl,1})}\,,\quad \mu(\Pi_{22})\sim \frac{m_Q}{z_Q(\la,\mu_{\rm gl,1})z^{\prime}_Q(\mu_{\rm gl,1},\mu_{\rm gl,2})}\gg \mu(\Pi_{11})\,.\label{(3.17)}
\eq
Finally, the mass scale of gluonia from the unhiggsed $SU(N_c-N_F)$ group is $\sim \lym^{\rm (br1)}$\,, where
\bq
(\lym^{\rm (br1)})^3=\langle S\rangle_{\rm br1}=\frac{\langle \Pi_1\rangle_{\rm br1}\langle \Pi_2\rangle_{\rm br1}}{\mph}\sim m_Q\langle\Pi_2\rangle_{\rm br1} \sim\la^3\Biggl (\frac{\la}{\mph}\Biggr)^{\frac{n_1}
{N_c-n_1}}\Biggl (\frac{m_Q}{\la}\Biggr)^{\frac{n_2-n_1}{N_c-n_1}}\,.\label{(3.18)}
\eq

{\,\,\,\bf 3.2.3}\,\,\,At scales $\la\ll\mu<\mu_{\rm gl,1}\sim \langle \Pi_1 \rangle^{1/2}\sim (m_Q\mph)^{1/2}$ (ignoring logarithmic factors) the light degrees of freedom include the $SU(N_c-n_1)$ gluons and active quarks ${\oq}_2,\, {\sq}^2$ with unhiggsed colors and  $n_2<(N_c-n_1)$ flavors, $n_1^2$ pions $\Pi_{11}$ and $2n_1 n_2$ hybrid pions $\Pi_{12}$ and $\Pi_{21}$ (in essence, these are the quarks ${\ov Q}_2,\, Q^2$ with higgsed colors in this case). The scale factor ${\Lambda_1}$ of the gauge coupling in this lower energy theory is
\bq
\Lambda^{{\rm b}^{\prime}_{\rm o}}_1\sim\la^{\bo}/\det \Pi_{11}\,,\quad {\rm b}^{\prime}_{\rm o}=3(N_c-n_1)-n_2\,,\quad  \bo=3N_c-N_F\,.\label{(3.19)}
\eq
The scale of the perturbative pole mass of ${\oq}_2,\, {\sq}^2$ quarks is $m_Q^{\rm pole}\sim m_Q$\,, while the scale of $\mu_{\rm gl,2}$ is $\mu_{\rm gl,2}\sim\langle{\oq}_2{\sq}^2\rangle^{1/2}=
\langle\Pi_2\rangle^{1/2}$\,, with $\langle\Pi_2\rangle\ll \langle\Pi_1\rangle$ given in \eqref{(3.15)}. Hence, the hierarchy at $\mo\ll\mph\ll{\tilde\mu}_{\Phi}$ looks as\, $m_Q\ll {\Lambda_1}\ll\mu_{\rm gl, 2}
\sim\langle\Pi_2\rangle^{1/2}$ and active ${\oq}_2,\, {\sq}^2$ quarks are also higgsed, while at $\mph\gg {\tilde\mu}_{\Phi}$ the hierarchy looks as $\langle\Pi_2\rangle^{1/2}\equiv\langle\Qt\rangle^{1/2}\ll {\Lambda_1}\ll m_Q$ and the active quarks ${\oq}_2,\, {\sq}^2$ become too heavy, they are not higgsed but are in the $\rm HQ_2$ (heavy quark) phase. The phase changes at
\bq
\langle \Pi_2 \rangle^{1/2} \sim m_Q\sim {\langle\Lambda_1\rangle}\sim\lym^{(\rm br1)} \,\,\ra\,\, {\tilde\mu}_{\Phi}\sim \la \Biggl (\frac{\la}{m_Q}\Biggr )^{\frac{ \bo-n_1}{n_1}}\gg \mo\,.\label{(3.20)}
\eq

Hence, we consider now this $Higgs_1-HQ_2$ phase realized at $\mph>{\tilde\mu}_{\Phi}$. For this it is convenient to retain all fields $\Phi$ although, in essence, they are too heavy and dynamically irrelevant. After integrating out all heavy higgsed gluons and ${\ov Q}_1, Q^1$ quarks, we write the Lagrangian at $\mu^2=\mu^2_{\rm gl,1}\sim N_c g^2(\mu=\mu_{\rm gl,1})z_Q(\la,\mu_{\rm gl,1})\langle\Pi_1\rangle$  in the form (see the Appendix A)
\bq
K=\Bigl [\,\frac{1}{f^2}{\rm Tr}(\Phi^\dagger\Phi)+z_Q(\la,\mu^2_{\rm gl,1})\Bigl (K_{\Pi}+K_{{\sq}_2}\Bigr )\,\Bigr ], \label{(3.21)}
\eq
\bbq
K_{{\sq}_2}={\rm Tr}\Bigl ({\sq}^{\dagger}_2 {\sq}^2 +({\sq}^2\ra
{\oq}_2 )\Bigr )\,, \quad K_{\Pi}= 2{\rm Tr}\sqrt{\Pi^{\dagger}_{11}\Pi_{11}}+K_{\rm hybr},
\eeq
\bbq
K_{\rm hybr}={\rm Tr}\Biggl (\Pi^{\dagger}_{12}\frac{1}{\sqrt{\Pi_{11}\Pi^{\dagger}_{11}}}\Pi_{12}+
\Pi_{21}\frac{1}{\sqrt{\Pi^{\dagger}_{11}\Pi_{11}}}\Pi^\dagger_{21}\Biggr ),
\eeq
\bbq
\cw=\Bigl [-\frac{2\pi}{\alpha(\mu_{\rm gl,1})}{\textsf S}\Bigr ]+\frac{\mph}{2}\Biggl [{\rm Tr}\, (\Phi^2) -\frac{1}{\nd}\Bigl ({\rm Tr}\,\Phi\Bigr)^2\Biggr ]+{\rm Tr}\Bigl ({\oq_2}m^{\rm tot}_{{\sq}_2}{\sq}^2\Bigr )+\cw_{\Pi},
\eeq
\bbq
\cw_{\Pi}= {\rm Tr}\Bigl (m_Q\Pi_{11}+m^{\rm tot}_{{\sq}_2}\,\Pi_{21}\frac{1}{\Pi_{11}}\Pi_{12}\Bigr )-
{\rm Tr}\Bigl (\Phi_{11}\Pi_{11}+\Phi_{12}\Pi_{21}+\Phi_{21}\Pi_{12} \Bigr ),
\quad m^{\rm tot}_{{\sq}_2}=(m_Q-\Phi_{22}).
\eeq

In \eqref{(3.21)}: $\oq_2,\, \sq^2$ and $\textsf V$ are the active ${\ov Q}_2, Q^2$ guarks and gluons with unhiggsed colors ($\textsf S$ is their field strength squared), $\Pi_{12}, \Pi_{21}$ are the hybrid pions (in essence, these are the ${\ov Q}_2, Q^2$ guarks with higgsed colors), $z_Q(\la,\mu^2_{\rm gl,1})\gg 1$ is the corresponding perturbative logarithmic renormalization factor of massless quarks, see \eqref{(3.2)}. Evolving now down in the scale and integrating $\oq_2,\, \sq^2$ quarks as heavy ones at $\mu<m^{\rm pole}_{\sq_2}$ and then unhiggsed gluons at $\mu<\lym^{(\rm br1)}$ one obtains the Lagrangian of pions and fions
\bq
K=\Bigl [\frac{1}{f^2}{\rm Tr}(\Phi^\dagger\Phi)+z_Q(\la,\mu^2_{\rm gl,1})K_{\Pi}\Bigr ],\, \label{(3.22)}
\eq
\bbq
W=(N_c-n_1)S+\frac{\mph}{2}\Biggl [{\rm Tr} (\Phi^2) -\frac{1}{\nd}\Bigl ({\rm Tr}\,\Phi\Bigr)^2\Biggr ]+W_{\Pi}\,,
\quad\quad S=\Biggl [\frac{\la^{\bo}\det m^{\rm tot}_{{\sq}_2}}{\det \Pi_{11}}\Biggr ]^{\frac{1}{N_c-n_1}}\,,
\eeq

We start with determining the masses of hybrids $\Pi_{12}, \Pi_{21}$ and $\Phi_{12}, \Phi_{21}$. They are mixed and their kinetic and mass terms look as
\bq
K_{\rm hybr}={\rm Tr}\Bigl [\phi^{\dagger}_{12}\phi_{12}+\phi^{\dagger}_{21}\phi_{21}+\pi^{\dagger}_{12}
\pi_{12}+\pi^{\dagger}_{21}\pi_{21} \Bigr ], \label{(3.23)}
\eq
\bbq
W_{\rm hybr}={\rm Tr}\Bigl (m_{\phi}\phi_{12}\phi_{21}+m_{\pi}\pi_{12}\pi_{21}-m_{\rm \phi\pi}(\phi_{12}\pi_{21}+\phi_{21}\pi_{12})\Bigr )\,,
\eeq
\bbq
m_{\phi}=f^2\mph,\quad m_{\pi}=\frac{m_Q-\langle\Phi_{2}\rangle}{z_Q}=\frac{\langle\Pi_1\rangle}{\mph z_Q}\sim\frac{m_Q}{z_Q}\ll m_{\phi}\,,\quad z_Q=z_Q(\la,\mu_{\rm gl,1})\,,
\eeq
\bq
m_{\rm \phi\pi}=\Bigl (\frac{f^2\langle\Pi_1\rangle}{z_Q}\Bigr )^{1/2},\quad m_{\rm \phi\pi}^2=m_{\phi}m_{\pi}\,. \label{(3.24)}
\eq

Hence, the scalar potential looks as
\bq
V_S=|m|^2\cdot |\Psi^{(-)}_{12}|^2+0\cdot |\Psi^{(+)}_{12}|^2 +(12\rightarrow 21),\quad |m|=(|m_{\phi}|+|m_{\pi}|)\,, \label{(3.25)}
\eq
\bbq
\Psi^{(-)}_{12}=\Bigl (c\,\phi_{12}-s\,\pi_{12} \Bigr ),\quad \Psi^{(+)}_{12}=\Bigl (c\,\pi_{12}+s\,\phi_{12} \Bigr ),\quad c=\Bigl (\frac{|m_{\phi}|}{|m|}\Bigr )^{1/2},\quad s=\Bigl (\frac{|m_{\pi}|}{|m|}\Bigr)^{1/2}\,.
\eeq
Therefore, the fields $\Psi^{(-)}_{12}$ and $\Psi^{(-)}_{21}$ are heavy, with the masses $|m|\approx |m_{\phi}|\gg\la$, while the fields  $\Psi^{(+)}_{12}$ and $\Psi^{(+)}_{21}$ are massless. But the mixing is really parametrically small, so that the heavy fields are mainly $\phi_{12}, \phi_{21}$ while the massless ones are mainly $\pi_{12}, \pi_{21}$.
\footnote{\,
Everywhere below in the text we neglect mixing when it is small.
}

And finally from \eqref{(3.22)}, the pole mass of  pions $\Pi_{11}$ is
\bq
\mu(\Pi_{11})\sim \frac{\langle\Pi_1\rangle}{z_Q(\la,\mu_{\rm gl,1})\mph}\sim\frac{m_Q}{z_Q(\la,\mu_{\rm gl,1})}\,. \label{(3.26)}
\eq

On the whole for this $Higgs_1-HQ_2$ phase the mass spectrum looks as follows at $\mph\gg{\tilde\mu}_{\Phi}$\,. a) The heaviest are $n_1(2N_c-n_1)$ massive gluons and the same number of their scalar superpartners with the masses $\mu_{\rm gl,1}$, see \eqref{(3.16)}, these masses originate from the higgsing of ${\ov Q}_1, Q^1$ quarks. b) There is a large number of 22-flavored hadrons made of weakly interacting and weakly confined non-relativistic $\oq_2, \sq^2$ quarks with unhiggsed colors (the tension of the confining string originated from the unbroken $SU(N_c-n_1)$ color group is $\sqrt\sigma\sim\lym^{(\rm br1)}\ll m^{\rm pole}_{\sq,2}$, see \eqref{(3.18)}, the scale of their masses is $m^{\rm pole}_{\sq,2}\sim m_Q/[z_Q(\la,\mu_{\rm gl,1})z^{\prime}_Q(\mu_{\rm gl,1}, m^{\rm pole}_{\sq,2})]$, where $z_Q\gg 1$ and $z^{\rm \prime}_Q\ll 1$ are the corresponding massless perturbative logarithmic renormalization factors. c) There are $n_1^2$ pions $\Pi_{11}$ with the masses \eqref{(3.26)}, $\mu(\Pi_{11})\ll m^{\rm pole}_{\sq,2}$. d) There is a large number of gluonia made of gluons with unhiggsed colors, the scale of their masses is $\sim\lym^{(\rm br1)}$, see \eqref{(3.18)}. e) The hybrids $\Pi_{12}, \Pi_{21}$ are massless.

All $N^2_F$ fions $\Phi_{ij}$ remain too heavy and dynamically irrelevant (see the footnote 3), their pole masses are $\mu^{\rm pole}_1(\Phi)\sim f^2\mph\gg\mu_{\rm gl,1}$.

\section{Quark and gluino condensates and multiplicities of vacua at $\mathbf{N_c<N_F<2N_c}$}

\hspace{3mm}  To obtain the numerical values of the quark condensates $\langle{\ov Q}_j Q^i\rangle=\delta
^i_j\langle ({\ov Q}Q)\rangle_i$ at $N_c<N_F<2N_c$ (but only for this purpose), the simplest way is to use the known {\it exact form} of the non-perturbative contribution to the superpotential in the standard SQCD with the quark superpotential $m_Q{\rm Tr}({\ov Q}Q)$ and without the fions $\Phi$. It seems clear that at sufficiently large values of $\mph$ among the vacua of the $\Phi$-theory there should be $N_c$ vacua of SQCD in which, definitely, all fions $\Phi$ are too heavy and dynamically irrelevant. Therefore, they all can be integrated out and {\it the exact} superpotential can be written as ($m_Q=m_Q(\mu=\la),\, \mph=\mph(\mu=\la)$, see section 2 above and sections 3 and 7 in \cite{ch1})
\bq
\cw=-\nd\Bigl (\frac{\det {\ov Q}Q}{\la^{\bo}}\Bigr )^{1/\nd}+m_Q{\rm Tr}\,\qq -\frac{1}{2\mph}
\Biggl [{\rm Tr}\, ({\qq})^2- \frac{1}{N_c}({\rm Tr}\,\qq)^2  \Biggr ]\,.\label{(4.1)}
\eq

Indeed, at sufficiently large $\mph$, there are $N_c$ vacuum solutions in \eqref{(4.1)} with the unbroken $SU(N_F)$ flavor symmetry. In these, the last term in \eqref{(4.1)} gives a small correction only and can be neglected and one obtains
\bq
\langle{\ov Q}_j Q^i\rangle_{SQCD}\approx\delta^i_j\frac{1}{m_Q}\Bigl (\lym^{(\rm SQCD)}\Bigr )^3=\delta^i_j\frac{1}{m_Q}\Bigl (\la^{\bo}m_Q^{N_F}\Bigr)^{1/N_c}\,.\label{(4.2)}
\eq

Now, using the holomorphic dependence of the exact superpotential on the chiral superfields $({\ov Q}_j Q^i)$ and the chiral parameters $m_Q$ and $\mph$, the exact form \eqref{(4.1)} can be used to find the values of the quark condensates $\langle{\ov Q}_j Q^i\rangle$ in all other vacua of the $\Phi$ - theory and at all other values of $\mph>\la$. It is worth recalling only that, in general, as in the standard SQCD \cite{ch1,ch3}:  \eqref{(4.1)} {\it is not the superpotential of the genuine low energy Lagrangian describing lightest particles, it determines only the values of the vacuum condensates} $\langle{\ov Q}_j Q^i\rangle$. (The genuine low energy Lagrangians in different vacua will be obtained below in sections 6-11, both in the direct and dual theories).  \\

\subsection{Vacua with the unbroken flavor symmetry}

One obtains from \eqref{(4.1)} that at $\la\ll\mph\ll \mo$ there are two groups of such vacua with parametrically different values of condensates, $\langle{\ov Q}_j Q^i\rangle_L=\delta^i_j\langle{\ov Q}
Q\rangle_L$ and $\langle{\ov Q}_j Q^i\rangle_S=\delta^i_j\langle{\ov Q} Q\rangle_S$.

{\bf a}) There are $(2N_c-N_F)$ L - vacua (L=large, see also the footnote 1) with
\bq
\langle\qq(\mu=\la)\rangle_L\sim \la^2\Biggl (\frac{\la}{\mph}\Biggr )^{\frac{\nd}{2N_c-N_F}}\ll \la^2\,.
\label{(4.3)}
\eq
In these quantum L -vacua the second term in the superpotential \eqref{(4.1)} gives numerically only a small correction.

{\bf b}) There are $(N_F-N_c)$ classical S - vacua (S=small) with
\bq
\langle\qq(\mu=\la)\rangle_S\approx -\frac{N_c}{\nd}\, m_Q\mph\,.\label{(4.4)}
\eq
In these S - vacua, the first non-perturbative term in the superpotential \eqref{(4.1)} gives only small corrections with $Z_{N_F-N_c}$ phases, but just these corrections determine the multiplicity of these $(N_F-N_c)$ nearly degenerate vacua. On the whole, there are
\bq
N_{\rm unbrok}=(2N_c-N_F)+(N_F-N_c)=N_c \label{(4.5)}
\eq
vacua with the unbroken flavor symmetry at $N_c<N_F<2N_c$.\\

One obtains from \eqref{(4.1)} that at $\mph\gg \mo$ the above $(2N_c-N_F)$ L - vacua and $(N_F-N_c)$ S - vacua degenerate into $N_c$ SQCD vacua \eqref{(4.2)}.

The value of $\mo$ is determined from the matching
\bbq
\Biggl [\langle\qq\rangle_L\sim \la^2\Biggl (\frac{\la}{\mo}\Biggr )^{\frac{\nd}{2N_c-N_F}}\Biggr ]\sim \Biggl [\langle\qq\rangle_S\sim m_Q\mo\Biggl ]\sim \Biggl [\langle\qq\rangle_{\rm SQCD}\sim \la^2\Bigl (\frac{m_Q}{\la}\Bigr )^{\frac{\nd}{N_c}}\Biggl ]\quad\ra
\eeq
\bq
\ra \mo\sim \la\Bigl (\frac{\la}{m_Q}\Bigr )^{\frac{2N_c-N_F}{N_c}}\gg \la\,.\label{(4.6)}
\eq

\subsection{Vacua with the spontaneously broken flavor symmetry}

In these, there are $n_1$ equal condensates $\langle{\ov Q}_1Q^1(\mu=\la)\rangle\equiv\langle\Qo
\rangle$ and $n_2\geq n_1$ equal condensates $\langle{\ov Q}_2 Q^2(\mu=\la)\rangle\equiv\langle
\Qt\rangle\neq\langle\Qo\rangle$. The simplest way to find the values of quark condensates in these vacua is to use \eqref{(2.5)}. We rewrite it here for convenience
\bbq
\langle\Qo+\Qt-\frac{1}{N_c}{\rm Tr}\,\qq\rangle_{\rm br}=m_Q\mph\,,
\eeq
\bq
\langle S\rangle_{\rm br}=\Bigl
(\frac{\det \langle\qq\rangle_{\rm br}=\langle\Qo\rangle^{\rm {n}_1}_{\rm br}\langle\Qt\rangle^{\rm {n}_2}_{\rm br}}{\la^{\bo}}\Bigr )^{1/\nd}=\frac{\langle\Qo\rangle_{\rm br}\langle\Qt\rangle_{\rm br}}{\mph}\,.\label{(4.7)}
\eq
Besides, the multiplicity of vacua will be shown below at given values of $n_1$ and $n_2\geq n_1$.\\

{\bf 4.2.1} The region $\la\ll\mph\ll\mo$.\\

{\bf a)} At $n_2\lessgtr N_c$, including $n_1=n_2=N_F/2$ for even $N_F$ but excluding $n_2=N_c$\,, there are $(2N_c-N_F){\ov C}^{\,n_1}_{N_F}$ Lt - vacua (Lt=L -type) with the parametric behavior of condensates (see the footnote 1)
\bq
(1-\frac{n_1}{N_c})\langle\Qo\rangle_{\rm Lt}\approx -(1-\frac{n_2}{N_c})\langle\Qt\rangle_{\rm Lt}\sim \la^2\Biggl (\frac{\la}{\mph}\Biggr )^{\frac{\nd}{2N_c-N_F}},\label{(4.8)}
\eq
i.e. as in the L - vacua above but $\langle\Qo\rangle_{\rm Lt}\neq\langle\Qt\rangle_{\rm Lt}$ here.

{\bf b)} At $n_2>N_c$ there are $(n_2-N_c)C^{n_1}_{N_F}$ $\rm br2$ - vacua (br2=breaking-2) with, see \eqref{(4.7)},
\bq
\langle\Qt\rangle_{\rm br2}\sim m_Q\mph\,,\quad \langle\Qo\rangle_{\rm br2}\sim \la^2\Bigl (\frac{\mph}{\la}\Bigr )^{\frac{n_2}{n_2-N_c}}\Bigl (\frac{m_Q}{\la}\Bigr )^{\frac{N_c-n_1}{n_2-N_c}},\quad
\frac{\langle\Qo\rangle_{\rm br2}}{\langle\Qt\rangle_{\rm br2}}\sim \Bigl (\frac{\mph}{\mo}\Bigr )^{\frac{N_c}{n_2-N_c}}\ll 1.\,\,\label{(4.9)}
\eq

{\bf c)} At $n_1=\nd,\, n_2=N_c$ there are $(2N_c-N_F)\cdot C^{n_1=\nd}_{N_F}$ 'special' vacua with, see \eqref{(4.7)},
\bq
\langle\Qo\rangle_{\rm spec}=\frac{N_c}{2N_c-N_F}(m_Q\mph)\,,\quad \langle\Qt\rangle_{\rm spec}\sim \la^2\Bigl (\frac{\la}{\mph}\Bigr )^{\frac{\nd}{2N_c-N_F}},\,\,\label{(4.10)}
\eq
\bbq
\frac{\langle\Qo\rangle_{\rm spec}}{\langle\Qt\rangle_{\rm spec}}\sim\Bigl (\frac{\mph}{\mo}\Bigr )^{\frac{N_c}{2N_c-N_F}}\ll 1\,.
\eeq

On the whole, there are (\,$\theta(z)$ is the step function\,)
\bq
N_{\rm brok}(n_1)=\Bigl [(2N_c-N_F)+\theta(n_2-N_c)(n_2-N_c)\Bigr ]{\ov C}^{\,n_1}_{N_F}=\label{(4.11)}
\eq
\bbq
=\Bigl [(N_c-\nd)+\theta(\nd-n_1)(\nd-n_1)\Bigr ]{\ov C}^{\,n_1}_{N_F}\,,
\eeq
( ${\ov C}^{\,n_1}_{N_F}$ differ from the standard $C^{\,n_1}_{N_F}$ only by ${\ov C}^{\,n_1={\rm k}}_{N_F=2{\rm k}}=C^{\,n_1={\rm k}}_{N_F=2{\rm k}}/2$, see \eqref{(3.13)}\, ) vacua with the broken flavor symmetry $U(N_F)\ra U(n_1)\times U(n_2)$, this agrees with \cite{CKM} (see also the related paper \cite{GK}, but the superpotential in \cite{GK} is somewhat different and this difference is crucial for the special vacua, see the Appendix B).\\

{\bf 4.2.2} The region $\mph\gg\mo$.\\

{\bf a)} At all values of $n_2\lessgtr N_c$, including $n_1=n_2=N_F/2$ at even $N_F$ and the `special' vacua with $n_1=\nd,\, n_2=N_c$, there are $(N_c-n_1){\ov C}^{\,n_1}_{N_F}$ $\rm br1$ - vacua (br1=breaking-1) with, see \eqref{(4.7)},
\bq
\langle\Qo\rangle_{\rm br1}\sim m_Q\mph\,,\quad \langle\Qt\rangle_{\rm br1}\sim \la^2\Bigl (\frac{\la}{\mph}\Bigr )^{\frac{n_1}{N_c-n_1}}\Bigl (\frac{\la}{m_Q}\Bigr )^{\frac{N_c-n_2}{N_c-n_1}}\,,\label{(4.12)}
\eq
\bbq
\frac{\langle\Qt\rangle_{\rm br1}}{\langle\Qo\rangle_{\rm br1}}\sim \Bigl (\frac{\mo}{\mph}\Bigr )^{\frac{N_c}{N_c-n_1}}\ll 1\,.
\eeq

{\bf b)} At $n_2<N_c$, including $n_1=n_2=N_F/2$, there are also $(N_c-n_2){\ov C}^{\,n_2}_{N_F}=(N_c-n_2){\ov C}^{\,n_1}_{N_F}$\,\, $\rm br2$ - vacua with, see \eqref{(4.7)},
\bq
\langle\Qt\rangle_{\rm br2}\sim m_Q\mph\,,\quad \langle\Qo\rangle_{\rm br2}\sim \la^2\Bigl (\frac{\la}{\mph}\Bigr )^{\frac{n_2}{N_c-n_2}}\Bigl (\frac{\la}{m_Q}\Bigr )^{\frac{N_c-n_1}{N_c-n_2}}\,,\label{(4.13)}
\eq
\bbq
\frac{\langle\Qo\rangle_{\rm br2}}{\langle\Qt\rangle_{\rm br2}}\sim \Bigl (\frac{\mo}{\mph}\Bigr )^{\frac{N_c}{N_c-n_2}}\ll 1\,.
\eeq

On the whole, there are
\bq
N_{\rm brok}(n_1)=\Bigl [(N_c-n_1)+\theta (N_c-n_2)(N_c-n_2)\Bigr ]{\ov C}^{\,n_1}_{N_F}= \label{(4.14)}
\eq
\bbq
=\Bigl [(N_c-\nd)+\theta (\nd-n_1)(\nd-n_1)\Bigr ]{\ov C}^{\,n_1}_{N_F}
\eeq
vacua. As it should be, the number of vacua at $\mph\lessgtr \mo$ is the same.\\

As one can see from the above, all quark condensates become parametrically the same at $\mph\sim\mo$. Clearly, this region $\mph\sim\mo$ is very special and most of the quark condensates change their parametric behavior and hierarchies at $\mph\lessgtr\mo$. For example, the br2 - vacua with $n_2<N_c\,,\,\,\langle\Qt \rangle\sim m_Q\mph\gg\langle\Qo\rangle$ at $\mph\gg\mo$ evolve into the L - type vacua with $\langle
\Qt\rangle\sim\langle\Qo\rangle\sim \la^2 (\la/\mph)^{\nd/(2N_c-N_F)}$ at $\mph\ll\mo$, while the br2 - vacua with $n_2>N_c\,,\,\,\langle\Qt\rangle\sim m_Q\mph\gg\langle\Qo\rangle$ at $\mph\ll\mo$ evolve into the br1 - vacua with $\langle\Qo\rangle\sim m_Q\mph\gg\langle\Qt\rangle$ at $\mph\gg\mo$, etc. The exception is the special vacua with $n_1=\nd,\, n_2=N_c$\,. In these, the parametric behavior $\langle\Qo\rangle\sim m_Q\mph, \,\langle\Qt\rangle\sim \la^2(\la/\mph)^{\nd/(2N_c-N_F)}$ remains the same but the hierarchy is reversed at $\mph\lessgtr\mo\, :\, \langle\Qo\rangle/\langle\Qt\rangle\sim (\mph/\mo)^{N_c/(2N_c-N_F)}$.\\

The total number of all vacua at $N_c<N_F<2N_c$ is
\bq
N_{\rm tot}=\Bigl ( N_{\rm unbrok}=N_c \Bigr )+\Bigl ( N_{\rm brok}^{\rm tot}=\sum_{n_1=1}^{[N_F/2]}N_{\rm brok}(n_1)
\Bigr )=\sum_{k=0}^{N_c}(N_c-k)C^{\,k}_{N_F}\,, \label{(4.15)}
\eq
this agrees with \cite{CKM}\,.
\footnote{\,
But we disagree with their `derivation' in section 4.3. There is no their ${\cal N}_2$ vacua with $\langle M^i_i\rangle\langle{\ov q}^i q_i\rangle/\la=\langle S\rangle=0,\,\, i=1,...N_F$ (no summation over $i$) in the $SU(N_c)$ dual theory at $m_Q\neq 0$. In all $N_{\rm tot}$ vacua in both direct and dual theories\,:\, $\langle\det M/\la^{\bo}\rangle^{1/\nd}=\langle\det {\ov Q}Q/\la^{\bo}\rangle^{1/\nd}=\langle S\rangle\neq 0$ at $m_Q\neq 0$ (see sections 6-11 below and the Appendix B). Really, the superpotential (4.48) in \cite{CKM} contains all $N_{\rm tot}={\cal N}_1+{\cal N}_2$ vacua.
}

Comparing this with the number of vacua \eqref{(3.13)},\eqref{(3.14)} at $N_F<N_c$ it is seen that, for both $N_{\rm unbrok}$ and $N_{\rm brok}^{\rm tot}$ separately, the multiplicities of vacua at $N_F<N_c$ and $N_F>N_c$  are not analytic continuations of each other.\\

The analog of \eqref{(4.1)} in the dual theory with $|\Lambda_q|=\la$, see \eqref{(2.7)}, is obtained by the replacement ${\ov Q} Q(\mu=\la)\ra M(\mu=\la)$, so that $\langle M(\mu=\la)\rangle=\langle {\ov Q} Q(\mu=\la)\rangle$ in all vacua and multiplicities of vacua are the same.

\section{Fions $\mathbf{\Phi}$ in the direct theory\,: one or three generations}

\hspace{3mm} At $N_c<N_F<2N_c$ and in the interval of scales $\mu_H<\mu<\la$ ( $\mu_H$ is the largest physical mass in the quark-gluon sector), the quark and gluon fields are effectively massless. Because the quark renormalization factor $z_Q(\la,\mu\ll\la)=(\mu/\la)^{\gamma_Q>0}\ll 1$ decreases in this case in a {\it power fashion} with  lowering energy due to the perturbative RG evolution, it is seen  from \eqref{(2.3)} that the role of the 4-quark term $({\ov Q}Q)^2/\mph$ increases with lowering energy. Hence, while it is irrelevant at the scale $\mu\sim\la$ because $\mph\gg \la$, the question is whether it becomes dynamically relevant in the range of energies $\mu_H\ll\mu\ll \la$. For this, we estimate the scale $\mu_o$ where it becomes relevant in the massless theory (see section 7 in \cite{ch1} for the perturbative strong coupling regime with $a(\mu\sim\la)\sim 1,\, a(\mu\ll\la)\sim (\la/\mu)^{\nu\,> 0}\gg 1$ at $N_c<N_F<3N_c/2$\,)
\bq
\frac{\mu_o}{\mph}\frac{1}{z^2_Q(\la,\mu_o)}=\frac{\mu_o}{\mph}\Bigl (\frac{\la}{\mu_o}\Bigr )^{2\gamma_Q}\sim 1\quad\ra \quad \frac{\mu_o}{\la}\sim \Bigl (\frac{\la}{\mph}\Bigr )^{\frac{1}{(2\gamma_Q-1)}}\,\,, \label{(5.1)}
\eq
\bbq
\gamma^{\rm conf}_Q=\frac{\bo}{N_F}\,\ra\,\frac{\mu^{\rm conf}_o}{\la}\sim \Bigl (\frac{\la}{\mph}\Bigr )^{\frac{N_F}{3(2N_c-N_F)}}\,\,,  \quad \gamma^{\rm strong}_Q=\frac{2N_c-N_F}{\nd}\,\ra\,
\frac{\mu^{\rm strong}_o}{\la}\sim \Bigl (\frac{\la}{\mph}\Bigr )^{\frac{\nd}{(5N_c-3N_F)}}\,\,.
\eeq

Hence, if $\mu_H\ll\mu_o$, then at scales $\mu<\mu_o$ the four-quark terms in the superpotential \eqref{(2.3)} cannot be neglected any more and we have to account for them. For this, we have to reinstate the fion fields $\Phi$ and to use the Lagrangian \eqref{(2.1)} in which the Kahler term at $\mu_H<\mu\ll\la$ looks as
\bq
K=\Bigl [\frac{z_{\Phi}(\la,\mu)}{f^2}{\rm Tr}\,(\Phi^\dagger \Phi)+z_Q(\la,\mu){\rm Tr}\Bigl (Q^\dagger Q+(Q\ra {\ov Q})\Bigr )\Bigr ],\,\, z_Q(\la,\mu)=\Bigl (\frac{\mu}{\la}\Bigr )^{\gamma_Q}\ll 1. \label{(5.2)}
\eq

We recall that even at those scales $\mu$ that the running perturbative mass of fions $\mu_{\Phi}(\mu)\equiv\mph/f^2 z_{\Phi}(\la,\mu)\gg \mu$ and so they are too heavy and dynamically irrelevant, the quarks and gluons remain effectively massless and active. Therefore, due to the Yukawa interactions of fions with quarks, the loops  of still active light quarks (and gluons interacting with quarks) still induce the running renormalization factor $z_{\Phi}(\la,\mu)$ of fions at all those scales until quarks are effectively massless, $\mu>\mu_H$. But, in contrast with a very slow logarithmic RG evolution at $N_F<N_c$ in section 3, the perturbative running mass of fions decreases now at $N_c<N_F<2N_c$ and $\mu<\la$ monotonically and {\it very quickly} with diminishing scale (see below), $\mph(\mu\ll \la)=\mph/f^2 z_\Phi(\la,\mu)\sim\mph(\mu/\la)^{|\gamma_{\Phi}|>1}\ll \mph$. Nevertheless, until $\mph(\mu)\gg \mu$, the fields $\Phi$ remain heavy and do not influence the RG  evolution. But,  when $\mu_H\ll\mu_o$ and $\mph(\mu)\sim\mph/z_{\Phi}(\la,\mu)$ is the main contribution to the fion mass
\footnote{\,
the cases when the additional contributions to the masses of fions from other perturbative or non-perturbative terms in the superpotential are not small in comparison with $\sim\mph/z_{\Phi}(\la,\mu)$ have to be considered separately
}
,
the quickly decreasing mass $\mph(\mu)$ becomes $\mu^{\rm pole}_2(\Phi)=\mph(\mu=\mu^{\rm pole}_2(\Phi))$ and $\mph(\mu<\mu^{\rm pole}_2(\Phi))< \mu$, so that\,: 1) there is a pole in the fion propagator at $p=\mu^{\rm pole}_2(\Phi)$ (ignoring here and below a nonzero fion width, in any case the nonzero width can have only massive particle), this is a second generation of fions (the first one is at $\mu^{\rm pole}_1(\Phi)\gg\la$, see Appendix A)\,; 2) the fields $\Phi$ {\it become effectively massless at $\mu<\mu^{\rm pole}_2(\Phi)$ and begin to influence the perturbative RG evolution}. In other words, the seemingly `heavy' fields $\Phi$ {\it return back}, they become effectively massless and dynamically {\it relevant}. Here and below the terms `relevant' and `irrelevant' (at a given scale $\mu$\,) will be used in the sense of whether the running mass $\sim\mph/z_{\Phi}(\la,\mu\ll\la)$ of fions at a given scale $\mu$ is $<\mu$, so that they are effectively massless and participate actively in interactions at this scale\,, or they remain too heavy with the running mass $>\mu$ whose interactions at this scale give only small corrections.

It seems clear that {\it the physical reason why the $4$-quark terms in the superpotential \eqref{(2.3)} become relevant at scales $\mu<\mu_o$ is that the fion field $\Phi$ which was too heavy and so dynamically irrelevant at $\mu>\mu_o,\, \mph(\mu>\mu_o)>\mu$\,, becomes effectively massless at $\mu<\mu_o,\, \mph(\mu<\mu_o)<\mu$\,, and begins to participate actively in the RG evolution, i.e. it becomes relevant}. In other words, the four quark term in \eqref{(2.3)} 'remembers' about fions and signals about the scale below which the fions become effectively massless, $\mu_o=\mu^{\rm pole}_2(\Phi)$. This allows us to find the value of $z_{\Phi}(\la,\mu_o)$,
\bbq
\frac{f^2\mph}{z_{\Phi}(\la,\mu_o)}\approx\mu_o\,,\quad z_{\Phi}(\la,\mu_o<\mu\ll\la)=1+f^2\Biggl [
\Bigl (\frac{\mu}{\la}\Bigr )^{\gamma_{\Phi}<\,0}-1\Biggr ]\approx
\eeq
\bq
\approx f^2\Bigl (\frac{\la}{\mu}\Bigr )^{2\gamma_Q>\,
0}\gg 1\,,\quad \gamma_{\Phi}=-2\gamma_Q<0\,. \label{(5.3)}
\eq

The perturbative running mass $\mph(\mu)\sim\mph/z_{\Phi}(\la,\mu\ll\la)\ll\mph$ of fions continues to decrease strongly with diminishing $\mu$ at all scales $\mu_H<\mu<\la$ until quarks remain effectively massless, and becomes frozen only at scales below the quark physical mass, when the heavy quarks decouple.

Hence, if $\mu_H\gg\mu_o$\,, there is no pole in the fion propagator at momenta $p<\la$ because the running fion mass is too large in this range of scales, $\mph(p>\mu_o)>p$. The fions remain dynamically irrelevant in this case at all momenta $p<\la$.

But when $\mu_H\ll\mu_o$, {\it there will be not only the second generation of fions at $p=\mu^{\rm pole}_2(\Phi)=\mu_o$ but also a third generation at $p\ll\mu_o$}. Indeed, after the heavy quarks decouple at momenta $p<\mu_H\ll\mu_o$ and the renormalization factor $z_{\Phi}(\la,\mu)$ of fions becomes frozen in the region of scales where the fions already became relevant, $z_{\Phi}(\la,\mu<\mu_H)\sim z_{\Phi}(\la,\mu\sim
\mu_H)$, the frozen value $\mph(\mu<\mu_H)$ of the running perturbative fion mass is now $\mph(\mu\sim\mu_H)\ll p_H=\mu_H$. Hence, {there is one more pole in the fion propagator} at $p=\mu^{\rm pole}_3(\Phi)\sim \mph(\mu\sim\mu_H)\ll \mu_H$.

On the whole, in a few words for the direct theory (see the footnote 6 for reservations).\\
{\bf a)} The fions remain dynamically irrelevant and there are no poles in the fion propagator at momenta $p<\la$ if $\mu_H\gg\mu_o$.\\
{\bf b)} If $\mu_H\ll\mu_o\ll\la$, there are two poles in the fion propagator at momenta $p\ll\la$\,:\, $\mu^{\rm pole}_2(\Phi)\sim \mu_o$ and $\mu^{\rm pole}_3(\Phi)\sim \mph/z_{\Phi}(\la,\mu_H)\ll\mu^{\rm pole}_2(\Phi)$ (here and everywhere below in similar cases, - up to corrections due to possible nonzero decay widths of fions). In other words, the fions appear in three generations in this case (we recall that there is always the largest pole mass of fions $\mu^{\rm pole}_1(\Phi)\gg\la$, see the appendix A). Hence, the fions are effectively massless and dynamically relevant in the range of scales $\mu^{\rm pole}_3(\Phi)<\mu<\mu^{\rm pole}_2(\Phi)$.

Moreover, once the fions become effectively massless and dynamically relevant with respect to internal interactions, they begin to contribute simultaneously to the external anomalies ( the 't Hooft triangles in the external background fields).

The case $\mu_H\sim\mu_o$ requires additional information. The reason is that at scales $\mu\lesssim\mu_H$, in addition to the canonical kinetic term $\Phi^{\dagger}_R p^2\Phi_R$ (R=renormalized) of fions, there are also terms $\sim \Phi^{\dagger}_R p^2(p^2/\mu_H^2)^k\Phi_R$ with higher powers of momenta induced by loops of heavy quarks (and gluons). If $\mu_H\ll\mu_o$, then the pole in the fion propagator at $p=\mu^{\rm pole}_2(\Phi)=\mu_o$ is definitely there and, because $\mph(\mu=\mu_H)\ll\mu_H$, these additional terms are irrelevant in the region $p\sim\mph(\mu=\mu_H)\ll\mu_H$ and the pole in the fion propagator at $p=\mu^{\rm pole}_3(\Phi)=\mph(\mu=\mu_H)\ll\mu_H$ is also guaranteed.  But $\mph(\mu\sim\mu_H)\sim\mu_H$ if $\mu_H\sim\mu_o$, and these additional terms become relevant. Hence, whether there is a pole in the fion propagator in this case or not depends on all these terms.\\

Now, if $\mu_H<\mu_o$ so that the fions become relevant at $\mu<\mu_o$, the question is\,: what are the values of the quark and fion anomalous dimensions, $\gamma_Q$ and $\gamma_\Phi$, in the massless perturbative regime at $\mu_H<\mu<\mu_o$\,?

To answer this question, we use the approach used in \cite{ch1} (see section 7). For this, we introduce first the corresponding massless Seiberg dual theory \cite{S2}. Our direct theory includes at $\mu_H<\mu<\mu^{\rm conf}_o$ not only the original effectively massless in this range of scales quark and gluon fields, but also the active $N_F^2$ fion fields $\Phi^j_i$ as they became now also effectively massless, so that the effective superpotential becomes nonzero and includes the Yukawa term ${\rm Tr}\,({\ov Q}\Phi Q)$. Then, the massless dual theory with the same 't Hooft triangles includes only the massless qual quarks ${\ov q},\, q$ with $N_F$ flavors and the dual $SU(\nd=N_F-N_c)$ gluons. Further, one equates two NSVZ $\,{\widehat\beta}_{ext}$ - functions of the external baryon and $SU(N_F)_{L,R}$ - flavor vector fields in the direct and dual theories,
\bq
\frac{d}{d\,\ln \mu}\,\frac{2\pi}{\alpha_{ext}}={\widehat\beta}_{ext}= -\frac{2\pi}{\alpha^2_{ext}}\,\beta_{ext}= \sum_i T_i\,\bigl (1+\gamma_i\bigr )\,,\label{(5.4)}
\eq
where the sum runs over all fields which are effectively massless at scales $\mu_H<\mu<\mu_o$, the unity in the brackets is due to one-loop contributions while the anomalous dimensions $\gamma_i$  of fields represent all higher-loop effects, $T_i$ are the coefficients. It is worth noting that these general NSVZ forms \eqref{(5.4)} of the external 'flavored' $\widehat\beta$-functions are independent of the kind of massless perturbative regime of the internal gauge theory, i.e. whether it is conformal, or the strong coupling or the IR free one.

The effectively massless particles in the direct theory here are the original quarks $Q,\,{\ov Q}$ and gluons and, in addition, the fions $\Phi^j_i$, while in the dual theory these are the dual quarks $q,\, {\ov q}$ and dual gluons only.

It is clear that, in comparison with the standard SQCD without the fion fields (see section 7 in \cite{ch1}), the addition of the fion fields with zero baryon charge does not influence ${\widehat\beta}_{ext}$ for the baryon charge, so that in the whole interval $\mu_H<\mu<\la$ it remains the same as in \cite{ch1}
\bq
N_F N_c\,\Bigl ( B_Q=1 \Bigr )^2\,(1+\gamma_Q)=N_F \nd \,\Bigl ( B_q=\frac{N_c}{\nd}
\Bigr )^2\,(1+\gamma_q)\,. \label{(5.5)}
\eq

The form of \eqref{(5.4)} for the $SU(N_F)_L$  flavor charge at scales $\mu_H<\mu<\mu_o$ where the fion fields became effectively massless and relevant differs from those in \cite{ch1}, now it looks as
\bq
N_c\,(1+\gamma_Q)+N_F\,(1+\gamma_\Phi)=\nd \,(1+\gamma_q)\,.\label{(5.6)}
\eq
In \eqref{(5.5)},\eqref{(5.6)}: the left-hand sides are from the direct theory while the right-hand sides are from the dual one, $\gamma_Q$ and $\gamma_\Phi$ are the anomalous dimensions of the quark $Q$ and fion $\Phi$\,, while $\gamma_q$ is the anomalous dimension of the dual quark $q$.

The massless dual theory is in the conformal regime at $3N_c/2<N_F<2N_c$\,, so that $\gamma^{\rm conf}_q=\rm{{\ov b}_o}/N_F=(3\nd-N_F)/N_F$. Therefore, one obtains from \eqref{(5.5)},\eqref{(5.6)} that $\gamma^{\rm conf}_Q=\bo/N_F=(3N_c-N_F)/N_F$ and $\gamma^{\rm conf}_\Phi=-2\gamma^{\rm conf}_Q$, i.e. while only the quark-gluon sector of the direct theory behaves conformally at scales $\mu^{\rm conf}_o<\mu< \la$ where the fion fields $\Phi$ remain heavy and irrelevant, the whole theory including the fields $\Phi$ becomes conformal at scales $\mu_H<\mu< \mu^{\rm conf}_o$ where fions become effectively massless and relevant.
\footnote{\,
This does not mean that nothing changes at all after the fion field $\Phi$ begins to participate actively in the perturbative RG evolution at $\mu_H<\mu<\mu^{\rm conf}_o$. In particular, the frozen fixed point values of the gauge and Yukawa couplings $a^*$ and $a_{f}^*$ will change.
}

In the region $N_c<N_F<3N_c/2$ the situation with \eqref{(5.5)},\eqref{(5.6)} is somewhat different. The massless direct theory is now in the strong gauge coupling regime starting from $\mu<\la$,\, $a(\mu\ll\la)\sim (\la/\mu)^{\nu\,>\,0}\gg 1$, see section 7 in \cite{ch1}, while the massless dual theory is in the IR free logarithmic regime. Therefore, $\gamma_q$ is logarithmically small at $\mu\ll\la,\, \gamma_q\ra 0$, and one obtains in this case from \eqref{(5.5)} for the baryon charge the same value of $\gamma^{\rm strong}_Q(\mu_H\ll\mu\ll\la)$ as in \cite{ch1}
\bq
\gamma^{\rm strong}_Q(\mu_H\ll\mu\ll\la)=\frac{2N_c-N_F}{\nd}\,,\quad a(\mu_H\ll\mu\ll\la)\sim (\la/\mu)^{\nu}\gg 1\,, \label{(5.7)}
\eq
\bq
\quad \nu=\frac{N_F}{N_c}\gamma^{\rm strong}_Q-3=\frac{3N_c-2N_F}{\nd}>0\,. \label{(5.8)}
\eq
In other words, the value of the quark anomalous dimension $\gamma^{\rm strong}_Q(\mu_H\ll\mu\ll\la)$ in the $\Phi$-theory is the same as in the standard SQCD, independently of whether the field $\Phi$ is relevant or not.

The value of $\gamma_{\Phi}$ at $\mu_H\ll\mu\ll\mu^{\rm strong}_o$ obtained from \eqref{(5.6)} will be $\gamma^{\rm strong}_{\Phi}=-(1+\gamma^{\rm strong}_Q)=-N_c/\nd$. But we know from the standard SQCD that the corresponding analog of \eqref{(5.6)} for the flavor charge is not fulfilled in the region $N_c<N_F<3N_c/2$, see section 7 in \cite{ch1}). Therefore, we will not use \eqref{(5.6)} in this region of $N_F/N_c$ in the $\Phi$-theory also. Instead, we will present now other arguments about the value of $\gamma^{\rm strong}_{\Phi}$ in the $\Phi$-theory at $N_c<N_F<3N_c/2$ and $\mu_H<\mu<\mu_o^{\rm strong}$ when the field $\Phi$ already became effectively massless.

First, we point out that the gauge coupling $a(\mu)$ entered already into a strong coupling regime in the range of scales $\mu_o^{\rm strong}<\mu<\la,\,\,\, \mu_o^{\rm strong}\sim\la(\la/\mph)^{\nd/(5N_c-3N_F)}\ll\la$, so that $a(\mu\sim\la)\sim 1$ while  $a(\mu_o^{\rm strong})\sim (\la/\mu_o^{\rm strong})^{\nu\,>\, 0}\gg 1$. At the same time the Yukawa coupling $a_f(\mu)\sim f^2/z_{\Phi}(\la,\mu)z_Q(\la,\mu)\sim (\la/\mu)^{2\gamma_Q+\gamma_{\Phi}}$ of the field $\Phi$ stays intact, $a_f(\mu\sim\la)\sim a_f(\mu\sim\mu_o^{\rm strong})\sim 1$, because $\gamma^{\rm strong}_{\Phi}=-2\gamma^{\rm strong}_{Q}$ at $\mu_o^{\rm strong}<\mu<\la$.

Consider now the Feynman diagrams contributing to the renormalization factors $z_{\Phi}(\mu)$ and $z_{Q}(\mu)$ at $\mu_H\ll\mu\ll\mu_o^{\rm strong}$. Order by order in the perturbation theory the extra loop with the exchange of the field $\Phi$ is $a_f(\mu)/a(\mu)\sim (\mu/\la)^{\nu\,>\,0}\ll 1$ times smaller than the extra loop but with the exchange of gluon and can be neglected. In effect, the field $\Phi$ in such a situation plays a role of the "external"\, background field which is "weakly coupled"\, in comparison with the "internal" very strong quark-gluon interactions. Therefore, the fact that the field $\Phi$ became effectively massless and formally relevant at $\mu<\mu_o^{\rm strong}$ is really of no importance for the RG-evolution, so that both $\gamma^{\rm strong}_{Q}$ and $\gamma^{\rm strong}_{\Phi}=-2\gamma^{\rm strong}_{Q}$ remain the same at $\mu\gtrless\mu_o^{\rm strong}$ (i.e. the Yukawa coupling $a_f(\mu)$ still stays at $a_f(\mu)\sim 1$ at $\mu_H<\mu<\mu_o^{\rm strong}$). As for $\gamma^{\rm strong}_{Q}$, this agrees with the fact that \eqref{(5.5)} remains the same at $\mu_o^{\rm strong}<\mu<\la$ and at $\mu_H<\mu<\mu_o^{\rm strong}$.\\

On the whole, according to the above considerations, the values of $\gamma^{\rm strong}_Q(\mu)$ and $\gamma^
{\rm strong}_{\Phi}(\mu)$ in the $\Phi$-theory are
\bq
\gamma^{\rm strong}_Q(\mu)=\frac{2N_c-N_F}{\nd}\,,\quad \gamma^{\rm strong}_{\Phi}(\mu)=-2\gamma^{\rm strong}_Q(\mu),\quad \mu_H<\mu<\la \label{(5.9)}
\eq
in the strong gauge coupling regime $a(\mu)\gg 1$ at $N_c<N_f<3N_c/2$ and in the whole range of scales $\mu_H<\mu<\la$ if $\mu_H<\mu_o^{\rm strong}$. If the largest mass $\mu_H$ in the quark-gluon sector is such that $\mu_o^{\rm strong}\ll\mu_H\ll\la$, then the form of the RG-evolution is those in \eqref{(5.9)} at $\mu_H<\mu<\la$ and changes at $\mu<\mu_H$. \\

In the rest of this paper the mass spectra of the direct and dual theories will be considered within the conformal window $3N_c/2<N_F<2N_c$ only.
\vspace*{6mm}

\addcontentsline{toc}{section}
 {\hspace*{3cm} Mass spectra at $\mathbf{\Large 3 N_c/2<N_F<2 N_c}$}

{ \bf\LARGE \hspace*{2cm} Mass spectra at $\mathbf{\Large 3 N_c/2<N_F<2 N_c}$\\
\hspace*{4cm}}
\vspace*{4mm}

Let us recall that, within the dynamical scenario used in this paper for the strong coupling regimes with the gauge coupling $a\sim 1$, the quarks can be  either in the HQ (heavy quark) phase where they are confined, or they are higgsed at the appropriate conditions. Besides, it is implied that no `unexpected' parametrically lighter particles (e.g. magnetic monopoles or dyons) are formed in ${\cal N}=1$ theories without colored adjoint superfields considered below in sections 6-11.

\section{Direct theory. Unbroken flavor symmetry}

\subsection{\quad  L - vacua}

\hspace*{4mm} The theory enters the conformal regime as the scale is decreased below $\la$.
In these $(2N_c-N_F)$ vacua with the unbroken flavor symmetry $U(N_F)$ the current quark mass at $\la\ll\mph\ll\mo$ looks as, see \eqref{(4.3)},\eqref{(6.2)}
\bq
\langle m^{\rm tot}_Q\rangle_L\equiv\langle m^{\rm tot}_Q(\mu=\la)\rangle_{L}=m_Q-\langle\Phi\rangle_L= m_Q+\frac{\nd}{N_c}\frac{\qql}{\mph}\,, \label{(6.1)}
\eq
\bbq
\qql\sim\la^2\Bigl (\frac{\la}{\mph}\Bigr )^{\frac{\nd}{2N_c-N_F}}\gg m_Q\mph,\quad
\langle m^{\rm tot}_Q\rangle_L\sim \la\Bigl (\frac{\la}{\mph}\Bigr )^{\frac{N_c}{2N_c-N_F}}\,,
\eeq
\bbq
\mql=\frac{\langle m^{\rm tot}_Q\rangle_{L}}{z_Q(\la,m^{\rm pole}_{Q,\,L})}\sim \la \Bigl (\frac{\la}{\mph}\Bigr )^{\frac{N_F}{3(2N_c-N_F)}}\sim\lym^{(L)}\,,
\quad z_Q(\la,\mu\ll\la)\sim\Bigl (\frac{\mu}{\la}\Bigr )^{\bo/N_F}\ll 1\,.
\eeq
We compare $\mql$ with the gluon mass due to possible higgsing of quarks. This last looks as
\bq
\mgl^2\sim \Bigl (a_{*}\sim 1\Bigr ) z_Q(\la,\mgl)\qql \quad \ra \quad \mgl\sim \mql\sim\lym^{(L)}=\langle S\rangle^{1/3}_L\,. \label{(6.2)}
\eq
Hence, qualitatively, the situation is the same as in the standard SQCD \cite{ch3}. And one can use here the same reasonings, see the footnote 3 in \cite{ch3}. In the case considered, there are only $(2N_c-N_F)$ these isolated L vacua with {\it unbroken flavor symmetry}. If quarks were higgsed in these L vacua, then {\it the flavor symmetry will be necessary broken spontaneously} due to the rank restriction because $N_F>N_c$ and there will appear the genuine exactly massless Nambu-Goldstone fields $\Pi$ (pions), so that there will be a continuous family of non-isolated vacua. This is "the standard point of tension" in the dynamical scenario $\#2$, see \cite{ch3}. Therefore, as in \cite{ch3}, assuming here and everywhere below in similar situations that this scenario $\#2$ is self-consistent, we conclude that $\mg=\mql/(\rm several)$, so that quarks are not higgsed but are in the HQ (heavy quark) phase and are confined.

Therefore (see sections 3, 4 in \cite{ch3}), after integrating out all quarks as heavy ones at $\mu<\mql$ and then all $SU(N_c)$ gluons at $\mu<\lym^{(L)}=\mql/(\rm several)$ via the Veneziano-Yankielowicz (VY) procedure \cite{VY},  we obtain the Lagrangian of fions
\bq
K=z_{\Phi}(\la,\mql){\rm Tr}\,(\Phi^\dagger\Phi)\,, \quad z_{\Phi}(\la,\mql)\sim \frac{1}{z^2_Q(\la,\mql)}\sim\Bigl (\frac{\la}{\mql}\Bigr )^{2\bo/N_F}\gg 1\,, \label{(6.3)}
\eq
\bbq
\cw=N_c S+\frac{\mph}{2}\Biggl [{\rm Tr}\,(\Phi^2)-\frac{1}{\nd}\Bigl ({\rm Tr}\,\Phi\Bigr )^2\Biggr ]\,,
\quad S=\Bigl (\la^{\bo}\det m^{\rm tot}_{Q}\Bigr )^{1/N_c}\,,\quad m^{\rm tot}_{Q}=(m_Q-\Phi)\,,
\eeq
and one has to choose the L - vacua in \eqref{(6.3)}.

There are two contributions to the mass of fions in \eqref{(6.3)}, the perturbative one from the term $\sim \mph\Phi^2$ and the non-perturbative one from $\sim S$, and both are parametrically the same, $\sim \lym^{(L)}\gg m_Q$. Therefore,
\bq
\mu(\Phi)\sim\frac{\mph}{z_{\Phi}(\la,\mql)}\sim\mql\sim\lym^{(L)}\,. \label{(6.4)}
\eq
Besides, see \eqref{(5.1)}, because
\bq
\mu^{\rm conf}_o\sim\la \Bigl (\frac{\la}{\mph}\Bigr )^{\frac{N_F}{3(2N_c-N_F)}}\sim\mql\sim\lym^{(L)}\,, \label{(6.5)}
\eq
and fions are dynamically irrelevant at $\mu^{\rm conf}_o<\mu<\la$ and can become relevant only at the scale $\mu<\mu^{\rm conf}_o$, it remains unclear in these
L - vacua whether there is a pole in the fion propagators at $p\sim\mu^{\rm conf}_o\sim\mql$.  May be yes but maybe not, see section 5.\\

On the whole for the mass spectrum in these L - vacua. The quarks ${\ov Q}, Q$ are confined and strongly coupled here, the coupling being $a_*\sim 1$. Parametrically, there is only one scale $\sim \lym^{(L)}$ in the mass spectrum at $\la\ll\mph\ll\mo$. And there is no parametrical guaranty that there is the second generation of fions with the pole masses $\mu_2^{\rm pole}(\Phi)\sim\lym^{(L)}$.

The condensate $\langle{\ov Q}Q\rangle_L$ and the quark pole mass $\mql$ become frozen at their SQCD values at $\mph\gg\mo,\, \langle{\ov Q}Q\rangle_{SQCD}\sim\la^2(m_Q/\la)^{\nd/N_c},\,\, m^{\rm pole}_{SQCD}\sim \lym^{(SQCD)}\sim\la(m_Q/\la)^{N_F/3N_c}$ \cite{ch3}, while $\mph$ increases and $\mu^{\rm conf}_o\ll m^{\rm pole}_{Q,SQCD}$ decreases, see \eqref{(5.1)}. Hence, the perturbative contribution $\sim\mph/z_{\Phi}(\la,\mql)\gg m^{\rm pole}_{Q,SQCD}$ to the fion mass becomes dominant at $\mph\gg\mo$ and the fion fields will be dynamically irrelevant at $\mu<\la$.\\

Finally, it is worth emphasizing for all what follows that, unlike the dual theory, {\it in all vacua of the direct theory the mass spectra remain parametrically the same at $\,\bd/N_F=O(1)$ or $\,\bd/N_F\ll 1$}.\\

\subsection{\quad  S - vacua}

In these $\nd$ vacua the quark mass at $\la\ll\mph\ll\mo$ looks as, see \eqref{(4.4)},
\bbq
\frac{\langle m^{\rm tot}_Q(\mu=\la)\rangle_S}{\la}\sim \frac{\langle S\rangle_S}{\la\langle{\ov Q}Q\rangle_S}\sim \Bigl (\frac{\langle{\ov Q}Q\rangle_S}{\la^2}\Bigr )^{N_c/\nd}\sim \Bigl (\frac{m_Q\mph}{\la^2}\Bigr )^{N_c/\nd}\,,
\eeq
\bq
\mqs\sim \la\Bigl (\frac{m_Q\mph}{\la^2}\Bigr )^{N_F/3\nd}\sim\lym^{(S)}=\langle S\rangle^{1/3}_S,\quad \la\ll \mph\ll\mo\,. \label{(6.6)}
\eq
This has to be compared with the gluon mass due to possible higgsing of quarks
\bq
\mgs^2\sim z_Q(\la,\mgs)\qqs\,\ra\,\mgs\sim\mqs\sim\lym^{(S)},\quad z_Q(\la,\mgs)\sim\Bigl (\frac{\mgs}{\la}\Bigr )^{\bo/N_F}. \label{(6.7)}
\eq

For the same reasons as in previous section, it is clear that quarks will not be higgsed in these vacua at $N_F>N_c$ (as otherwise the flavor symmetry will be broken spontaneously). Hence, as in \cite{ch3}, we assume here also that the pole mass of quarks is the largest physical mass, i.e. $\mu_H=\mqs=(\rm several)\mgs$.

But, in contrast with the L - vacua, the fion fields {\it become dynamically relevant in these S - vacua} at scales $\mu<\mu^{\rm conf}_o$, see \eqref{(5.1)}, if
\bq
\mu^{\rm conf}_o\sim\la\Bigl (\frac{\la}{\mph}\Bigr )^{\frac{N_f}{3(2N_c-N_F)}}\gg\mqs \quad\ra \quad {\rm i.e.\,\, at}\quad  \la\ll\mph\ll\mo\,.\label{(6.8)}
\eq

Therefore, there is the second generation of $N_F^2$ fions with the pole masses
\bq
\mu_2^{\rm pole}(\Phi)\sim \mu_o^{\rm conf}\gg\mqs\sim\lym^{(S)}\,.\label{(6.9)}
\eq

Nevertheless, see section 5, the theory remains in the conformal regime and the quark and fion anomalous dimensions remain the same in the whole range of $\mqs<\mu<\la$ of scales, but fions become effectively massless at $\mu<\mu^{\rm conf}_o$ and begin to contribute to the 't Hooft triangles.

The RG evolution of the quark and fion fields becomes frozen at scales $\mu<\mqs$ because the heavy quarks decouple. Proceeding as before, i.e. integrating out first all quarks as heavy ones at $\mu<\mqs=(\rm several) \lym^{(S)}$ and then all $SU(N_c)$ gluons at $\mu<\lym^{(S)}$, one obtains the Lagrangian of fions as in \eqref{(6.3)}, with a replacement $z_Q(\la,\mql)\ra z_Q(\la,\mqs)$ (and the S-vacua have to be chosen therein).

Because fions became relevant at $\mqs\ll\mu\ll\mu^{\rm conf}_o$, one could expect that their running mass will be much smaller than $\mqs$. This is right, but only for $\mu^{\rm pert}_{\Phi}\sim \mph/z_Q(\la,\mqs)\ll\mqs$. But there is also additional {\it non-perturbative contribution} to the fion mass originating from the region of scales $\mu\sim\mqs$ and it is dominant in these S - vacua,
\bq
\mu(\Phi)\sim\frac{1}{z_{\Phi}(\la,\mqs)}\,\frac{\langle S\rangle_S}{\langle m^{\rm tot}_Q\rangle^2_S}\sim\mqs\,,\quad
z_{\Phi}(\la,\mqs)\sim \Bigl (\frac{\la}{\mqs}\Bigr )^{2\bo/N_F}\,.\label{(6.10)}
\eq
Therefore, despite the fact that the fions are definitely dynamically relevant in the range of scales $\mqs\ll\mu\ll\mu^{\rm conf}_o\ll\la$ at $\la\ll\mph\ll\mo$, whether there is the third generation of fions, i.e. whether there is a pole in the fion propagator at $p=\mu_3^{\rm pole}(\Phi)\sim\mqs\sim\lym^{(S)}$ remains unclear. \\

On the whole for the mass spectra in these S - vacua. The largest are the masses of the second generation fions, $\mu_2^{\rm pole}(\Phi)\sim\la\Bigl (\la/\mph\Bigr )^{N_F/3(2N_c-N_F)}\gg\mqs$. The scale of all other masses is $\sim\mqs\sim\lym^{(S)}$, see \eqref{(6.6)}. There is no parametrical guaranty that there is the third generation of fions with the pole masses $\mu_3^{\rm pole}(\Phi)\sim\lym^{(S)}$. May be yes, but maybe not.\\

The vacuum condensates $\langle{\ov Q}Q\rangle_{S}$ and $\mqs$ evolve into their independent of $\mph$ SQCD-values at $\mph\gg\mo$,
\bq
\langle{\ov Q}Q\rangle_{SQCD}\sim \la^2\Bigl (\frac{m_Q}{\la}\Bigr )^{\nd/N_c}\,,\quad m^{\rm pole}_{Q,SQCD}\sim\la\Bigl (\frac{m_Q}{\la}\Bigr )^{N_F/3N_c}\,,\label{(6.11)}
\eq
and the perturbative contribution $\sim \mph/z_Q(\la,m^{\rm pole}_{Q,SQCD})$ to the fion mass becomes dominant. Hence, because $m^{\rm pole}_{Q,SQCD}\gg\mu^{\rm conf}_o$, the fions fields become dynamically irrelevant at all scales $\mu<\la$ when $\mph\gg\mo$.\\

\section{Dual theory. Unbroken flavor symmetry}

\subsection{\quad  L - vacua,\,\,\, $\bd/N_F\ll 1$}

Let us recall, see \eqref{(2.7)} and section 4 in \cite{ch3}, that the Lagrangian of the dual theory at $\mu=\la$ and $0<\bd/N_F\ll 1,\,\,\bd=3\nd-N_F$, looks as
\bbq
K= {\rm Tr}\Bigl ( q^\dagger q +(q\ra\ov q) \Bigr )+{\rm Tr}\,\frac{M^{\dagger}M}{Z_q^2\la^2}\,,\,\,
\cw=\, -\,\frac{2\pi}{\ov \alpha(\mu=\la)}\, {\ov s}+\cw_M+\cw_q,\,
Z_q\sim\exp \Bigl\{-\frac{\nd}{7\bd}\Bigr\}\ll 1\,,
\eeq
\bq
\cw_M=m_Q{\rm Tr}\,M -\frac{1}{2\mph}\Biggl [{\rm Tr}\, (M^2)- \frac{1}{N_c}({\rm Tr}\, M)^2 \Biggr ]\,,\quad
\cw_q= -\,\frac{1}{Z_q\la}\,\rm {Tr} \Bigl ({\ov q}\,M\, q \Bigr )\,. \label{(7.1)}
\eq

Because $\la^2/\mph\ll\la$, the mions are effectively massless and dynamically relevant at $\mu\sim\la$ (and so in some range of scales below $\la$). By definition, $\mu\sim\la$ is such a scale that the dual theory already entered sufficiently deep the conformal regime, i.e. the dual gauge coupling ${\ov a}(\mu=\la)=\nd{\ov\alpha}(\mu=\la)/2\pi$ is sufficiently close to its small frozen value,  $\ov\delta=[{\ov a}_*-{\ov a}(\mu\sim\la)]/{\ov a}_*\ll 1$, and $\ov\delta$ is neglected everywhere below in comparison with 1 for simplicity (and the same for the Yukawa coupling ${\ov a}_f=\nd{\ov\alpha}_f/2\pi$), see \cite{ch3} and the Appendix therein). The fixed point value of the dual gauge coupling is ${\ov a}_*\approx 7\bd/3\nd\ll 1$ \cite{KSV}.

We recall also that the mion condensates are matched to the condensates of direct quarks {\it in all vacua}, $\langle M^i_j(\mu=\la)\rangle=\langle{\ov Q}_j Q^i(\mu=\la)\rangle$\,. Hence, in these L - vacua
\bq
\langle M\rangle_L\sim\la^2\Bigl (\frac{\la}{\mph}\Bigr )^{\frac{\nd}{2N_c-N_F}}\,,\quad\langle N\rangle_L\equiv\langle{\ov q}q(\mu=\la)\rangle=
\frac{Z_q\la\langle S\rangle_L}{\langle M\rangle_L}\sim Z_q\la^2 \Bigl (\frac{\la}{\mph}\Bigr )^{\frac{N_c}{2N_c-N_F}}\,,\label{(7.2)}
\eq
\bbq
Z_q\sim\exp \Bigl\{-\frac{1}{3{\ov a}_{*}}\Bigr\}\sim \exp \Bigl\{-\frac{\nd}{7\bd}\Bigr\}\ll 1,
\eeq
and here and everywhere below, as in \cite{ch3}, a parametric dependence on the small parameter $\bd/N_F\ll 1$ is traced with an exponential accuracy only (i.e. powers of $\bd/N_F$ are not traced, only powers of $Z_q$).

The current mass $\mu_{q,L}\equiv \mu_{q,L}(\mu=\la)$ of dual quarks ${\ov q},\,q$ and their pole mass in these $(2N_c-N_F)$ L - vacua are, see \eqref{(7.1)},
\bbq
\frac{\mu_{q,L}}{\la}=\frac{\langle M\rangle_L}{Z_q\la^2}\sim\frac{1}{Z_q}\Bigl (\frac{\la}{\mph}\Bigr )^{\frac{\nd}{2N_c-N_F}}\,,\quad\mu^{\rm pole}_{q,L}=\frac{\mu_{q,L}}{z_q(\la,\mu^{\rm pole}_{q,L})}\,,\quad z_q(\la,\mu^{\rm pole}_{q,L})\sim\Bigl (\frac{\mu^{\rm pole}_{q,L}}{\la}\Bigr )^{\bd/N_F}\,,
\eeq
\bq
\mu^{\rm pole}_{q,L}\sim\la\Bigl (\frac{\mu_{q,L}}{\la}\Bigr )^{N_F/3\nd}\sim\frac{\la}{Z_q}\Bigl (\frac{\la}{\mph}\Bigr )^{\frac{N_F}{3(2N_c-N_F)}}\sim\frac{1}{Z_q}\lym^{(L)}\gg\lym^{(L)} \,, \quad \la\ll\mph\ll\mo\,,\label{(7.3)}
\eq
\bbq
\frac{\mu_{q,L}}{\la}\sim\frac{1}{Z_q}\Bigl (\frac{m_Q}{\la}\Bigr )^{\frac{\nd}{N_c}}\,,\quad \mu^{\rm pole}_{q,L}\sim\frac{\la}{Z_q}\Bigl (\frac{m_Q}{\la}
\Bigr )^{\frac{N_F}{3N_c}}\sim\frac{1}{Z_q}\lym^{(SQCD)}\gg \lym^{(SQCD)}\,,\quad \mph\gg\mo\,,
\eeq
while the gluon mass due to possible higgsing of dual quarks looks at $\la\ll\mph\ll\mo$ as
\bq
{\ov\mu}_{\rm gl, L}\sim\Bigl [{\ov a}_*\langle N\rangle_{L}\, z_q(\la,\mug)\Bigr]^{1/2}\sim Z_q^{1/2}\la
\Bigl (\frac{\la}{\mph}\Bigr )^{N_F/3(2N_c-N_F)}\sim Z_q^{3/2}\mu^{\rm pole}_{q,L}\ll \mu^{\rm pole}_{q,L}\,. \label{(7.4)}
\eq
Therefore, the dual quarks are definitely in the HQ phase in these L - vacua at $\bd/\nd\ll 1$.

With decreasing scale the perturbative running mass $\mu_{M}(\mu)$ of mions
\bq
\mu_{M}(\mu)\sim\frac{Z_q^2\la^2}{\mph z_M(\mu)}=\frac{Z_q^2\la^2}{\mph}\Bigl (\frac{\mu}{\la}\Bigr )^{2\bd/N_F} \label{(7.5)}
\eq
decreases but more slowly than the scale $\mu$ itself because $\gamma_M= - (2{\rm\bd}/N_F),\,\,|\gamma_M|<1$ at $3/2<N_F/N_c<2$\,, and $\mu_{M}(\mu)$ becomes frozen at $\mu<\mu^{\rm pole}_{q,L}$,\, $\mu_M(\mu<\mu^{\rm pole}_{q,L})=\mu_M(\mu=\mu^{\rm pole}_{q,L})$.

After integrating out all dual quarks as heavy ones at $\mu<\mu^{\rm pole}_{q,L}$ and then all $SU(\nd)$ gluons at $\mu<\lym^{(L)}$ via the Veneziano-Yankielowicz (VY) procedure \cite{VY}, the Lagrangian of mions looks as
\bbq
K=\frac{z^{(L)}_M(\la,\mu^{\rm pole}_{q,L})}{Z_q^2\la^2}{\rm Tr}\,(M^\dagger M)\,,\quad z^{(L)}_M(\la,\mu^{\rm pole}_{q,L})\sim\Bigl (\frac{\la}{\mu^{\rm pole}_{q,L}}\Bigr )^{2\bd/N_F}\,,\quad
S=\Biggl (\,\frac{\det M}{\la^{\bo}}\,\Biggr )^{1/\nd},
\eeq
\bq
W= -\nd S+m_Q{\rm Tr}\,M -\frac{1}{2\mph}\Biggl [{\rm Tr}\, (M^2)- \frac{1}{N_c}({\rm Tr}\, M)^2 \Biggr ]\,. \label{(7.6)}
\eq

There are two contributions to the mass of mions in \eqref{(7.6)}, the perturbative one from the term $\sim M^2/\mph$ and non-perturbative one from $\sim S$.
Both are parametrically the same and the total contribution looks as
\bq
\mu^{\rm pole}(M)\sim \frac{Z_q^2\la^2}{z_M(\la,\mu^{\rm pole}_{q,L})\mph}\sim Z_q^2\lym^{(L)}\ll \lym^{(L)}\ll \mu^{\rm pole}_{q,L}\,, \label{(7.7)}
\eq
and this parametrical hierarchy guarantees that the mass $\mu^{\rm pole}(M)$ in \eqref{(7.7)} is indeed the {\it pole mass} of mions.\\

On the whole, the mass spectrum in these dual L - vacua looks as follows at $\la\ll\mph\ll\mo$. a) There is a large number of heaviest flavored hadrons made of weakly interacting and weakly confined (the tension of the confining string originating from the unbroken $SU(\nd)$ SYM is $\sqrt\sigma\sim\lym^{(L)}\ll\mu^{\rm pole}_{q,L}$) non-relativistic quarks ${\ov q}, q$ with the pole masses $\mu^{\rm pole}_{q,L}/\lym^{(L)}\sim \exp (\nd/7\bd)\gg 1$. The mass spectrum of low-lying flavored mesons is Coulomb-like with parametrically small mass differences $\Delta\mu_H/\mu_H=O(\bd^2/\nd^2)\ll 1$. b) A large number of gluonia made of $SU(\nd)$ gluons with the mass scale $\sim\lym^{(L)}\sim\la (\la/\mph)^{N_F/3(2N_c-N_F)}$. c) $N_F^2$ lightest mions with parametrically smaller masses $\mu^{\rm pole}(M)/\lym^{(L)}\sim \exp (-2\nd/7\bd)\ll 1$.

At $\mph\gg\mo$ these L - vacua evolve into the vacua of the dual SQCD theory (dSQCD), see section 4 in \cite{ch3}.\\

\subsection{\quad  S - vacua,\,\,\, $\bd/N_F\ll 1$}

The current mass $\mu_{q,S}\equiv \mu_{q,S}(\mu=\la)$ of dual quarks ${\ov q},\,q$ at the scale $\mu=\la$ in these $(N_F-N_c)$ dual S-vacua is, see \eqref{(4.4)},
\bq
\mu_{q,S}=\frac{\langle M\rangle_S=\langle\qq\rangle_S}{Z_q\la}\sim\frac{m_Q\mph}{Z_q\la}\,,\quad Z_q\sim \exp \Bigl\{-\frac{\nd}{7\bd}\Bigr\}\ll 1\,.\label{(7.8)}
\eq

In comparison with the L - vacua in section 7.1, a qualitatively new element here is that $\mu^{\rm pole}(M)$ is the largest mass, $\mu^{\rm pole}(M)\gg\mu^{\rm pole}_{q,S}$, in the wide region $\la\ll\mph\ll Z_q^{\,3/2}\mo$ (see \eqref{(7.15)} below). In this region: a) the mions are effectively massless and dynamically relevant at scales $\mu^{\rm pole}(M)\ll\mu\ll\la$,\,\, b) there is a pole in the mion propagator at the momentum $p=\mu^{\rm pole}(M)$,
\bbq
\mu^{\rm pole}(M)=\frac{Z_q^2\la^2}{z_M(\la,\mu^{\rm pole}(M))\mph}\,,\quad z_M(\la,\mu^{\rm pole}(M))\sim\Bigl (\frac{
\la}{\mu^{\rm pole}(M)}\Bigr )^{2\bd/N_F}\,,
\eeq
\bq
\mu^{\rm pole}(M)\sim Z_q^2\la\Bigl (\frac{\la}{\mph}\Bigr )^{\frac{N_F}{3(2N_c-N_F)}}\,,\quad Z_q^2\sim \exp\{-\frac{2\nd}{7\bd}\}\ll 1\,. \label{(7.9)}
\eq

The mions then become too heavy and dynamically irrelevant at $\mu\ll\mu^{\rm pole}(M)$. Due to this, they decouple from the RG evolution of dual quarks and gluons and from the 't Hooft triangles, and (at $\mph$ not too close to $\mo$ to have enough "time" to evolve, see \eqref{(7.14)}\,) the remained dual theory of $N_F$ quarks ${\ov q}, q$ and $SU(\nd)$ gluons evolves into a new conformal regime with a new smaller value of the frozen gauge coupling, ${\ov a}^{\,\prime}_*\approx\bd/3\nd={\ov a}_*/7\ll 1$. It is worth noting that, in spite of that  mions are dynamically irrelevant at $\mu<\mu^{\rm pole}(M)$, their renormalization factor $z_M(\mu<\mu^{\rm pole}(M))$ still runs in the range of scales $\mu^{\rm pole}_{q,S}<\mu<\mu^{\rm pole}(M)$ being induced by loops of still effectively massless dual quarks and gluons.

The next physical scale is the perturbative pole mass of dual quarks
\bbq
\mu^{\rm pole}_{q,S}=\frac{\langle M\rangle_S}{Z_q\la}\,\frac{1}{z_q(\la,\mu^{\rm pole}_{q,S})}\,,\quad
z_q(\la,\mu^{\rm pole}_{q,S})=\Bigl (\frac{\mu^{\rm pole}_{q,S}}{\la}\Bigr )^{\bd/N_F}{\ov\rho}_S\,,
\eeq
\bbq
{\ov\rho}_S=\Bigl (\frac{{\ov a}_*}{{\ov a}^{\,\prime}_*}\Bigr )^{\frac{\nd}{N_F}}\exp\Bigl\{\frac{\nd}{N_F}
\Bigl (\frac{1}{{\ov a}_*}-\frac{1}{{\ov a}^{\,\prime}_*}\Bigr )\Bigr\}\sim \frac{{\ov Z}_q}{Z_q}\ll 1\,,\quad
{\ov Z}_q\sim \exp \Bigl\{-\frac{\nd}{\bd}\Bigr\}\sim \Bigl (Z_q\Bigr )^{7}\ll Z_q\,,
\eeq
\bq
\mu^{\rm pole}_{q,S}\sim \frac{1}{{\ov Z}_q}\lym^{(S)}\gg\lym^{(S)}\,,\quad \lym^{(S)}=\la\Bigl (\frac{m_Q\mph}{\la^2}
\Bigr )^{N_F/3\nd}\,. \label{(7.10)}
\eq

This has to be compared with the gluon mass due to possible higgsing of ${\ov q}, q$
\bq
{\ov\mu}^{\,2}_{\rm gl, S}\sim  z_q(\la,{\ov\mu}_{\rm gl, S})\langle{\ov q}q\rangle_S\,\,\ra\,\,{\ov\mu}_{\rm gl, S}\sim{\ov Z}_q^{\,1/2}\lym^{(S)}\ll\lym^{(S)}\ll\mu^{\rm pole}_{q,S}. \label{(7.11)}
\eq
The parametric hierarchy in \eqref{(7.11)} guarantees that the dual quarks are in the HQ phase in these S - vacua.

Hence, after integrating out all quarks at $\mu<\mu^{\rm pole}_{q,S}$ and, finally, $SU(\nd)$ gluons at $\mu<\lym^{(S)}$, the Lagrangian looks as in \eqref{(7.6)} but with a replacement $z^{(L)}_M (\la,\mu^{\rm pole}_{q,L})\ra z^{(S)}_M (\la,\mu^{\rm pole}_{q,S})$,
\bq
z^{(S)}_M(\la,\mu^{\rm pole}_{q,S})=\frac{a_f(\mu=\la)}{a_f(\mu=\mu^{\rm pole}_{q,S})}\frac{1}{z^2_q(\la,\mu^{\rm pole}_{q,S})}\sim\frac{1}{z^2_q(\la,\mu^{\rm pole}_{q,S})}\sim\frac{Z_q^{\,2}}{{\ov Z}_q^{\,2}}\Bigl (\frac{\la}{\mu^{\rm pole}_{q,S}}\Bigr )^{2\bd/N_F}\,. \label{(7.12)}
\eq
The contribution of the term $\sim M^2/\mph$ in the superpotential \eqref{(7.6)} to the frozen low energy value $\mu(M)$ of the running mion mass is dominant at $\mph/\mo\ll 1$ and is
\bq
\hspace*{-5mm}\mu(M)=\frac{Z_q^2\la^2}{z^{(S)}_M(\la,\mu^{\rm pole}_{q,S})\mph}\sim
\frac{{\ov Z}_q^{\,2}\la^2}{\mph}\Bigl (\frac{\mu^{\rm pole}_{q,S}}{\la}\Bigr )^{\frac{2\bd}{N_F}}\ll\mu^{\rm pole}(M)\,. \label{(7.13)}
\eq
The requirement of self-consistency looks in this case as
\bq
\frac{\mu(M)}{\mu^{\rm pole}_{q,S}}\sim {\ov Z}_q^{\,3}\Bigl (\frac{\mo}{\mph}\Bigr )^{N_c/\nd}\gg 1\quad\ra\quad
\frac{\mph}{\mo}\ll {\ov Z}^{\, 3/2}_q\sim\exp \Bigl\{-\frac{3\nd}{2\bd}\Bigr\}\ll Z_q^{3/2}\,, \label{(7.14)}
\eq
the meaning of \eqref{(7.14)} is that only at this condition the range of scales between $\mu^{\rm pole}(M)$ in \eqref{(7.9)} and $\mu^{\rm pole}_{q,S}\ll\mu^{\rm pole}(M)$ in \eqref{(7.10)} is sufficiently large that theory has enough "time" to evolve from ${\ov a}_*=7\bd/3\nd$ to ${\ov a}^{\,\prime}_*=\bd/3\nd$. There is no pole in the mion propagator at the momentum $p=\mu(M)\gg\mu^{\rm pole}_{q,S}$.

The opposite case with $\mu^{\rm pole}_{q,S}\gg\mu^{\rm pole}(M)$ is realized if the ratio $\mph/\mo$ is still $\ll 1$ but is much larger than $Z_q^{\,3/2}\gg{\ov Z}_q^{\,3/2}$, see \eqref{(7.15)} below. In this case the theory at $\mu^{\rm pole}_{q,S}<\mu<\la$ remains in the conformal regime with ${\ov a}_*=7\bd/3\nd$ and the largest mass is $\mu^{\rm pole}_{q,S}$. One has in this case instead of \eqref{(7.9)},\eqref{(7.10)},\eqref{(7.14)}
\bbq
{\ov\rho}_S\sim 1\,,\quad \mu^{\rm pole}_{q,S}\sim\frac{1}{Z_q}\lym^{(S)}\,,\quad \frac{\mu^{\rm pole}
(M)}{\mu^{\rm pole}_{q,S}}\sim Z^3_q\Bigl (\frac{\mo}{\mph}\Bigr )^{N_c/\nd}\,,
\eeq
\bq
\frac{\mu^{\rm pole}(M)}{\mu^{\rm pole}_{q,S}}\ll 1\quad\ra\quad Z_q^{\,3/2}\ll\frac{\mph}{\mo}\ll 1\,. \label{(7.15)}
\eq

On the whole, the mass spectrum in these $\nd$ dual S - vacua looks as follows at $\la\ll\mph\ll{\ov Z}_q^{\,3/2}\mo$. a) The heaviest are $N_F^2$ mions with the pole masses \eqref{(7.9)}. b) There is a large number of flavored hadrons made of weakly interacting and weakly confined (the tension of the confining string is $\sqrt\sigma\sim\lym^{(S)}\ll\mu^{\rm pole}_{q,S}\ll\mu^{\rm pole}(M)$) non-relativistic dual quarks ${\ov q}, q$ with the perturbative pole masses \eqref{(7.10)}. The mass spectrum of low-lying flavored mesons is Coulomb-like with parametrically small mass differences $\Delta\mu_H/\mu_H=O(\bd^2/\nd^2)\ll 1$. b) A large number of gluonia made of $SU(\nd)$ gluons with the mass scale $\sim\lym^{(S)}\sim\la (m_Q\mph/\la^2)^{N_F/3\nd}$.

The mions with the pole masses \eqref{(7.9)} remain the heaviest ones, $\mu^{\rm pole}(M)\gg\mu^{\rm pole}_{q,S}$, at values $\mph$ in the range ${\ov Z}_q^{\,3/2}\mo\ll\mph\ll Z_q^{\,3/2}\mo$, while the value $\mu^{\rm pole}_{q,S}$ varies in a range $\lym^{(S)}/Z_q\ll\mu^{\rm pole}_{q,S}\ll\lym^{(S)}/{\ov Z}_q$\,. Finally, in a close vicinity of $\mo,\,\, Z_q^{\,3/2}\mo\ll\mph\ll\mo$, the perturbative pole mass of quarks, $\mu^{\rm pole}_{q,S}\sim\lym^{(S)}
/Z_q\gg\lym^{(S)}$, becomes the largest one, while the pole masses of mions $\mu^{\rm pole}(M)\ll\mu^{\rm pole}_{q,S}$ become as in \eqref{(7.15)}.

At $\mph\gg\mo$ these S - vacua evolve into the vacua of dSQCD, see section 4 in \cite{ch3}.

\section{Direct theory. Broken flavor symmetry, $\mathbf{\la\ll\mph\ll\mo}$}

\subsection{\quad  L - type vacua}

\hspace*{4mm} The quark condensates are parametrically the same as in the L - vacua with unbroken flavor symmetry in section 6.1,
\bq
(1-\frac{n_1}{N_c})\langle\Qo\rangle_{Lt}\approx -(1-\frac{n_2}{N_c})\langle\Qt\rangle_{Lt},\quad \langle S\rangle=\frac{\langle\Qo\rangle\langle\Qt\rangle}{\mph}\,, \label{(8.1)}
\eq
\bbq
\langle\Qo\rangle_{Lt}\sim\langle\Qt\rangle_{Lt}\sim\la^2\Bigl (\frac{\la}{\mph}\Bigr )^{\frac{\nd}{2N_c-N_F}}\,.
\eeq

All quarks are in the HQ  phase and are confined and the Lagrangian of fions looks as in \eqref{(6.3)}, but one has to choose the L - type vacua with the broken flavor symmetry in \eqref{(6.3)}. Due to this, see \eqref{(2.5)}, the masses of hybrid fions $\Phi_{12}, \Phi_{21}$ are qualitatively different, they are the Nambu-Goldstone particles here and are massless. The "masses" of $\Phi_{11}$ and $\Phi_{22}$ are parametrically as in \eqref{(6.4)},
\bq
\mu(\Phi_{11})\sim\mu(\Phi_{22})\sim\frac{\mph}{z_{\Phi}(\la, m^{\rm pole}_Q)}\sim m^{\rm pole}_{Q,1}\sim m^{\rm pole}_{Q,2}\sim\lym^{(L)}\sim\la \Bigl (\frac{\la}{\mph}\Bigr )^{\frac{N_F}{3(2N_c-N_F)}}\,, \label{(8.2)}
\eq
and hence there is no guaranty that these are the pole masses of fions, see section 5. May be yes, but maybe not.

On the whole, there are only two characteristic scales in the mass spectra in these L - type vacua. The hybrid fions $\Phi_{12}, \Phi_{21}$ are massless
while all other masses are $\sim\lym^{(L)}$.\\

\subsection{\quad  $\rm br2$ vacua}

\hspace*{4mm} The condensates of quarks look as
\bq
\langle\Qt\rangle_{\rm br2}\approx \Bigl (\rho_2=-\frac{n_2-N_c}{N_c}\Bigr )m_Q\mph,\,\,\, \langle\Qo\rangle_{\rm br2}\sim \la^2\Bigl(\frac{\mph}{\la}\Bigr )^{\frac{n_2}{n_2-N_c}}\Bigl (\frac{m_Q}{\la}\Bigr )^{\frac{N_c-n_1}{n_2-N_c}},\label{(8.3)}
\eq
\bbq
\frac{\langle\Qo\rangle_{\rm br2}}{\langle\Qt\rangle_{\rm br2}}\sim \Bigl (\frac{\mph}{\mo}\Bigr)^{\frac{N_c}{n_2-N_c}}\ll 1
\eeq
in these vacua with $n_2>N_c\,, 1\leq n_1<\nd$\,. Hence, the largest among the masses smaller than $\la$ are the masses of the $N_F^2$ second generation fions, see \eqref{(5.1)},
\bq
\mu^{\rm pole}_2(\Phi_i^j)=\mu_o^{\rm conf}\sim \la\Bigl (\frac{\la}{\mph}\Bigr )^{\frac{N_F}{3(2N_c-N_F)}}\,, \label{(8.4)}
\eq
while some other possible characteristic masses look here as
\bq
\langle m^{\rm tot}_{Q,1}\rangle_{\rm br2}=\frac{\langle\Qt\rangle_{\rm br2}}{\mph}\sim m_Q\,,\quad m^{\rm pole}_{Q,1}\sim\la\Bigl(\frac{m_Q}{\la}\Bigr )^{N_F/3N_c}\gg {\tilde m}^{\rm pole}_{Q,2}\,, \label{(8.5)}
\eq
\bbq
\mgt^2\sim z_Q(\la,\mgt)\langle\Qt\rangle_{\rm br2},\quad z_Q(\la,\mgt)\sim\Bigl (\frac{\la}{\mgt}\Bigr )^{\frac{\bo}{N_F}}\ll 1\,,
\eeq
\bq
\mgt\sim\la\Bigl(\frac{m_Q\mph}{\la^2}\Bigr )^{N_F/3\nd}\gg\mgo\,,\quad \frac{\mgt}{m^{\rm pole}_{Q,1}}\sim
\Bigl (\frac{\mph}{\mo}\Bigr )^{\frac{N_F}{3\nd}}\ll 1\,, \label{(8.6)}
\eq
where $m^{\rm pole}_{Q,1}$ and ${\tilde m}^{\rm pole}_{Q,2}$ are the pole masses of quarks ${\ov Q}_1, Q^1$ and ${\ov Q}_2, Q^2$ and $\mgo,\, \mgt$ are the gluon masses due to possible higgsing of these quarks. Hence, the largest mass is $m^{\rm pole}_{Q,1}$ and the overall phase is $HQ_1-HQ_2$.

The lower energy theory at $\mu<m^{\rm pole}_{Q,1}$ has $N_c$ colors and $N_F^\prime =n_2>N_c$ flavors of quarks ${\ov Q}_2, Q^2$. In the range of scales $m^{\rm pole}_{Q,2}<\mu<m^{\rm pole}_{Q,1}$, it will remain in the conformal regime at $n_1<\bd=(2N_F-3N_c)/2$, while it will be in the strong coupling regime at $n_1>\bd/2$, with the gauge coupling $a(\mu\ll m^{\rm pole}_{Q,1})\gg 1$. We do not consider the strong coupling regime in this paper and for this reason we take $\bd/\nd=O(1)$ in this subsection and consider $n_1<\bd/2$ only.

After the heaviest quarks ${\ov Q}_1, Q^1$ decouple at $\mu<m^{\rm pole}_{Q,1}$, the pole mass of quarks ${\ov Q}_2, Q^2$ in the lower energy theory looks as
\bq
m^{\rm pole}_{Q,2}=\frac{1}{z^{\,\prime}_Q(m^{\rm pole}_{Q,1},m^{\rm pole}_{Q,2})}\Biggl(\,\frac{\langle\Qo\rangle_{\rm br2}}{\langle\Qt\rangle_{\rm br2}}\, m^{\rm pole}_{Q,1}\,\Biggr )\sim\lym^{(\rm br2)}\,, \label{(8.7)}
\eq
\bbq
z^{\,\prime}_Q(m^{\rm pole}_{Q,1},m^{\rm pole}_{Q,2})\sim\Bigl (\frac{m^{\rm pole}_{Q,2}}{m^{\rm pole}_{Q,1}}\Bigr )^{\frac{3N_c-n_2}{n_2}}\ll 1\,.
\eeq

Hence, after integrating out  quarks ${\ov Q}_1, Q^1$ at $\mu<m^{\rm pole}_{Q,1}$ and then  quarks ${\ov Q}_2, Q^2$ and $SU(N_c)$ gluons at $\mu<\lym^{(\rm br2)}$, the Lagrangian of fions looks as
\bq
K=z_{\Phi}(\la,m^{\rm pole}_{Q,1})\,{\rm Tr}\,\Bigl [\,\Phi_{11}^\dagger \Phi_{11}+\Phi_{12}^\dagger \Phi_{12}+\Phi_{21}^\dagger \Phi_{21}+z^{\,\prime}_{\Phi}(m^{\rm pole}_{Q,1},m^{\rm pole}_{Q,2})\Phi_{22}^\dagger \Phi_{22}\,\Bigr ]\,, \label{(8.8)}
\eq
\bbq
z_{\Phi}(\la,m^{\rm pole}_{Q,1})\sim\Bigl (\frac{\la}{m^{\rm pole}_{Q,1}}\Bigr )^{\frac{2(3N_c-N_F)}
{N_F}}\gg 1\,,\quad z^{\,\prime}_{\Phi}(m^{\rm pole}_{Q,1},m^{\rm pole}_{Q,2})\sim\Bigl (\frac{m^{\rm pole}_{Q,1}}{m^{\rm pole}_{Q,2}}\Bigr )^{\frac{2(3N_c-n_2)}{n_2}}\gg 1\,,
\eeq
\bq
\cw=N_c S+\cw_{\Phi}\,,\quad m^{\rm tot}_Q= (m_Q-\Phi)\,,\quad \label{(8.9)}
\eq
\bbq
S=\Bigl (\la^{\bo}\det m^{\rm tot}_Q\Bigr )^{1/N_c}\,,\quad \cw_{\Phi}=\frac{\mph}{2}\Bigl ({\rm Tr}\,(\Phi^2)-
\frac{1}{\nd}({\rm Tr}\,\Phi)^2 \,\Bigr ).
\eeq
From \eqref{(8.8)},\eqref{(8.9)}, the main contribution to the mass of the third generation fions $\Phi_{11}$ gives the term $\sim\mph\Phi^2_{11}$,
\bq
\mu^{\rm pole}_3(\Phi_{11})\sim\frac{\mph}{z_{\Phi}(\la,m^{\rm pole}_{Q,1})}\sim\Bigl (\frac{\mph}{\mo}\Bigr )m^{\rm pole}_{Q,1}\,, \label{(8.10)}
\eq
while the third generation hybrid fions $\Phi_{12}, \Phi_{21}$ are massless, $\mu^{\rm pole}_3(\Phi_{12})=\mu^{\rm pole}_3(\Phi_{21})=0$. As for the third generation fions $\Phi_{22}$, the main contribution to their masses comes from the non-perturbative term $\sim S$ in the superpotential \eqref{(8.9)}
\bq
\mu_3(\Phi_{22})\sim\frac{\langle S\rangle}{\langle m^{\rm tot}_{Q,2}\rangle^2}\frac{1}{z_{\Phi}(\la,m^{\rm pole}_{Q,1}) z^{\,\prime}_{\Phi}(m^{\rm pole}_{Q,1},m^{\rm pole}_{Q,2})}\sim m^{\rm pole}_{Q,2}\sim\lym^{(\rm br2)}. \label{(8.11)}
\eq
In such a situation there is no guaranty that there is a pole in the propagator of $\Phi_{22}$ at the momentum $p\sim m^{\rm pole}_{Q,2}$. May be yes but maybe not, see section 5.\\

\subsection{\quad Special vacua,\,\, $n_1=\nd,\,\, n_2=N_c$}

\hspace*{4mm} In these vacua at $\la\ll\mph\ll\mo$, see \eqref{(4.7)},\eqref{(4.10)},
\bbq
\langle\Qo\rangle_{\rm spec}=\frac{N_c}{2N_c-N_F}(m_Q\mph)\,,\,\, \langle\Qt\rangle_{\rm spec}=\la^2\Bigl (\frac{\la}{\mph}\Bigr )^{\frac{\nd}{2N_c-N_F}}\,,
\eeq
\bq
\frac{\langle\Qo\rangle_{\rm spec}}{\langle\Qt\rangle_{\rm spec}}\sim\Bigl (\frac{\mph}{\mo}\Bigr )^{\frac{N_c}{2N_c-N_F}}\ll 1\,. \label{(8.12)}
\eq

The most important possible masses look here as follows
\bbq
\langle m^{\rm tot}_{Q,1}\rangle=\frac{\langle\Qt\rangle_{\rm spec}}{\mph}\sim\la\Bigl (\frac{\la}{\mph}\Bigr )^{\frac{N_c}{2N_c-N_F}}\,\,\ra\,\, m^{\rm pole}_{Q,1}\sim\la\Bigl (\frac{\la}{\mph}\Bigr )^{\frac{N_F}{3(2N_c-N_F)}}\gg m^{\rm pole}_{Q,2}\,,
\eeq
\bbq
\mu^2_{\rm gl,2}\sim (a_*\sim 1)\langle\Qt\rangle_{\rm spec}\Bigl (\frac{\mu_{\rm gl,2}}{\la}\Bigr )^{\frac{\bo}{N_F}}\,\,\ra\,\, \mu_{\rm gl,2}\sim \la\Bigl (\frac{\la}{\mph}\Bigr )^{\frac{N_F}{3(2N_c-N_F)}}\sim  m^{\rm pole}_{Q,1}\gg\mu_{\rm gl,1}\,,
\eeq
where $\mu_{\rm gl,2}$ is the gluon mass due to possible higgsing of ${\ov Q}_2, Q^2$ quarks. Therefore, the overall phase is $HQ_1-Higgs_2$ and the whole gauge group is higgsed at $\mu\sim\mu_{\rm gl,2}$. Supposing that $m^{\rm pole}_{Q,1}=(\rm several)\mu_{\rm gl,2}$ and integrating out first the quarks ${\ov Q}_1, Q^1$ as heavy ones at $\mu<m^{\rm pole}_{Q,1}$ and then all higgsed gluons and their superpartners at $\mu<\mu_{gl,2}$, the Lagrangian takes the form
\bq
K={\rm Tr}\,\Biggl [\, z_{\Phi}(\Phi^\dagger\Phi)+ z_Q\Biggl ( 2\sqrt {\Pi^\dagger_{22}\Pi_{22} }+B^{\dagger}_2 B_2+{\ov B}^{\,\dagger}_2{\ov B}_2\, \Biggr )\,\Biggr ]\,, \label{(8.13)}
\eq
\bbq
z_Q=z_Q(\la,m^{\rm pole}_{Q,1})=\Bigl (\frac{m^{\rm pole}_{Q,1}}{\la}\Bigr )^{\bo/N_F}\,,\,\, z_{\Phi}=z_{\Phi}(\la,m^{\rm pole}_{Q,1})=1/z^2_Q\,,
\eeq
\bbq
\cw=\cw_{\rm non-pert}+\cw_{\Phi}+{\rm Tr}\,\Pi_{22}\Bigl (m^{\rm tot}_{Q,2}-\Phi_{21}\frac{1}{m^{\rm tot}_{Q,1}}\Phi_{12}\Bigr ),\quad \cw_{\Phi}=\frac{\mph}{2}\Biggl [{\rm Tr}\, (\Phi^2) -\frac{1}{\nd}\Bigl ({\rm Tr}\,\Phi\Bigr)^2\Biggr ],
\eeq
\bbq
m^{\rm tot}_{Q,1}=m_Q-\Phi_{11}\,,\,\,  m^{\rm tot}_{Q,2}=m_Q-\Phi_{22}\,,
\eeq
where for the non-perturbative term we use the form proposed in \cite{S1}
\bq
\cw_{\rm non-pert}=A\Bigl [1-\frac{\det\Pi_{22}}{\lambda^{2N_c}}+\frac{{\ov B}_2 B_2}{\lambda^2}\Bigr ],\,\langle A\rangle=\langle S\rangle,\,\lambda^2=\Bigl (\la^{\bo}\det m^{\rm tot}_{Q,1}\Bigr )^{\frac{1}{N_c}},\,\langle\lambda^2\rangle=\langle\Qt\rangle,\,\, \label{(8.14)}
\eq
in which $A$ is the auxiliary field.

From \eqref{(8.13)},\eqref{(8.14)}, the hybrids $\Phi_{12}, \Phi_{21}$ are massless, the baryons ${\ov B}_2,\, B_2$ are light
\bq
\mu(B_2)=\mu({\ov B}_2)\sim \frac{m_Q}{z_Q}\sim m_Q\Bigl (\frac{\mph}{\la}\Bigr )^{\frac{\bo}{3(2N_c-N_F)}}\ll \mu_{\rm gl,2}\,, \label{(8.15)}
\eq
while all other masses are parametrically $\sim\mu_{\rm gl,2}\sim m^{\rm pole}_{Q,1}$ (the pion masses increased due to their mixing with the fions). Besides, in particular, because $\mu_o^{\rm conf}\sim m^{\rm pole}_{Q,1}$ in these special vacua, there is no warranty that these nonzero masses of fions $\Phi_{11}$ and $\Phi_{22}$ are the pole masses. Maybe yes, but maybe not (see section 5).

On the whole, there are three scales in the mass spectrum\,:\, the hybrid fions $\Phi_{12}, \Phi_{21}$ are massless, the baryons have small masses \eqref{(8.15)}, while all other masses are $\mu_{\rm gl,2}\sim m^{\rm pole}_{Q,1}\sim\la(\la/\mph)^{N_F/3(2N_c-N_F)}$ in these special vacua at $\la\ll\mph\ll\mo$.\\

\section{Dual theory. Broken flavor symmetry, $\mathbf{\la\ll\mph\ll\mo}$}

\subsection{\quad L - type vacua, $\,\,\bd/N_F\ll 1$}

\hspace*{4mm} The condensates of mions and dual quarks look here as
\bbq
\langle M_1+M_2-\frac{1}{N_c}{\rm Tr}\,M\rangle_{Lt}=m_Q\mph\quad\ra\quad \frac{\langle M_1\rangle_{Lt}}{\langle M_2\rangle_{Lt}}\approx -\,\frac{N_c-n_1}{N_c-n_2}\,,
\eeq
\bbq
\langle M_1\rangle_{Lt}\langle\qo\rangle_{Lt}=\langle M_2\rangle_{Lt}\langle\qt\rangle_{Lt}=Z_q\la\langle S\rangle_{Lt},\quad\langle S\rangle_{Lt}=\frac{\langle M_1\rangle_{Lt}\langle M_2\rangle_{Lt}}{\mph}\,.
\eeq

I.e., all condensates are parametrically the same as in the L - vacua with unbroken flavor symmetry in section 7.1 and the overall phase is also $HQ_1-HQ_2$. The pole masses of dual quarks are as in \eqref{(7.3)}, the Lagrangian of mions is as in \eqref{(7.6)} and the pole masses of mions $M_{11}$ and $M_{22}$ are as in \eqref{(7.7)}. But the masses of hybrid mions $M_{12}$ and $M_{21}$ are qualitatively different here. They are the Nambu-Goldstone particles now and are exactly massless, $\mu(M_{12})=\mu (M_{21})=0$.\\

\subsection{\quad  $\rm br2$ vacua, $\,\,\bd/N_F=O(1)$}

\hspace*{4mm} In these vacua with $n_2>N_c\,, 1\leq n_1<\nd$ the condensates of mions and dual quarks look as
\bbq
\langle M_1\rangle_{\rm br2}=\langle\Qo\rangle_{\rm br2}\sim \la^2\Bigl(\frac{\mph}{\la}\Bigr )^{\frac{n_2}{n_2-N_c}}\Bigl (\frac{m_Q}{\la}\Bigr )^{\frac{N_c-n_1}{n_2-N_c}},
\eeq
\bq
\langle M_2\rangle_{\rm br2}=\langle\Qt\rangle_{\rm br2}\approx -\,\frac{n_2-N_c}{N_c}\, m_Q\mph\,,\quad
\frac{\langle M_1\rangle_{\rm br2}}{\langle M_2\rangle_{\rm br2}}\sim \Bigl (\frac{\mph}{\mo}\Bigr )^{\frac{N_c}{n_2-N_c}}\ll 1\,, \label{(9.1)}
\eq
\bbq
\langle\qo\rangle_{\rm br2}=\langle{\ov q}^1 q_1(\mu=\la)\rangle_{\rm br2}=\frac{\la\langle S\rangle_{\rm br2}}{\langle M_1\rangle_{\rm br2}}=\frac{\la\langle M_2\rangle_{\rm br2}}{\mph}\sim m_Q\la\gg\langle \qt\rangle_{\rm br2}\,.
\eeq

From these, the heaviest are $N_F^2$ mions $M^i_j$ with the pole masses
\bq
\hspace*{-4mm}\mu^{\rm pole}(M)=\frac{\la^2/\mph}{z_M(\la,\mu^{\rm pole}(M))}\sim \la\Bigl (\frac{\la}{\mph}\Bigr )^{\frac{N_F}{3(2N_c-N_F)}}\,,\label{(9.2)}
\eq
\bbq
z_M(\la,\mu^{\rm pole}(M))\sim\Bigl (\frac{\la}{\mu^{\rm pole}(M)}\Bigr )^{\frac{2\bd}{N_F}}\gg 1,
\quad \bd=3\nd-N_F\,,
\eeq
while some other possible characteristic masses look as
\bq
\mu_{q,2}=\frac{\langle M_2\rangle}{\la}\sim\frac{m_Q\mph}{\la},\quad {\tilde\mu}^{\rm pole}_{q,2}\sim\la
\Bigl (\frac{m_Q\mph}{\la^2}\Bigr )^{N_F/3\nd}\gg\mu^{\rm pole}_{q,1}\,, \label{(9.3)}
\eq
\bbq
{\ov\mu}_{\rm gl,1}\sim\la\Bigl (\frac{\langle\qo\rangle}{\la^2}\Bigr )^{N_F/3N_c}\sim \la\Bigl(\frac{m_Q}{\la}\Bigr )^{N_F/3N_c}\gg{\ov\mu}_{\rm gl,2}\,,\quad\frac{{\ov\mu}_{\rm gl,1}}{{\tilde\mu}^{\rm pole}_{q,2}}
\sim \Bigl(\frac{\mo}{\mph}\Bigr )^{N_F/3\nd}\gg 1\,,
\eeq
where $\mu^{\rm pole}_{q,1}$ and ${\tilde\mu}^{\rm pole}_{q,2}$ are the perturbative pole masses of quarks ${\ov q}^1, q_1$ and ${\ov q}^2, q_2$ and ${\ov\mu}_{\rm gl,1},\, {\ov\mu}_{\rm gl,2}$ are the gluon masses due to possible higgsing of these quarks. Hence, the largest mass is ${\ov\mu}_{\rm gl,1}$ and the overall phase is $Higgs_1-HQ_2$.

After integrating out all higgsed gluons and quarks ${\ov q}^1, q_1$, we write the dual Lagrangian at $\mu=\muo$ as
\bbq
K= z_M(\la,\muo){\rm Tr}\,\frac{M^\dagger M}{\la^2}+ z_q(\la,\muo){\rm Tr}\,\Bigl [\,2\sqrt{N_{11}^\dagger N_{11}}+K_{\rm hybr}+\Bigl ({\dq}^{\dagger}_2 {\dq}_2 +({\dq}_2\ra {\odq}_2 )\Bigr )\,\Bigr ]\,,
\eeq
\bq
K_{\rm hybr}=\Biggl (N^{\dagger}_{12}\frac{1}{\sqrt{N_{11} N^{\dagger}_{11}}} N_{12}+
N_{21}\frac{1}{\sqrt{N^{\dagger}_{11} N_{11}}} N^\dagger_{21}\Biggr ),\quad z_q(\la,\muo)=\Bigl (\frac{\muo}{\la}\Bigr )^{\bd/N_F}\,,\label{(9.4)}
\eq
\bbq
z_M(\la,\muo)=1/z^2_q(\la,\muo),\quad \cw=\Bigl [-\frac{2\pi}{{\ov\alpha}(\mu)}{\ov{\textsf s}}
\Bigr ]-\frac{1}{\la}{\rm Tr}\,\Bigl ({\odq}_2 M_{22}\dq_2\Bigr )- \cw_{MN}+\cw_{M},
\eeq
\bbq
\cw_{MN}=\frac{1}{\la}{\rm Tr}\,\Bigl (M_{11}N_{11}+M_{21} N_{12}+N_{21}
M_{12}+M_{22} N_{21}\frac{1}{N_{11}} N_{12}\Bigr )\,,
\eeq
where the nions (dual pions) $N_{11}$ originate from higgsing of ${\ov q}^1, q_1$ dual quarks while $\odq^2, \dq_2$ are the active quarks ${\ov q}^2, q_2$ with unhiggsed colors, $\ov{\textsf s}$ is the field strength of unhiggsed dual gluons and the hybrid nions $N_{12}$ and $N_{21}$ are, in essence, the quarks ${\ov q}^2, q_2$ with higgsed colors, $\cw_M$ is given in \eqref{(2.7)}. The lower energy theory at $\mu<\muo$ has $\nd^{\,\prime}=\nd-n_1$ colors and $n_2>N_c$ flavors, $\bd^{\,\prime}=\bd-2n_1<\bd$. We consider here only the case $\bd^{\,\prime}>0$ when it remains in the conformal window. In this case the value of the pole mass $\qtp$ in this lower energy theory is
\bq
\qtp\sim \frac{\langle M_2\rangle}{\la}\frac{1}{z_q(\la,\muo) z^{\,\prime}_q(\muo,\qtp)}\sim\lym^{(\rm br2)}\,, \quad z^{\,\prime}_q(\muo,\qtp)\sim\Bigl (\frac{\qtp}{\muo}\Bigr )^{\bd^{\,\prime}/n_2}\ll 1\,. \label{(9.5)}
\eq

The fields $N_{11}, N_{12}, N_{21}$ and $M_{11}, M_{12}, M_{21}$ are frozen and do not evolve at $\mu<\muo$.  After integrating out remained unhiggsed quarks $\odq^2, \dq_2$ as heavy ones and unhiggsed gluons at $\mu<\lym^{(\rm br2)}$ the Lagrangian of mions and nions looks as, see \eqref{(9.4)},
\bbq
K=z_M(\la,\muo){\rm Tr}\,K_M+z_q(\la,\muo)\Bigl [\,2\sqrt{N_{11}^\dagger N_{11}}+K_{\rm hybr}\,\Bigr ],\,\, z^{\,\prime}_M (\muo,\qtp)\sim\Bigl (\frac{\muo}{\qtp}\Bigr )^{\frac{2\bd^{\,\prime}}{n_2}}\gg 1,
\eeq
\bq
K_M=\frac{1}{\la^2}\Bigl (M_{11}^\dagger M_{11}+M_{12}^\dagger M_{12}+M_{21}^\dagger M_{21}+
z^{\,\prime}_M (\muo,\qtp)M_{22}^\dagger M_{22}\Bigr )\,, \label{(9.6)}
\eq
\bbq
\cw=-\nd^{\,\prime}S-\cw_{MN}+\cw_M\,,\quad S=\Bigl (\lym^{(\rm br2)}\Bigr )^3 \Biggl (\det\frac{\langle N_1\rangle}{N_{11}}\det\frac{M_{22}}{\langle M_2\rangle}\Biggr )^{1/\nd^{\,\prime}},\quad \lym^{(\rm br2)}\sim\Bigl (m_Q\langle M_1\rangle\Bigr )^{1/3}.
\eeq

From \eqref{(9.6)}, the "masses" of mions look as
\bq
\mu(M_{11})\sim\mu(M_{12})\sim\mu(M_{21})\sim\frac{\la^2}{z_M(\la,\muo)\mph}\sim\Bigl (\frac{\mo}{\mph}\Bigr )\muo\gg\muo\,,\label{(9.7)}
\eq
\bq
\mu(M_{22})\sim\frac{\la^2}{z_M(\la,\muo)z^\prime_M (\muo,\qtp)\mph}\sim \Bigl (\frac{\mo}{\mph}\Bigr )^{\frac{3N_c-n_2}{3(n_2-N_c)}}\,\muo\gg \muo\,,\label{(9.8)}
\eq
while the pole masses of nions $N_{11}$ are
\bq
\mu^{\rm pole}(N_{11})\sim \frac{\mph\langle N_1\rangle_{\rm br2}}{z_q(\la,\muo)\la^2}\sim\Bigl (\frac{\mph}{\mo}\Bigr )\muo\,,\label{(9.9)}
\eq
and the hybrid nions $N_{12}, N_{21}$ are massless, $\mu(N_{12})=\mu(N_{21})=0$. The mion "masses" \eqref{(9.7)},\eqref{(9.8)} are not the pole masses but simply the low energy values of mass terms in their propagators, the only pole masses are given in \eqref{(9.2)}.\\

\subsection{\quad   $\rm br2$ vacua, $\bd/N_F\ll 1$}

Instead of \eqref{(9.2)}, the pole mass of mions is parametrically smaller now, see \eqref{(7.1)},
\bq
\mu^{\rm pole}(M)=\frac{Z_q^{\, 2}\la^2/\mph}{z_M(\la,\mu^{\rm pole}(M))}\sim Z_q^{\, 2}\la\Bigl (\frac{\la}{\mph}\Bigr )^{\frac{N_F}{3(2N_c-N_F)}},\quad \frac{\mu^{\rm pole}(M)}{\mu^{\rm pole}_2(\Phi)}\sim Z_q^{\, 2}\ll 1\,, \label{(9.10)}
\eq
while instead of \eqref{(9.3)} we have now
\bq
\mu_{q,2}=\frac{\langle M_2\rangle}{Z_q\la}\sim\frac{m_Q\mph}{Z_q\la},\quad {\tilde\mu}^{\rm pole}_{q,2}\sim\frac{\la}{Z_q}\Bigl (\frac{m_Q\mph}{\la^2}\Bigr )^{N_F/3\nd}\gg\mu^{\rm pole}_{q,1}\,,\label{(9.11)}
\eq
\bq
{\ov\mu}_{\rm gl,1}\sim \la\Bigl (\frac{\langle N_1\rangle}{\la^2}\Bigr )^{N_F/3N_c}\sim Z_q^{1/2}\la\Bigl(\frac{m_Q}{\la}\Bigr )^{N_F/3N_c}\gg{\ov\mu}_{\rm gl,2}\,,\quad \frac{{\ov\mu}_{\rm gl,1}}{m^{\rm pole}_{Q,1}}\sim Z_q^{1/2}\ll 1\,,\label{(9.12)}
\eq
\bq
\quad\frac{{\ov\mu}_{\rm gl,1}}{{\tilde\mu}^{\rm pole}_{q,2}}\sim Z_q^{3/2}\Bigl(\frac{\mo}{\mph}\Bigr )^{N_F/3\nd}\gg 1\,,\quad \la\ll {\mph}\ll Z_q^{\,3/2}\mo\,,\quad Z_q\sim\exp\{-\frac{\nd}{7\bd}\}\ll 1\,.\label{(9.13)}
\eq

Hence, at the condition \eqref{(9.13)}, the largest mass is ${\ov\mu}_{\rm gl,1}$ and the overall phase is also $Higgs_1-HQ_2$. But now, at $\bd/\nd\ll 1$, it looks unnatural to require $\bd^{\, \prime}=(\bd-2n_1)>0$. Therefore, with $n_1/\nd=O(1)$, the lower energy theory at $\mu<{\ov\mu}_{\rm gl,1}$ has $\bd^{\, \prime}<0$ and is in the logarithmic IR free regime in the range of scales $\qtp<\mu<{\ov\mu}_{\rm gl,1}$. Then instead of \eqref{(9.5)} (ignoring all logarithmic renormalization factors),
\bq
\lym^{(\rm br2)}\ll\qtp\sim \frac{\langle M_2\rangle_{\rm br2}}{Z_q\la}\frac{1}{z_q(\la,\muo)}\sim\frac
{\mph}{Z_q^{3/2}\mo}\,{\ov\mu}_{\rm gl,1}\ll {\ov\mu}_{\rm gl,1}\,.\label{(9.14)}
\eq

The Lagrangian of mions and nions has now the form \eqref{(9.6)} with accounting additionally for $Z_q$ factors, and with a replacement $z^{\,\prime}_M (\muo,\qtp)\sim 1$, and so $\mu(M_{22})\sim\mu(M_{11})\sim\mu(M_{12})\sim\mu(M_{21})$ now, see \eqref{(9.7)},\eqref{(9.8)},\eqref{(9.13)},
\bq
\mu(M^i_j)\sim\frac{Z_q^2\la^2}{z_M(\la,\muo)\mph}\sim Z_q^{3/2}\Bigl (\frac{\mo}{\mph}\Bigr )\muo\gg\muo\,,\label{(9.15)}
\eq
while, instead of \eqref{(9.9)}, the mass of nions looks now as
\bq
\mu^{\rm pole}(N_{11})\sim \frac{\mph\langle N_1\rangle_{\rm br2}}{z_q(\la,\muo)\la^2}\sim Z_q^{1/2}\Bigl (\frac{\mph}{\mo}\Bigr )\,\muo\,.\label{(9.16)}
\eq

On the whole for the mass spectra in this case. a) The heaviest are $N_F^2$ mions with the pole masses \eqref{(9.10)} (the 'masses' \eqref{(9.15)} are not the pole masses but simply the low energy values of mass terms in the mion propagators). \, b) The next are the masses \eqref{(9.12)} of $n_1(2\nd-n_1)$ higgsed gluons and their superpartners. \, c) There is a large number of flavored hadrons, mesons and baryons, made of non-relativistic and weakly confined (the string tension is $\sqrt{\sigma}\sim\lym^{(\rm br2)}\ll\qtp$\,) quarks $\odq^2, \dq_2$ with unhiggsed colors. The mass spectrum of low-lying flavored mesons is Coulomb-like with parametrically small mass differences, $\Delta\mu_H/\mu_H=O(\bd^{\,2}/N^2_F)\ll 1$.\, d) A large number of gluonia made of $SU(\nd-n_1)$ gluons with the mass scale $\sim\lym^{(\rm br2)}$.\, e) $n_1^2$ nions $N_{11}$ with the masses \eqref{(9.16)}.\, f) The hybrid nions $N_{12}, N_{21}$ are the Nambu-Goldstone particles here and are massless.\\

\subsection{\quad  Special vacua,\,\, $n_1=\nd,\,\, n_2=N_c$}

\hspace*{4mm}The most important possible masses look here as follows,
\bq
\langle M_1\rangle_{\rm spec}=\frac{N_c}{2N_c-N_F}(m_Q\mph)\,,\quad \langle M_2\rangle_{\rm spec}=\la^2\Bigl (\frac{\la}{\mph}\Bigr )^{\nd/(2N_c-N_F)}\gg
\langle M_1\rangle_{\rm spec}\,,\label{(9.17)}
\eq
\bbq
\mu_{q,2}=\frac{\langle M_2\rangle}{\la}\,,\quad \mu^{\rm pole}_{q,2}\sim\la\Bigl (\frac{\langle M_2\rangle}{\la^2}\Bigr )^{N_F/3\nd}\sim\la\Bigl (\frac{\la}{\mph}\Bigr )^{N_F/3(2N_c-N_F)}\gg \mu^{\rm pole}_{q,1}\,,
\eeq
\bbq
{\ov\mu}_{\rm gl,1}\sim \la\Bigl (\frac{\langle N_1\rangle}{\la^2}\Bigr )^{N_F/3N_c}\sim \mu^{\rm pole}_{q,2}\gg {\ov\mu}_{\rm gl,2}\,,
\eeq
where ${\ov\mu}_{\rm gl,1}$ is the gluon mass due to possible higgsing of ${\ov q}^1, q_1$ quarks. Therefore, the overall phase is $Higgs_1-HQ_2$
\footnote{\,
taking $\bd/\nd\ll 1$ and using the results from \cite{ch3} we obtain $\mu^{\rm pole}_{q,2}/{\ov\mu}_{\rm gl,1}\sim\exp \{3\nd/14\bd\}\gg 1$.
}
and the whole dual gauge group will be higgsed.

We proceed now as in the section 8.3. I.e., after integrating out first the quarks ${\ov q}^2, q_2$ as heavy ones at $\mu<\mu^{\rm pole}_{q,2}$ and then all higgsed dual gluons and their superpartners at $\mu<{\ov\mu}_{\rm gl,1}$, the lower energy Lagrangian takes the form
\bbq
K={\rm Tr}\,\Biggl [\, z_M\frac{M^\dagger M}{\la^2}+ z_q\Biggl (\sqrt{N^\dagger_{11} N_{11}}+b^{\dagger}_1 b_1+
{\ov b}^{\,\dagger}_1 {\ov b}_1\Biggr )\,\Biggr ]\,,
\eeq
\bbq
z_q=z_q(\la,\mu^{\rm pole}_{q,2})=\Bigl (\frac{\mu^{\rm pole}_{q,2}}{\la}\Bigr )^{\bd/N_F}\,,\quad z_M=z_M(\la,\mu^{\rm pole}_{q,2})=1/z^2_q\,,
\eeq
\bq
\cw=\cw_{\rm non-pert}-\cw_M -\frac{1}{\la}{\rm Tr}\, N_{11}\Bigl (M_{11}-M_{12}\frac{1}{M_{22}}M_{21}\Bigr )\,,\label{(9.18)}
\eq
\bbq
\cw_M=\frac{1}{2\mph}\Biggl [{\rm Tr} (M^2) -\frac{1}{N_c}\Bigl ({\rm Tr}\,M\Bigr)^2\Biggr ]+m_Q{\rm Tr}\,M\,,
\eeq
where the non-perturbative term looks here as
\bq
\cw_{\rm non-pert}={\ov A}\,\Bigl [\,1-\frac{\det N_{11}}{{\ov\lambda}^{2\nd}}+\frac{{\ov b}_1 b_1}{{\ov\lambda}^2}\,
\Bigr ]\,,\quad \langle \ov A\rangle=\langle S\rangle=\frac{\langle M_1\rangle\langle M_2\rangle}{\mph}\,,\label{(9.19)}
\eq
\bbq
{\ov\lambda}^2=\Biggl (\la^{\bd}\det \frac{M_{22}}{\la}\Biggr )^{1/\nd}\,,\quad \langle{\ov\lambda}^2\rangle=
\langle N_1\rangle=\langle m^{\rm tot}_{Q,1}\rangle\la=\frac{\langle M_2\rangle\la}{\mph}\,,
\eeq
and $\ov A$ is the auxiliary field.

From \eqref{(9.18)},\eqref{(9.19)}\,:\, the hybrids $M_{12}, M_{21}$ are massless, the baryons ${\ov b}_1,\, b_1$ are light
\bq
\mu(b_1)=\mu ({\ov b}_1)\sim \frac{\langle M_1\rangle}{z_q\la}\sim m_Q\Bigl (\frac{\mph}{\la}\Bigr )^{\frac{\bo}{3(2N_c-N_F)}}\ll {\ov\mu}_{\rm gl,1}\,,\label{(9.20)}
\eq
while all other masses are $\sim{\ov\mu}_{\rm gl,1}\sim \mu^{\rm pole}_{q,2}$ (the nion masses increased due to their mixing with the mions). Besides, in particular, because $\mu_o^{\rm conf}\sim \mu^{\rm pole}_{q,2}$ in these special vacua, there is no warranty that these nonzero masses of mions $M_{11}$ and $M_{22}$ are the pole masses. Maybe so but maybe not (see section 5).

On the whole, there are three scales in the mass spectrum\,:\, the hybrid mions $M_{12}, M_{21}$ are massless, the baryon masses are \eqref{(9.20)}, while all other masses are $\sim {\ov\mu}_{\rm gl,1}\sim\mu^{\rm pole}_{q,2}\sim\\
\sim\la(\la/\mph)^{N_F/3(2N_c-N_F)}$ in these special vacua at $\la\ll\mph\ll{\ov\mu}^{(\rm DC)}_{\Phi}$.\\

\section{\hspace*{-2mm} Direct theory. Broken flavor symmetry, $\mathbf{\mo\ll\mph\ll\frac{\la^2}{m_Q}}$}

\subsection{\quad  $\rm br1$ vacua}

\hspace*{4mm} The values of quark condensates are here
\bq
\langle\Qo\rangle_{\rm br1}\approx\frac{N_c}{N_c-n_1}\, m_Q\mph\,,\quad \langle\Qt\rangle_{\rm br1}\sim \la^2\Bigl (\frac{\la}{\mph}\Bigr )^{\frac{n_1}{N_c-n_1}}\Bigl (\frac{m_Q}{\la}\Bigr )^{\frac{n_2-N_c}{N_c-n_1}}\,, \label{(10.1)}
\eq
\bbq
\frac{\langle\Qt\rangle_{\rm br1}}{\langle\Qo\rangle_{\rm br1}}\sim\Bigl (\frac{\mo}{\mph}\Bigr )^{\frac{N_c}{N_c-n_1}}\ll 1\,.
\eeq
From these, the values of some potentially relevant masses look as
\bbq
\mgo^2\sim \Bigl (a_*\sim 1\Bigr )z_Q(\la,\mgo)\langle\Qo\rangle_{\rm br1} \,,\quad z_Q(\la,\mgo)\sim
\Bigl (\frac{\mgo}{\la}\Bigr )^{\bo/N_F}\,,
\eeq
\bq
\mgo\sim\la \Bigl (\frac{m_Q\mph}{\la^2}\Bigr )^{N_F/3\nd}\gg\mgt\,, \label{(10.2)}
\eq
\bbq
\langle m^{\rm tot}_{Q,2}\rangle=\frac{\langle\Qo\rangle_{\rm br1}}{\mph}\sim m_Q\,,\quad  {\tilde m}^{\rm pole}_{Q,2}=\frac{\langle m^{\rm tot}_{Q,2}\rangle_{\rm br1}}{z_Q(\la,{\tilde m}^{\rm pole}_{Q,2})}\,,
\eeq
\bq
{\tilde m}^{\rm pole}_{Q,2}\sim\la\Bigl (\frac{m_Q}{\la}\Bigr )^{N_F/3N_c}\gg m^{\rm pole}_{Q,1},\quad\,\,\, \frac{{\tilde m}^{\rm pole}_{Q,2}}{\mgo}\sim\Bigl (\frac{\mo}{\mph}\Bigr )^{N_F/3\nd}\ll 1\,. \label{(10.3)}
\eq
Hence, the largest mass is $\mgo$ due to higgsing of ${\ov Q}_1, Q^1$ quarks and the overall phase is $Higgs_1-HQ_2$.

The lower energy theory at $\mu<\mgo$ has $N_c^\prime=N_c-n_1$ colors and $n_2\geq N_f/2$ flavors. At $2n_1<\bo$ it remains in the conformal window with ${\rm b}_o^\prime>0$, while at $2n_1>\bo,\,\,{\rm b}_o^\prime<0$ it enters the logarithmic IR free perturbative regime.

We start with $\bo^\prime>0$. Then the value of the pole mass of quarks $\oqt,\,\sqt$ with unhiggsed colors looks as
\bbq
\ma=\frac{\langle m^{\rm tot}_{Q,2}\rangle_{\rm br1}}{z_Q(\la,\mgo)z_Q^{\,\prime}(\mgo,
\ma)}\,,\quad z_Q^{\,\prime}(\mgo,\ma)\sim\Bigl (\frac{\ma}{\mgo}\Bigr )^{{\rm b}_o^\prime/n_2}\,,
\eeq
\bq
\ma\sim\la\Bigl (\frac{\la}{\mph}\Bigr )^{\frac{n_1}{3(N_c-n_1)}}\Bigl (\frac{m_Q}{\la}\Bigr )^{\frac{n_2-n_1}{3(N_c-n_1)}}\sim\lym^{(\rm br1)}\,.\label{(10.4)}
\eq
It is technically convenient to retain all fion fields $\Phi$ although, in essence, they are too heavy and dynamically irrelevant at $\mph\gg\mo$. After integrating out all heavy higgsed gluons and quarks ${\ov Q}_1, Q^1$, we write the Lagrangian at $\mu=\mu_{\rm gl,1}$  in the form
\bbq
K=\Bigl [\,z_{\Phi}(\la,\mgo){\rm Tr}(\Phi^\dagger\Phi)+z_Q(\la,\mu^2_{\rm gl,1})\Bigl (K_{{\sq}_2}+K_{\Pi}\Bigr )\,\Bigr ],\quad z_{\Phi}(\la,\mgo)=1/z^2_Q(\la,\mgo)\,,
\eeq
\bq
K_{{\sq}_2}={\rm Tr}\Bigl ({\sq}^{\dagger}_2 {\sq}^2 +({\sq}^2\ra
{\oq}_2 )\Bigr )\,, \quad K_{\Pi}= 2{\rm Tr}\sqrt{\Pi^{\dagger}_{11}\Pi_{11}}+K_{\rm hybr}, \label{(10.5)}
\eq
\bbq
K_{\rm hybr}={\rm Tr}\Biggl (\Pi^{\dagger}_{12}\frac{1}{\sqrt{\Pi_{11}\Pi^{\dagger}_{11}}}\Pi_{12}+
\Pi_{21}\frac{1}{\sqrt{\Pi^{\dagger}_{11}\Pi_{11}}}\Pi^\dagger_{21}\Biggr ),
\eeq
\bbq
\cw=\Bigl [-\frac{2\pi}{\alpha(\mu_{\rm gl,1})}{\textsf S}\Bigr ]+\frac{\mph}{2}\Biggl [{\rm Tr}\, (\Phi^2) -\frac{1}{\nd}\Bigl ({\rm Tr}\,\Phi\Bigr)^2\Biggr ]+{\rm Tr}\Bigl ({\oq_2}m^{\rm tot}_{{\sq}_2}{\sq}^2\Bigr )+\cw_{\Pi},
\eeq
\bbq
\cw_{\Pi}= {\rm Tr}\Bigl (m_Q\Pi_{11}+m^{\rm tot}_{{\sq}_2}\,\Pi_{21}\frac{1}{\Pi_{11}}\Pi_{12}\Bigr )-
{\rm Tr}\Bigl (\Phi_{11}\Pi_{11}+\Phi_{12}\Pi_{21}+\Phi_{21}\Pi_{12} \Bigr ),
\quad m^{\rm tot}_{{\sq}_2}=(m_Q-\Phi_{22}),
\eeq
In \eqref{(10.5)}: $\oqt,\, \sqt$ and $\textsf V$ are the active ${\ov Q}_2, Q^2$ quarks and gluons with unhiggsed colors ($\textsf S$ is their field strength squared), $\Pi_{12}, \Pi_{21}$ are the hybrid pions (in essence, these are the quarks ${\ov Q}_2, Q^2$ with higgsed colors), $z_Q(\la,\mu^2_{\rm gl,1})$ is the corresponding perturbative renormalization factor of massless quarks, see \eqref{(10.2)}, while $z_{\Phi}(\la,\mgo)$ is that of fions. Evolving now down in the scale and integrating out at $\mu<\lym^{(\rm br1)}$ quarks  $\oq_2,\, \sq^2$ as heavy ones and unhiggsed gluons via the VY-procedure, we obtain the Lagrangian of pions and fions
\bbq
K=\Bigl [z_{\Phi}(\la,\mgo){\rm Tr}\Bigl(\Phi^\dagger_{11}\Phi_{11}+\Phi^\dagger_{12}\Phi_{12}+
\Phi^\dagger_{21}\Phi_{21}+z^{\,\prime}_{\Phi}(\mgo,\ma)\Phi^\dagger_{22}\Phi_{22}\Bigr )+z_Q(\la,\mu^2_{\rm gl,1})K_{\Pi}\Bigr ],\,
\eeq
\bq
\cw=(N_c-n_1)S+W_{\Phi}+W_{\Pi}\,,\quad S=\Biggl [\frac{\la^{\bo}\det m^{\rm tot}_{{\sq}_2}}{\det \Pi_{11}}\Biggr ]^{\frac{1}{N_c-n_1}}\,, \label{(10.6)}
\eq
\bbq
\cw_{\Phi}=\frac{\mph}{2}\Biggl [{\rm Tr} (\Phi^2) -\frac{1}{\nd}\Bigl ({\rm Tr}\,\Phi\Bigr)^2\Biggr ],\quad
z^{\,\prime}_{\Phi}(\mgo,\ma)\sim\Bigl (\frac{\mgo}{\ma}\Bigr )^{2{\rm b}_o^\prime/n_2}\,.
\eeq
We obtain from \eqref{(10.6)} that all fions are heavy with the "masses"
\bq
\mu(\Phi_{11})\sim\mu(\Phi_{12})\sim\mu(\Phi_{21})\sim\frac{\mph}{z_{\Phi}(\la,\mgo)}\sim\Bigl (\frac{\mph}{\mo}\Bigr )^{N_c/\nd}\mgo\gg\mgo\,, \label{(10.7)}
\eq
\bq
\mu(\Phi_{22})\sim\frac{\mph}{z_{\Phi}(\la,\mgo)z^{\,\prime}_{\Phi}(\mgo,\ma)}\sim
\Bigl (\frac{\mph}{\mo}\Bigr )^{\frac{N_c}{N_c-n_1}}\, \ma\gg \ma\,. \label{(10.8)}
\eq
These are not the pole masses but simply the low energy values of mass terms in their propagators. All fions are dynamically irrelevant at all scales $\mu<\la$. The mixings of $\Phi_{12}\leftrightarrow\Pi_{12},\, \Phi_{21}\leftrightarrow\Pi_{21}$ and $\Phi_{11}\leftrightarrow\Pi_{11}$ are parametrically small and are neglected. We obtain then for the masses of pions $\Pi_{11}$
\bq
\mu(\Pi_{11})\sim\Bigl (\frac{\mo}{\mph}\Bigr )^{\frac{N_c(\bo-2n_1)}{3\nd(N_c-n_1)}}\,\lym^{(\rm br1)}\sim\Bigl (\frac{\mo}{\mph}\Bigr )^{\frac{N_c(\bo-2n_1)}{3\nd(N_c-n_1)}}\, \ma\ll \ma\,, \label{(10.9)}
\eq
and, finally, the hybrids $\Pi_{12}, \Pi_{21}$ are massless, $\mu(\Pi_{12})=\mu(\Pi_{21})=0$.

At $2n_1>\bo$ the RG evolution at $\ma<\mu<\mgo$ is only slow logarithmic (and is neglected). We replace then $z^{\,\prime}_Q(\mgo,\ma)\sim 1$ in \eqref{(10.4)} and $z^{\,\prime}_{\Phi}(\mgo,\ma)\sim 1$ in \eqref{(10.8)} and obtain
\bq
\mu(\Phi_{22})\sim\mu(\Phi_{11})\sim\Bigl (\frac{\mph}{\mo}\Bigr )^{N_c/\nd}\mgo\gg\mgo\,, \label{(10.10)}
\eq
\bbq
\mu(\Pi_{11})\sim \ma\sim\frac{m_Q}{z_Q(\la,\mu^2_{\rm gl,1})}\sim\la\Bigl (
\frac{\la}{\mph}\Bigr )^{\bo/3\nd}\Bigl (\frac{m_Q}{\la}\Bigr )^{2\,\bd/3\nd}\sim\Bigl (\frac{\mo}{\mph}
\Bigr )^{N_c/\nd}\mgo\ll\mgo.
\eeq
\bq
\frac{\lym^{(\rm br1)}}{\ma}\sim\Bigl (\frac{\mo}{\mph}\Bigr )^{\Delta}\ll 1,\quad\Delta=\frac{N_c(2n_1-\bo)}{3\nd(N_c-n_1)}>0\,.  \label{(10.11)}
\eq
\vspace*{2mm}

\subsection{\quad  $\rm br2$ and $\rm special$ vacua}

\hspace*{4mm} At $n_2<N_c$ there are also $\rm br2$ - vacua. All their properties can be obtained by a replacement $n_1\leftrightarrow n_2$ in formulas of the preceding section 10.1. The only difference is that, because $n_2\geq N_F/2$ and so $2n_2>\bo$, there is no analog of the conformal regime at $\mu<\mu_{\rm gl,1}$ with $2n_1<\bo$. I.e. at $\mu<\mu_{\rm gl,2}$ the lower energy theory will be always in the perturbative IR free logarithmic regime and the overall phase will be $Higgs_2-HQ_1$.

As for the special vacua, all their properties can also be obtained with $n_1=\nd,\, n_2=N_c$ in formulas of the preceding section 10.1.\\

\section{\hspace*{-2mm} Dual theory. Broken flavor symmetry, $\mathbf{\mo\ll\mph\ll\frac{\la^2}{m_Q}}$}

\subsection{\quad $\rm br1$ vacua, $\,\,\bd/N_F\ll 1$}

We recall that condensates of mions and dual quarks in these vacua are
\bq
\langle M_1\rangle_{\rm br1}\approx\frac{N_c}{N_c-n_1}\, m_Q\mph\,,\quad \langle M_2\rangle_{\rm br1}\sim \la^2\Bigl (\frac{\la}{\mph}\Bigr )^{\frac{n_1}{N_c-n_1}}\Bigl (\frac{m_Q}{\la}\Bigr )^{\frac{n_2-N_c}{N_c-n_1}}\,,  \label{(11.1)}
\eq
\bbq
\frac{\langle M_2\rangle_{\rm br1}}{\langle M_1\rangle_{\rm br1}}\sim\Bigl (\frac{\mo}{\mph}\Bigr )^{\frac{N_c}{N_c-n_1}}\ll 1\,,
\eeq
\bbq
\langle N_2\rangle_{\rm br1}\equiv\langle {\ov q}^2 q_2(\mu=\la)\rangle_{\rm br1}=Z_q\frac{\langle M_1\rangle_{\rm br1}\la}{\mph}\sim Z_q m_Q\la\gg\langle N_1\rangle_{\rm br1}\,,
\eeq
and some potentially relevant masses look here as
\bq
\langle\qo\rangle=\langle{\ov q}^1 q_1(\mu=\la)\rangle=\frac{\langle M_1\rangle_{\rm br1}}{Z_q\la}\sim\frac{m_Q\mph}{Z_q\la}\,,\quad
\frac{\langle\qt\rangle}{\langle\qo\rangle}= \frac{\langle M_2\rangle_{\rm br1}}{\langle M_1\rangle_{\rm br1}}\ll 1\,,\label{(11.2)}
\eq
\bbq
Z_q\sim\exp \Bigl\{-\frac{1}{3{\ov a}_{*}}\Bigr\}\sim \exp \Bigl\{-\frac{\nd}{7\bd}\Bigr\}\ll 1\,,
\eeq
\bq
\qop\sim \frac{\la}{Z_q}\Bigl (\frac{m_Q\mph}{\la^2}\Bigr )^{N_F/3\nd}\gg\qtp\,,\quad
\frac{\lym^{(\rm br1)}}{\qop}\sim Z_q\Bigl (\frac{\mo}{\mph}\Bigr )^{\frac{n_2 N_c}{3\nd(N_c-n_1)}}\ll 1\,,\label{(11.3)}
\eq
\bbq
\mut\sim\la\Bigl (\frac{\langle N_2\rangle}{\la^2}\Bigr )^{N_F/3N_c}\sim Z_q^{1/2}\la
\Bigl (\frac{m_Q}{\la}\Bigr )^{N_F/3N_c}\gg\muo\,,
\eeq
\bq
\frac{\mut}{\qop}\sim Z_q^{3/2}\Bigl (\frac{\mo}{\mph}\Bigr )^{N_F/3\nd}\ll 1\,.\label{(11.4)}
\eq
Hence, the largest mass is $\qop$ while the overall phase is $HQ_1-HQ_2$. We consider below only the case $n_1<\bo/2$, so that the lower energy theory with $\nd$ colors and $N^\prime_F=n_2$ flavors at $\mu<\qop$ remains in the conformal window.

After integrating out the heaviest quarks ${\ov q}^1, q_1$ at $\mu<\qop$ and ${\ov q}^2, q_2$ quarks at $\mu
<\qtp$ and, finally, all $SU(\nd)$ dual gluons at $\mu<\lym^{(\rm br1)}$, the Lagrangian of mions looks as
\bq
K=\frac{z_M(\la,\qop)}{Z^2_q\la^2}\,{\rm Tr}\Bigl [\, M_{11}^\dagger M_{11}+M_{12}^\dagger M_{12}+M_{21}^\dagger M_{21}+z^{\,\prime}_{M}(\qop,\qtp)
M_{22}^\dagger M_{22} \Bigr ]\,, \label{(11.5)}
\eq
\bbq
\cw=-\nd S+\cw_M\,,\quad\quad S=\Bigl (\frac{\det M}{\la^{\bo}}\Bigr )^{1/\nd}\,,
\quad \lym^{(\rm br1)}=\langle S\rangle^{1/3}\sim\Bigl (m_Q\langle M_2\rangle\Bigr )^{1/3}\,.
\eeq
\bbq
\cw_M=m_Q{\rm Tr} M-\frac{1}{2\mph}\Bigl [\,{\rm Tr}(M^2)-\frac{1}{N_c}({\rm Tr} M)^2 \Bigr ]\,,
\quad z_M(\la,\qop)\sim \Bigl (\frac{\la}{\qop}\Bigr )^{2\,\bd/N_F}\gg 1\,.
\eeq

From \eqref{(11.5)}\,: the hybrids $M_{12}$ and $M_{21}$ are massless, $\mu(M_{12})=\mu(M_{21})=0$, while the pole mass of $M_{11}$ is (compare with \eqref{(10.9)}\,)
\bq
\mu^{\rm pole}(M_{11})\sim\frac{Z^2_q\la^2}{z_M(\la,\qop)\mph}\,,\quad \frac{\mu^{\rm pole}(M_{11})}{\lym^
{(\rm br1)}}\sim Z^2_q\Bigl (\frac{\mo}{\mph}\Bigr )^{\frac{N_c(\bo-2n_1)}{3\nd(N_c-n_1)}}\ll 1\,.\label{(11.6)}
\eq

The parametric behavior of $\qtp$ and $z^{\,\prime}_{M}(\qop,\qtp)$ depends on the value $\mph\lessgtr{\tilde\mu}_{\Phi,1}$ (see below). We consider first the case $\mph\gg{\tilde\mu}_{\Phi,1}$ so that, by definition, the lower energy theory with $\nd$ colors and $n_2$ flavors had enough "time" to evolve and entered already the new conformal regime at $\qtp<\mu\ll\qop$, with ${\rm\ov b\,}^\prime_o/\nd=(3\nd-n_2)/\nd=O(1)$ and ${\ov a\,}^\prime_*=O(1)$. Hence, when the quarks ${\ov q}^2, q_2$ decouple as heavy ones at $\mu<\qtp$, the coupling ${\ov a}_{YM}$ of the remained $SU(\nd)$ Yang-Mills theory is ${\ov a}_{YM}\sim {\ov a\,}^\prime_*=O(1)$ and this means that $\qtp\sim\lym^{(\rm br1)}$. This can be obtained also in a direct way. The running mass of quarks ${\ov q}^2, q_2$ at $\mu=\qop$ is, see \eqref{(11.1)}-\eqref{(11.3)},
\bq
\mu_{q,2}(\mu=\qop)=\frac{\langle M_2\rangle_{\rm br1}}{\langle M_1\rangle_{\rm br1}}\,\qop\,,\quad \qtp=\frac{\mu_{q,2}
(\mu=\qop)}{z^{\,\prime}_q(\qop,\qtp)}\sim \lym^{(\rm br1)}\sim\Bigl (m_Q\langle M_2\rangle\Bigr )^{1/3}\,,\label{(11.7)}
\eq
\bbq
z^{\,\prime}_q(\qop,\qtp)=\Bigl (\frac{\qtp}{\qop}\Bigr )^{\frac{{\rm\ov b\,}^\prime_o}{n_2}}\rho\,,\quad \rho=\Bigl (\frac{{\ov a\,}_*}{{\ov a\,}^\prime_*}\Bigr )^{\frac{\nd}{n_2}}\exp\Bigl\{\frac{\nd}{n_2}\Bigl (\frac{1}{{\ov a\,}_*}-\frac{1}{{\ov a\,}^\prime_*}\Bigr )\Bigr \}
\sim \exp\Bigl\{\frac{\nd}{n_2}\frac{1}{{\ov a\,}_*}\Bigr\}\gg 1\,.
\eeq
We obtain from \eqref{(11.5)} that the main contribution to the mass of mions $M_{22}$ originates from the non-perturbative term $\sim S$ in the superpotential and, using \eqref{(11.5)},\eqref{(11.7)},
\bq
z^{\,\prime}_{M}(\qop,\qtp)=\frac{{\ov a}_f(\mu=\qop)}{{\ov a}_f(\mu=\qtp)}\Bigl (\frac{1}{z^{\,\prime}_q
(\qop,\qtp)}\Bigr )^2\sim\Bigl (\frac{1}{z^{\,\prime}_q(\qop,\qtp)}\Bigr )^2\,,\label{(11.8)}
\eq
\bq
\mu(M_{22})\sim\frac{Z_q^2\la^2}{z_M(\la,\qop) z^{\,\prime}_{M}(\qop,\qtp)}\Biggl (\frac{\langle S\rangle}{\langle M_2\rangle^2}=\frac{\langle M_1\rangle}{\langle M_2\rangle}\frac{1}{\mph}\Biggr )_{\rm br1}\,\,\sim \lym^{(\rm br1)}\sim\qtp\,.\label{(11.9)}
\eq

We consider now the region $\mo\ll\mph\ll{\tilde\mu}_{\Phi,1},\, 2n_1\lessgtr\bo$ where, by definition, $\qtp$ is too close to $\qop$, so that the range of scales $\qtp<\mu<\qop$ is too small and the lower energy theory at $\mu<\qop$ has no enough "time" to enter a new regime (conformal at $2n_1<\bo$ or strong coupling one at $2n_1>\bo$) and remains in the weak coupling logarithmic regime. Then, ignoring logarithmic effects in renormalization factors, $z^{\,\prime}_q(\qop,\qtp)\sim z^{\,\prime}_{M}(\qop,\qtp)\sim 1$, and keeping as always only the exponential dependence on $\nd/\bd$\,:
\bbq
\qtp\sim\frac{\langle M_2\rangle_{\rm br1}}{\langle M_1\rangle_{\rm br1}}\,\qop\,,\quad\quad \frac{\lym^{(\rm br1)}}{\qtp}\ll 1\quad\ra\quad \mo\ll\mph\ll{\tilde\mu}_{\Phi,1}\,,
\eeq
\bq
{\tilde\mu}_{\Phi,1}\sim\exp\Bigl\{\frac{(N_c-n_1)}{2n_1}\frac{1}{{\ov a\,}_*}\Bigr\}\mo \gg \mo\,.\label{(11.10)}
\eq

The pole mass of mions $M_{22}$ looks in this case as
\bq
\frac{\mu^{\rm pole}(M_{22})}{\mu^{\rm pole}(M_{11})}\sim\frac{\langle M_1\rangle_{\rm br1}}{\langle M_2\rangle_{\rm br1}}\gg 1,\quad\frac{\mu^{\rm pole}(M_{22})}{\lym^{(\rm br1)}}\sim Z^2_q\Bigl (\frac{\mph}{\mo}\Bigr )^{\frac{2n_1 N_c}{3\nd(N_c-n_1)}}\ll 1\,.\label{(11.11)}
\eq

On the whole, see \eqref{(11.10)}, the mass spectrum at $\mo\ll\mph\ll{\tilde\mu}_{\Phi,1}$ and $2n_1\lessgtr\bo$ looks as follows. a) There is a large number of heaviest hadrons made of weakly coupled (and weakly confined, the tension of the confining string is $\sqrt{\sigma}\sim\lym^{(\rm br1)}\ll\qop)$ non-relativistic quarks ${\ov q}^1, q_1$, the scale of their masses is $\qop$, see \eqref{(11.3)}.\, b) The next physical mass scale is due to $\qtp\,:\,\, \lym^{(\rm br1)}\ll\qtp\ll\qop$. Hence, there is also a large number of hadrons made of weakly coupled and weakly confined non-relativistic quarks ${\ov q}^2, q_2$, the scale of their masses is $\qtp$, see \eqref{(11.10)}, and a large number of heavy hybrid hadrons with the masses $\sim (\qop+\qtp)$. Because all quarks are weakly coupled and non-relativistic in all three flavor sectors, $"11",\, "22"$ and $"12+21"$, the mass spectrum of low-lying flavored mesons is Coulomb-like with parametrically small mass differences $\Delta\mu_H/\mu_H=O(\bd^2/\nd^2)\ll 1$.\, c) A large number of gluonia made of $SU(\nd)$ gluons, with the mass scale $\sim\lym^{(\rm br1)}\sim\Bigl (m_Q\langle M_2\rangle\Bigr )^{1/3}$, see \eqref{(11.5)},\eqref{(11.1)}.\,    d) $n^2_2$ mions $M_{22}$ with the pole masses $\mu^{\rm pole}(M_{22})\ll\lym^{(\rm br1)}$, see \eqref{(11.11)}.\, e) $n^2_1$ mions $M_{11}$ with the pole masses $\mu^{\rm pole}(M_{11})\ll\mu^{\rm pole}(M_{22})$, see \eqref{(11.6)},\eqref{(11.11)}.\, f) $2n_1n_2$ hybrids $M_{12}, M_{21}$ are massless, $\mu(M_{12})=\mu(M_{21})=0$.

The pole mass of quarks ${\ov q}^2, q_2$ is smaller at ${\tilde\mu}_{\Phi,1}\ll\mph\ll \la^2/m_Q$ and $2n_1<\bo$, and stays at $\qtp\sim\lym^{(\rm br1)}$, while the mass of mions $M_{22}$ is larger and also stays at $\mu(M_{22})\sim\lym^{(\rm br1)}$.\\

\subsection{\quad  $\rm br2$ and $\rm special$ vacua, $\,\bd/N_F\ll 1$}

\hspace*{4mm} The condensates of mions look in these br2 - vacua as in \eqref{(11.1)} with the exchange $1\leftrightarrow 2$. The largest mass  is $\qtp$,
\bq
\qtp\sim \frac{\la}{Z_q}\Bigl (\frac{m_Q\mph}{\la^2}\Bigr )^{N_F/3\nd}\gg\qop\,,\quad
\frac{\lym^{(\rm br2)}}{\qtp}\sim Z_q\Bigl (\frac{\mo}{\mph}\Bigr )^{\frac{n_1 N_c}{3\nd(N_c-n_2)}}\ll 1\,,\label{(11.12)}
\eq
and the overall phase is $HQ_1-HQ_2$. After decoupling the heaviest quarks ${\ov q}^2, q_2$ at $\mu<\qtp$ the lower energy theory remains in the weak coupling logarithmic regime at, see \eqref{(11.10)},
\bq
\frac{\lym^{(\rm br2)}}{\qop}\ll 1 \quad\ra\quad \mo\ll\mph\ll {\tilde\mu}_{\Phi,2}\,,\quad
\frac{{\tilde\mu}_{\Phi,2}}{\mo}\sim\exp\Bigl\{\frac{(N_c-n_2)}{2n_2}\frac{1}{{\ov a\,}_*}\Bigr\}\gg 1\,.\label{(11.13)}
\eq
Hence, the  mass spectra in this range $\mo<\mph\ll {\tilde\mu}_{\Phi,2}$ can be obtained from corresponding formulas in section 11.1 by the replacements $n_1\leftrightarrow n_2$.

But because $n_2\geq N_F/2$, the lower energy theory with $1<n_1/\nd<3/2$ is in the strong coupling regime at $\mph\gg{\tilde\mu}_{\Phi,2}$, with ${\ov a}(\mu)\gg 1$ at $\lym^{(\rm br2)}\ll\mu\ll\qtp$. We do not consider the strong coupling regime in this paper. \\

As for the special vacua, the overall phase is also $HQ_1-HQ_2$ therein.  The mass spectra are obtained by substituting $n_1=\nd$ into the formulas of section 11.1. At $5/3<N_F/N_c<2$ and $\mph\gg{\tilde\mu}_{\Phi,1}$
the lower energy theory in these special vacua enters the strong coupling regime at $\lym^{(\rm spec)}\ll\mu\ll\qop$.

\section{ Broken $\mathcal{N}=2$ SQCD}

\hspace*{3mm} We consider now ${\cal N}=2$ SQCD with $SU(N_c)$ colors, $N_F$ flavors of light quarks, the scale factor $\Lambda_2$ of the gauge coupling, and with ${\cal N}=2$ broken down to ${\cal N}=1$ by the large mass parameter $\mu_X\gg \Lambda_2$ of the adjoint field $X=X^A T^A,\, {\rm Tr}\,(T^A T^B)=\delta^{AB}/2$. At very high scales $\mu\gg\mu_X$ the Lagrangian looks as (the exponents with gluons are implied in the Kahler term K)
\bq
K=\frac{1}{g^2(\mu,\Lambda_2)}{\rm Tr}\,(X^\dagger X)+{\rm Tr}\,({\textbf{Q}}^\dagger \textbf{Q}+\textbf{Q}\ra \ov{ \textbf{Q}})\,,\quad \label{(12.1)}
\eq
\bbq
\cw=-\frac{2\pi}{\alpha(\mu,\Lambda_2)}S+\mu_X {\rm Tr}\,(X^2)+\sqrt{2}\,{\rm Tr}\,(\ov {\textbf{Q}}X
\textbf{Q})+ m\,{\rm Tr}\,(\ov {\textbf{Q}}\textbf{Q}).
\eeq

The Konishi anomalies look here as
\bbq
\langle X^A\frac{\partial \cw}{\partial X^A}\rangle=\mu_X\langle X^A X^A \rangle+{\rm Tr}\langle J^A X^A \rangle=2N_c\langle S\rangle,\quad J^{a,i}_j={\sqrt 2}\,(\ov{\textbf{Q}}_j T^a \textbf{Q}^i)\,,\quad {\rm Tr}\,\Bigl (T^A T^B \Bigr )=\frac{1}{2}\,\delta^{AB}\,,
\eeq
\bbq
\langle\ov{\textbf{Q}}_i\frac{\partial \cw}{\partial \ov{\textbf{Q}}_i}\rangle=\langle J^{A,i}_i X^A\rangle+m\langle\ov {\textbf{Q}}_i \textbf{Q}^i \rangle=\langle S\rangle\,, \quad \rm {no \,\,\, summation\,\,\,  over\,\,\, i}\,.
\eeq
From these
\bbq
\langle{\rm Tr}\,X^2 \rangle=\frac{1}{2}\langle X^A X^A\rangle=\frac{1}{2\mu_X}\Bigl [(2N_c-N_F)
\langle S\rangle+m\langle{\rm Tr}\,\ov{\textbf{Q}} \textbf{Q}\rangle \Bigr ]\,.
\eeq

The running mass of $X$ is $\mu_X(\mu)=g^2(\mu)\mu_X$, so that at scales $\mu<\mu^{\rm pole}_X=g^2(\mu^{\rm pole}_X)\mu_X$ the field $X$ decouples from the dynamics and the RG evolution becomes those of ${\cal N}=1$ SQCD. The matching of ${\cal N}=2$ and ${\cal N}=1$ couplings at $\mu=\mu^{\rm pole}_X$ looks as ($\Lambda_2$ and $\la$ are the scale factors of ${\cal N}=2$ and ${\cal N}=1$ gauge couplings, $\la$ is held fixed when $\mu_X\gg\la$ is varied, ${\rm b}_2=2N_c-N_F\,,\,\bo=3N_c-N_F$)
\bbq
\frac{2\pi}{\alpha(\mu=\mu^{\rm pole}_X,\Lambda_2)}=\frac{2\pi}{\alpha(\mu=\mu^{\rm pole}_X,\la)}\,,
\quad \mu_X\gg\la\gg\Lambda_2\,,
\eeq
\bq
{\rm b}_2\ln\frac{\mu^{\rm pole}_X}{\Lambda_2}=\bo\ln\frac{\mu^{\rm pole}_X}{\la}-N_F\ln z_Q(\la,\mu^{\rm pole}_X)+N_c\ln\frac{1}{g^2(\mu=\mu^{\rm pole}_X)}\,,\label{(12.2)}
\eq
\bbq
\la^{\bo}=\frac{\Lambda^{{\rm b}_2}_2\mu^{N_c}_X}{z^{N_F}_Q (\la,\mu^{\rm pole}_X)}= z^{N_F}_Q (\mu^{\rm pole}
_X,\la)\Lambda^{{\rm b}_2}_2\mu^{N_c}_X\,,\quad z_Q(\la,\mu=\mu^{\rm pole}_X)
\sim\Bigl (\ln\frac{\mu^{\rm pole}_X}{\la}\Bigr )^{\frac{N_c}{\bo}}\gg 1.
\eeq

Although the field $X$ becomes too heavy and does not propagate any more at $\mu<\mu^{\rm pole}_X$, the loops of light quarks and gluons which are still active at $\la<\mu<\mu^{\rm pole}_X$ if the next largest physical mass $\mu_H$ is below $\la$, and if $\mu_H>\la$ they induce him at $\mu_H<\mu<\mu^{\rm pole}_{X}$ a non-trivial logarithmic renormalization factor $z_X(\mu^{\rm pole}_X,\mu<\mu^{\rm pole}_X)\ll 1$.

Therefore, finally, at scales $\la\ll\mu\ll\mu^{\rm pole}_X$ if $\mu_H<\la$ and at $\mu_H\ll\mu\ll\mu^{\rm pole}_X$ if $\mu_H>\la$, the Lagrangian of the broken ${\cal N}=2$ - theory with $0<N_F<2N_c$ can
be written as
\bq
K=\frac{z_X(\mu^{\rm pole}_X,\mu)}{g^2(\mu^{\rm pole}_X)}\,{\rm Tr}\,(X^\dagger X)+z_Q(\mu^{\rm pole}_X,\mu)\,{\rm Tr}\,({\textbf{Q}}
^\dagger \textbf{Q}+\textbf{Q}\ra \ov{\textbf{Q}})\,,\label{(12.3)}
\eq
\bbq
\cw=-\frac{2\pi}{\alpha(\mu,\la)}S+\mu_X {\rm Tr}\,(X^2)+\sqrt{2}\,{\rm Tr}\,(\ov{\textbf{Q}} X \textbf{Q})+ m\,{\rm Tr}\,(\ov{\textbf{Q}}\textbf{Q})\,.
\eeq
\bq
z_X(\mu^{\rm pole}_X,\mu)\sim\Biggl (\,\frac{\ln\, (\mu/\la)}{\ln\, (\mu^{\rm pole}_X/\la)}\,\Biggr)^{{\rm b}_2/\bo}\ll 1\,, \label{(12.4)}
\eq
\bbq
z_Q(\mu^{\rm pole}_X,\mu)=z_Q(\mu^{\rm pole}_X,\la)z_Q(\la,\mu),\quad z_Q(\la,\mu)\sim \Bigl (\ln\frac{\mu}
{\la}\Bigr)^{N_c/\bo}\gg 1\,.
\eeq

In all cases when the field $X$ remains too heavy and dynamically irrelevant, it can be integrated out in \eqref{(12.3)} and one obtains
\bq
K=z_Q(\mu^{\rm pole}_X,\mu)\,{\rm Tr}\,({\textbf{Q}}^\dagger \textbf{Q}+\textbf{Q}\ra \ov{\textbf{Q}})\,,\label{(12.5)}
\eq
\bbq
\cw_Q=-\frac{2\pi}{\alpha(\mu,\la)}S+ m\,{\rm Tr}(\ov{\textbf{Q}} \textbf{Q})-\frac{1}{2\mu_X}\Biggl ({\rm Tr}\,(\ov{\textbf{Q}}\textbf{Q})^2-\frac{1}{N_c}\Bigl({\rm Tr}\,\ov{\textbf{Q}} \textbf{Q} \Bigr)^2 \Biggr ).
\eeq

Now we redefine the normalization of the quarks fields
\bq
\textbf{Q}=\frac{1}{z^{1/2}_Q(\mu^{\rm pole}_X,\la)}\,Q\,,\quad \ov{\textbf{Q}}=\frac{1}{z^{1/2}_Q(\mu^{\rm pole}_X,\la)}\,{\ov Q}\,,\label{(12.6)}
\eq
\bq
K=z_Q(\la,\mu){\rm Tr}\Bigl (\,Q^\dagger Q+(Q\ra {\ov Q})\,\Bigr ),\quad \cw=-\frac{2\pi}{\alpha(\mu,\la)}S+W_Q\,,\label{(12.7)}
\eq
\bq
\cw_Q=\frac{m}{z_Q(\mu^{\rm pole}_X,\la)}\,{\rm Tr}({\ov Q} Q)-\frac{1}{2 z^2_Q(\mu^{\rm pole}_X,\la)\mu_X}
\Biggl ({\rm Tr}\,({\ov Q}Q)^2-\frac{1}{N_c}\Bigl({\rm Tr}\,{\ov Q} Q \Bigr)^2 \Biggr ).\label{(12.8)}
\eq
Comparing this with \eqref{(2.3)} and choosing
\bq
\frac{m}{z_Q(\mu^{\rm pole}_X,\la)}=m_Q\ll\la\,, \quad z^2_Q(\mu^{\rm pole}_X,\la)\mu_X=\mph\gg\la \label{(12.9)}
\eq
it is seen that with this matching the $\Phi$ - theory and the broken ${\cal N}=2$ SQCD will be equivalent.

Therefore, at large $\mu_X\gg\Lambda_2$ and until both $X$ and $\Phi$ fields remain dynamically irrelevant, all results obtained above for the $\Phi$ - theory will be applicable to the broken ${\cal N}=2$ SQCD as well. Besides, the $\Phi$ and $X$ fields remain dynamically irrelevant in the same region of parameters, i.e. at $N_F<N_c$ and at $\mph>\mo$ or $\mu_H>\mu_o$ at $\mph<\mo$ if $N_F>N_c$\,, see \eqref{(5.1)}.

Moreover, some general properties of both theories such as {\it the multiplicity of vacua with unbroken or broken flavor symmetry and the values of vacuum condensates of corresponding chiral superfields} (i.e. $\langle{\ov Q}_j Q^i\rangle$ and $\langle S\rangle$, see section 4) {\it are the same in these two theories, independently of whether the fields $\Phi$ and $X$ are irrelevant or relevant}.

Nevertheless, once the fields $\Phi$ and $X$ become relevant (e.g. at $\mu_X\ll\Lambda_2$), the phase states, the RG evolution, the mass spectra etc., {\it become very different in these two theories}. The properties of the $\Phi$ - theory were described in detail above in the text. In general, if $X$ is sufficiently light and dynamically relevant, the dynamics of the softly broken ${\cal N}=2$ SQCD  becomes complicated and is outside the scope of this paper.\\

Finally, we trace now a transition to the slightly broken ${\cal N}=2$ theory with small $\mu_X\ll\Lambda_2$
and fixed $\Lambda_2$. For this, we write first the appropriate form of the effective superpotential obtained from \eqref{(12.7)},\eqref{(12.8)}
\bbq
\cw^{\rm eff}_Q=-\nd S+\frac{m}{z_Q(\mu^{\rm pole}_X,\la)}\,{\rm Tr}({\ov Q} Q)-\frac{1}{2 z^2_Q(\mu^{\rm pole}_X,\la)\mu_X}\Biggl ({\rm Tr}\,({\ov Q}Q)^2-\frac{1}{N_c}\Bigl({\rm Tr}\,{\ov Q} Q \Bigr)^2 \Biggr ),
\eeq
\bq
S=\Biggl (\,\,\frac{\det {\ov Q}Q}{\la^{\bo}}\,\,\Biggr )^{1/\nd}\,,\quad \la^{\bo}=z^{N_F}_Q(\mu^{\rm pole}_X,\la)\Lambda^{b_2}_2\mu^{N_c}_X \label{(12.10)}
\eq
and restore now the original normalization of the quark fields $\ov{\textbf{Q}}, \textbf{Q}$ appropriate for the slightly broken ${\cal N}=2$ theory with varying $\mu_X\ll\Lambda_2$ and fixed $\Lambda_2$, see \eqref{(12.6)},
\bq
\hspace*{-1cm} \cw^{\rm eff}_Q=-\nd S+m\,{\rm Tr}(\ov{\textbf{Q}} \textbf{Q})-\frac{1}{2 \mu_X}\Biggl ({\rm Tr}\,(\ov{\textbf{Q}}\textbf{Q})^2-\frac{1}{N_c}\Bigl({\rm Tr}\,\ov{\textbf{Q}} \textbf{Q} \Bigr)^2 \Biggr ),\quad S=\Bigl (\frac{\det \ov{\textbf{Q}}\textbf{Q}}{\Lambda^{{\rm b}_2}_2\mu^{N_c}_X} \Bigr )^{1/\nd}. \label{(12.11)}
\eq

One can obtain now from \eqref{(12.11)} the values of the quark condensates $\langle\ov{\textbf{Q}}_j \textbf{Q}^i\rangle$ at fixed $\Lambda_2$ and small $\mu_X\ll\Lambda_2$. Clearly, in comparison with $\langle{\ov Q}_j Q^i\rangle$ in section 4, the results for $\langle\ov{\textbf{Q}}_j \textbf{Q}^i\rangle$ are obtained by the replacement\,:\, $m_Q\ra m\,,\, \mph\ra\mu_X\,,\,\la^{\bo}\ra\Lambda^{{\rm b}_2}_2\mu^{N_c}_X$, while the multiplicities of vacua are the same. From \eqref{(12.11)}, the dependence of $\langle\ov{\textbf{Q}}_j \textbf{Q}^i\rangle$ and $\langle S\rangle$ on $\mu_X$ is trivial in all vacua, $\sim \mu_X$.

With the above replacements, the expressions for $\langle{\ov Q}_j Q^i\rangle$ in section 4 in the region $\la\ll\mph\ll\mo$ correspond here to the hierarchy $m\ll\Lambda_2$, while those in the region $\mph\gg\mo$ correspond here to $m\gg\Lambda_2$. In the language of \cite{APS} used in \cite{CKM} (see sections 6-9 therein), the correspondence between the $r$ - vacua \cite{APS,CKM} of the slightly broken ${\cal N}=2$ theory with $0< \mu_X/\Lambda_2\ll 1,\,\, 0< m/\Lambda_2\ll 1$ and our vacua in section 4 looks as
\footnote{\,
This correspondence is based on comparison of multiplicities of our vacua at $\mph\ll\mo$ described in section 4 and those of $r$ - vacua at $m\ll\Lambda_2$ and $\mu_X\ll\Lambda_2$ as these last are given in \cite{CKM}.
}
\,: \,a)\, \,$r=n_1$,\, b)\, our L - vacua with the unbroken or the L - type ones with spontaneously broken flavor symmetry correspond, respectively, to the first group of vacua of the non-baryonic  branches with $r=0$ and $r\geq 1,\, r\neq\nd$ in \cite{CKM}\,,\,\, c)\, our S - vacua with the unbroken flavor symmetry and $\rm br2$ - vacua with the spontaneously broken flavor symmetry correspond to the first type from the second group of vacua of the baryonic branches  with, respectively, $r=0$ and $1\leq r<\nd$ in \cite{CKM},\,\, d)\, our special vacua with $n_1=\nd,\, n_2=N_c$ correspond to the second type of vacua from this group, see \cite{CKM}.

\section{ Conclusions}

\hspace {4mm} We described above in the text the mass spectra at $0<N_F<2N_c$ of the $\Phi$-theory which is the standard ${\cal N}=1$ SQCD with $SU(N_c)$ colors and $N_F$ flavors of light quarks and with added $N^2_F$ colorless but flavored fields $\Phi^j_i$, with Yukawa interactions with quarks.

At $0<N_F<N_c$ this theory is in the weak coupling regime, so that calculations of its mass spectra in various vacua in section 3 is straightforward and does not require any additional assumptions.

The calculations of values of quark and gluino condensates in multiple vacua of this $\Phi$-theory at $N_c<N_F<2N_c$ and multiplicities of various vacua were presented in section 4. The values of these condensates constitute a base for further calculations of mass spectra.

A qualitatively new phenomenon appearing in this $\Phi$-theory in the conformal window $3N_c/2<N_F<3N_c$ due to the strong power-like RG evolution of the quark and $\Phi$ renormalization factors of their Kahler terms was described in section 5. At the appropriate values of the Lagrangian parameters $\mph\gg\la$ and $m_Q\ll\la$, the seemingly heavy and dynamically irrelevant fields $\Phi^j_i$ "return back" and there appear two additional generations of light $\Phi$-particles.\\

At present, the calculations of mass spectra of the direct $\Phi$ - theory (and its Seiberg's dual variant, the $d\Phi$ -theory) in conformal window in the strongly coupled regime can not be performed directly (i.e. without additional assumptions about dynamics of these theories). Therefore, these mass spectra were calculated in sections 6-11 within the dynamical scenario introduced in \cite{ch3}. We recall that this scenario {\it assumes} that in such ${\cal N}=1$ SQCD-like theories quarks may be in two standard phases only. These are: a) the heavy quark (HQ) phase where they are not higgsed but confined, and b) the Higgs phase where they are higgsed and so not confined.

Moreover, this scenario includes the assumption that two above phases are realized in a 'standard way' even in a strong coupling regime with $a=(N_cg^2/8\pi^2)\sim 1$. This means that, unlike e.g. ${\cal N}=2$ SQCD with its very special properties, in these ${\cal N}=1$ SQCD-like theories without adjoint colored scalar superfields, the additional non-standard parametrically lighter particles (e.g. parametrically lighter magnetic monopoles or dyons) do not appear in the spectrum even in the strong coupling region $a\sim 1$, in comparison with that in the weak coupling one (see also the footnote \ref{(f1)}).

In comparison with the standard ${\cal N}=1$ SQCD with the superpotential $W=m_Q{\rm Tr}({\ov Q}Q)$ and the only small parameter $m_Q/\la\ll 1$ which serves as the infrared regulator, the $\Phi$ - theory considered in this paper includes two independent competing small parameters which serve as infrared regulators, $m_Q/\la\ll 1$ and $\la/\mph\ll 1$. Due to this the dynamics of this theory is much richer. Two main qualitatively new elements in this direct $\Phi$ - theory are\,:

a) the appearance of a large number of vacua with the spontaneously broken global flavor symmetry, $U(N_F)\ra U(n_1)\times U(n_2)$, and as a result, with a number of exactly massless Nambu-Goldstone particles in the mass spectrum\,;

b)\, in a number of cases with $N_F>N_c$\,, due to their Yukawa interactions with the light quarks, the seemingly heavy and dynamically irrelevant fion fields $\Phi$\, '{\it return back\,}' and there appear two additional generations of light $\Phi$ - particles with $\mu^{\rm pole}(\Phi)\ll\la$, see section 5.\\

This is not a purpose of these conclusions to repeat in a shorter form all results obtained above in the main text for the phase states and mass spectra of the direct and dual theories at different values of $\mph/\la\gg 1$. We will try only to formulate here in a few words the most general qualitative property of ${\cal N}=1$ SQCD-like theories which emerged from the studies in \cite{ch3} and in this paper. This is {\it the extreme sensitivity of their dynamical behavior in the IR region of momenta, of their mass spectra and even the phase states, to the values of small parameters in the Lagrangian which serve as infrared regulators}.

As was shown above in the main text, similarly to the standard ${\cal N}=1$ SQCD with the superpotential $W=m_Q{\rm Tr}({\ov Q}Q)$ \cite{ch3}, the direct $\Phi$ -theory and its Seiberg's dual variant, the $d\Phi$ - theory, are (within the dynamical scenario introduced in \cite{ch3} and used in this paper) also not equivalent as their mass spectra are {\it parametrically different} .
\footnote{\,
But, similarly to the standard ${\cal N}=1$ SQCD with the superpotential $W=m_Q{\rm Tr}({\ov Q}Q)$ \cite{ch3}, to see clearly {\it the parametric differences} in mass spectra of the direct $\Phi$ and dual $d\Phi$ theories, one needs to use the additional small parameter $0<\bd/N_F=(2N_F-3N_c)/N_F\ll 1$.
}

At present, unfortunately, no way is known to obtain {\it direct solutions} (i.e. without any additional assumptions) of ${\cal N}=1$ SQCD-like theories in the strong coupling region. Therefore, to calculate the mass spectra in such theories one has to introduce and use some assumptions about the dynamics of these theories when they are in the strong coupling region. In other words, one has to rely on a definite dynamical scenario. Therefore, clearly, the results obtained in \cite{ch3} and in this paper are not the direct proofs (i.e. without any additional assumptions) that the Seiberg hypothesis \cite{S1,S2} about an equivalence of the direct and dual theories is not correct. Still, strictly speaking, both possibilities remain open: it may be correct, but maybe not. Finally, the Seiberg hypothesis is based mainly on matching of the 't Hooft triangles in the direct and dual theories in those ranges of scales where all particles are effectively massless, and on some suitable correspondences of their behavior in the superconformal regime. Clearly, these are the necessary conditions. But they may be not sufficient.

The dynamical scenario introduced in \cite{ch3} and used in this paper looks self-consistent and not in contradiction with any proved result. Therefore, it seems possible at present. And, in particular, {\it all Seiberg's checks of duality in the conformal window are fulfilled} in this scenario. Nevertheless, as shown in \cite{ch3} and in this paper, in spite of that, {\it the mass spectra of the direct and dual theories are parametrically different}. This demonstrates clearly that, indeed, those checks on which the Seiberg hypothesis is based, although necessary, may well be insufficient. This does not mean, of course, that the scenario introduced in \cite{ch3} is right. But, nevertheless, this implies that it {\it may be right}. Therefore, what is still missing in this story at present is a proof that the dynamical scenario from \cite{ch3} is right, or the opposite proof that Seiberg's hypothesis about a complete equivalence of the direct and dual theories is right.

From our standpoint, a new and practically most important thing at present is a very ability to calculate the mass spectra of various ${\cal N}=1$ SQCD-like theories in the strong coupling regimes, even within a given dynamical scenario. It seems clear that further developments of the theory or lattice calculations will allow to find a unique right scenario in each such theory. Time will show, as always, what hypotheses are right and what are not.\\

The $\Phi$-theory with $\mph\gg\la$ considered in this paper is tightly connected with the $X$-theory which is the ${\cal N}=2$ SQCD broken down to ${\cal N}=1$ by the large mass parameter $\mu_X\gg\Lambda_2$ of the adjoint colored superfields $X^A$. The multiplicity of vacua and the values of the quark and gluino condensates, $\langle{\ov Q}_j Q^i\rangle$ and $\langle S\rangle$, are the same in both theories (under the appropriate matching of parameters, see section 12). Moreover, in all those cases when the fields $\Phi$ are dynamically irrelevant in the $\Phi$-theory, the fields $X$ are also dynamically irrelevant in the $X$-theory and these two theories are equivalent (up to inessential small power corrections). We have described in section 12 the connections between the values of the quark and gluino condensates in different vacua in the broken ${\cal N}=2$ SQCD at $m\gtrless\Lambda_2$, with those in the direct $\Phi$-theory with large varying $\mph\gg\la$.

But even in those cases when both fields $\Phi$ and $X^A$ are irrelevant, this does not mean that these two theories are simply equivalent to the standard ${\cal N}=1$ SQCD with small unimportant corrections. First, the whole physics in a large number of additional vacua with the spontaneously broken flavor symmetry is completely different. And second, even in vacua with the unbroken flavor symmetry, these theories evolve to the standard ${\cal N}=1$ SQCD with small corrections not simply at $\mph=(\rm{several})\la$, as one can naively expect, but only {\it at parametrically larger values} $\mph\gtrsim\mo=\la (\la/m_Q)^{(2N_c-N_F)/N_c}\gg\la$.

But e.g. when the corresponding mass parameters $\mph$ and $\mu_X$ are small and both fields $\Phi$ and $X$ are dynamically relevant, the phase states, the mass spectra, etc. become very different in these $\Phi$ and $X$ - theories.\\

{\appendix
\section{The RG flow in the $\mathbf \Phi$ - theory at $\mathbf{\mu>\la}$ }

{\bf \quad A.1\quad} We first consider the $\Phi$ - theory at $N_c<N_F<2N_c$ where it is taken as UV-free. We start with the canonically normalized Kahler term $K$ at the very high scale $\mu\sim \mu_{\rm UV}$ and the running couplings and mass parameters
\bbq
K={\rm Tr}\,\Bigl ({\widehat\Phi}^\dagger {\widehat\Phi}\Bigr )+{\rm Tr}\Bigl (\,{\widehat Q}^\dagger {\widehat Q}+
( {\widehat Q}\ra {\widehat{\ov Q}})\,\Bigr )\,,\quad \cw=-\frac{2\pi}{\alpha(\mu)}S+\cw_{\Phi}+\cw_Q\,,\quad
\eeq
\bq
\cw_{\Phi}=\frac{\mph(\mu)}{2}\Biggl [{\rm Tr}\,({\widehat\Phi}^2)-\frac{1}{\nd}\Bigl ({\rm Tr}\,{\widehat\Phi}\Bigr )^2\Biggr ]\,,\quad \cw_Q=- f(\mu){\rm Tr}\,\Bigl ( {\widehat{\ov Q}}{\widehat\Phi}{\widehat Q}\Bigr )+{\rm Tr}\,\Bigl ( {\widehat{\ov Q}}\, m_ Q(\mu) {\widehat Q}\Bigr )\,. \label{(A.1)}
\eq
Now, instead of running parameters, we introduce $\mu$-independent ones, $\la,\,\, \mph$ and $m_Q$ ($\mph\gg\la$ and $m_Q\ll\la$ in the main text),
\bq
\frac{1}{a(\mu)}=\frac{2\pi}{N_c\alpha(\mu)}=\frac{\bo}{N_c}\ln\frac{\mu}{\la}-\frac{N_F}{N_c}\ln z_Q(\la,\mu)+\ln\frac{1}{a(\mu)}+C_a\,,\quad \bo=3N_c-N_F\,,
\label{(A.2)}
\eq
\bbq
a_f(\mu)=\frac{N_c f^2(\mu)}{2\pi}=\frac{a_f=N_c f^2/2\pi}{z_{\Phi}(\la,\mu)z^2_Q(\la,\mu)}\,,\quad
\mph(\mu)\equiv\frac{f^2\mph}{z_{\Phi}(\la,\mu)}\,,\quad m_Q(\mu)\equiv\frac{m_Q}{z_Q(\la,\mu)}\,,
\eeq
where $z_Q(\la,\mu\gg\la)\gg 1$ and $z_{\Phi}(\la,\mu)$ are the perturbative renormalization factors (logarithmic in this case) in the theory with {\it all fields massless}, $a_f$ is taken as $a_f\sim 1/(\rm several)$ and $C_a$ is also $O(1)$ (it will be omitted for simplicity). Therefore, after redefinitions of the quark and $\Phi$ fields, the Lagrangian at the very high scale can be rewritten as
\bq
K=z_{\Phi}(\la,\mu)\frac{1}{f^2}{\rm Tr}\,(\Phi^\dagger\Phi)+z_Q(\la,\mu){\rm Tr}\Bigl (\,Q^\dagger Q+(Q\ra {\ov Q})\,\Bigr )\,,\label{(A.3)}
\eq
\bbq
\cw_{\Phi}=\frac{\mph}{2}\Biggl [{\rm Tr}\,(\Phi^2)-\frac{1}{\nd}\Bigl ({\rm Tr}\,\Phi\Bigr )^2\Biggr ]\,,
\quad \cw_Q=-{\rm Tr}\,\Bigl ( {\ov Q}\Phi Q\Bigr )+{\rm Tr}\,\Bigl ( {\ov Q}m_Q Q\Bigr )\,.
\eeq

From \eqref{(A.2)}
\bq
\frac{d a_f(\mu)}{d\ln\mu}=\beta_f=-a_f(\mu)\Bigl (2\gamma_Q(\mu)+\gamma_{\Phi}(\mu)\Bigr ),\quad \gamma_Q=\frac{d\ln z_Q(\mu)}{d\ln\mu},\quad \gamma_{\Phi}=\frac{d\ln z_{\Phi}(\mu)}{d\ln\mu}\,.\label{(A.4)}
\eq
In the approximation of leading logarithms at large $\mu$
\bq
\gamma_Q(\mu)\approx \frac{2 C_F}{N_c}a(\mu)-\frac{N_F}{N_c}a_f(\mu),\quad \gamma_{\Phi}(\mu)\approx -a_f(\mu)\,,\quad\frac{2 C_F}{N_c}=\frac{N_c^2-1}{N_c^2}\approx 1\,. \label{(A.5)}
\eq
From \eqref{(A.4)},\eqref{(A.5)}, there is the UV free solution
\bq
a(\mu)\approx \frac{N_c}{\bo}\frac{1}{\ln (\mu/\la)}\,,\quad a_f(\mu)\sim a_f\Bigl (\frac{1}
{\ln(\mu/\la)}\Bigr )^{\frac{2N_c}{\bo}}\ll a(\mu),\quad 1<\frac{2N_c}{\bo}<2\,,\label{(A.6)}
\eq
\bq
z_Q(\la,\mu)\sim \Bigl (\ln\frac{\mu}{\la}\Bigr )^{N_c/\bo}\gg 1,\quad z_{\Phi}(\la,\mu)\sim 1\,.\label{(A.7)}
\eq
It is seen from \eqref{(A.6)} that the Yukawa coupling $a_f(\mu)$ is parametrically small in comparison with the gauge coupling $a(\mu)$ and, up to small corrections, it has no effect on the RG evolution at large $\mu$.

The first physical mass parameter which influences the RG flow with lowering the scale $\mu$ is $\mu^{\rm pole}_{1}(\Phi)=\mph(\mu=\mu^{\rm pole}_{1}(\Phi)\,)=f^2\mph/z_{\Phi}(\la,\mu^{\rm pole}_{1}(\Phi))\sim f^2\mph\gg\la$, so that $\mu_{\Phi}(\mu)$ becomes $\mu_{\Phi}(\mu)\sim f^2\mph>\mu$ at $\mu<\mu^{\rm pole}_{1}(\Phi)$ and the fields $\Phi$ become too heavy. They do not propagate any more and do not influence the RG evolution until $\mu_{\Phi}(\mu)>\mu$. Nevertheless, the anomalous dimension $\gamma_{\Phi}(\mu)$ remains small but nonzero even at $\mu<\mu^{\rm pole}_{1}(\Phi)$ due to loops of still active light quarks (and gluons interacting with quarks) and, instead of \eqref{(A.5)}, the anomalous dimensions look at $\mu<\mu^{\rm pole}_{1}(\Phi)$ as
\bq
\gamma_Q(\mu)\approx a(\mu),\quad \gamma_{\Phi}(\mu)\approx -a_f(\mu)\,,\label{(A.8)}
\eq
while \eqref{(A.6)},\eqref{(A.7)} remain the same. Hence, although the heavy fields $\Phi_{ij}$ decouple at $\la<\mu<\mu^{\rm pole}_{1}(\Phi)$, the RG flow remains parametrically the same because their role even at $\mu>\mu^{\rm pole}_{1}(\Phi)$ was small.

Therefore, finally, at scales $\la<\mu<\mu^{\rm pole}_{1}(\Phi)$ if there is no physical masses $\mu_H>\la$ and at $\mu_H<\mu<\mu^{\rm pole}_{1}(\Phi)$ if $\mu_H>\la$, the Lagrangian of the $\Phi$ - theory with $N_c<N_F<2N_c$ light flavors can be written as
\bbq
K=\frac{1}{f^2}{\rm Tr}\,(\Phi^\dagger \Phi)+z_Q(\la,\mu){\rm Tr}\Bigl (\,Q^\dagger Q+(Q\ra {\ov Q})\,\Bigr )\,,\quad W=-\frac{2\pi}{\alpha(\mu,\la)}S+W_{\Phi}+W_Q\,,\quad
\eeq
\bq
\cw_{\Phi}=\frac{\mph}{2}\Biggl [{\rm Tr}\,(\Phi^2)-\frac{1}{\nd}\Bigl ({\rm Tr}\,\Phi\Bigr )^2\Biggr ]\,,\quad \cw_Q={\rm Tr}\,\Bigl ( {\ov Q}\,m^{\rm tot}_Q Q\Bigr )\,,\quad m^{\rm tot}_Q=m_Q-\Phi\,,\label{(A.9)}
\eq
with $z_Q(\la,\mu)$ given in \eqref{(A.7)}.\\

{\bf \quad A.2\quad} We consider now the case $1\leq N_F<N_c$\,. Although the $\Phi$ - theory is not UV free in this case and requires UV completion at $\mu>\mu_{\rm UV}$, the RG flow at $\mu_H<\mu\ll\mu_{\rm UV}$ is very specific (see below, the quarks are really higgsed in this case at $\mu_H=\mu_{\rm gl},\,\,\la\ll\mu_{\rm gl}\ll\mph\ll\mu_{UV}$, see section 3). We take from the beginning $a_f$ in \eqref{(A.2)} to be sufficiently small, $a_f\ll 1$, and calculate the behavior of $a(\mu)$ and $a_f(\mu)$ at $\la\ll\mu\ll\mu_{UV}$ in the massless theory which follows from their definitions in \eqref{(A.2)}. Then, by definition, in the theory with $\la\ll\mu^{\rm pole}_1(\Phi)\ll\mu_{UV}$, the behavior  of $a(\mu)$ and $a_f(\mu)$ at $\mu^{\rm pole}_1(\Phi)\ll\mu\ll\mu_{UV}$ will be the same while, in general, it can be different at $\mu<\mu^{\rm pole}_1(\Phi)$.

There is the same solution \eqref{(A.6)} also at $1\leq N_F<N_c$, with a difference that $2/3<2N_c/\bo<1$ now and $a_f\ll 1$. Hence, starting with $\mu>\la,\,\, a_f(\mu)$ begins first {\it to decrease with increasing} $\mu$, but more slowly now than $a(\mu)\sim 1/\ln(\mu/\la)$. Due to this, $\beta_f(\mu)$ in \eqref{(A.4)} changes a sign at $\mu\sim\ov\mu$,
\bq
a_f({\ov\mu})\sim a({\ov\mu})\,\,\ra \,\, \ln\frac{{\ov\mu}}{\la}\sim \Bigl (\frac{1}{a_f}\Bigr )^{\frac{\bo}{N_c-N_F}}\gg 1,\,\, a_f({\ov\mu})\sim\frac{1}{\ln ({\ov\mu}/\la)}\sim \Bigl (a_f\Bigr )^{\frac{\bo}{N_c-N_F}}\ll a_f\ll 1 \label{(A.10)}
\eq
and then $a_f(\mu)$ begins to grow
\bq
a_f(\mu>\ov\mu)\sim \frac{1}{\ln(\mu_{UV}/\mu)}\,,\quad \ln\Bigl (\frac{\mu_{UV}}{\ov\mu}\Bigr )\sim \Bigl (\frac{1}{a_f}\Bigr )^{\frac{\bo}{N_c-N_F}}\gg 1
\label{(A.11)}
\eq
with further increasing $\mu>{\ov\mu}$. Therefore, $z_{\Phi}(\la,\mu<{\ov\mu})\sim 1$ in the massless theory.

For our purposes in section 3 it will be sufficient to have $\mu_{\rm gl}\ll\mu^{\rm pole}_1(\Phi)\sim a_f\mph\ll{\ov\mu}\ll\mu_{\rm UV}$. This leads to a sufficiently weak logarithmic restriction
\bq
\frac{1}{a_f}\gg \Bigl (\ln\frac{\mph}{\la}\Bigr )^{\frac{N_c-N_F}{\bo}},\quad  0<\frac{N_c-N_F}{\bo}<\frac{1}{3}\,,\label{(A.12)}
\eq
and then $z_{\Phi}(\la,\mu<\mu^{\rm pole}_1(\Phi))$ remains $\sim 1$ also in the $\Phi$ - theory with massive fields $\Phi$.

\section{There are no vacua with $\mathbf{\langle S\rangle=0}$ in $SU(N_c)$ theories at $\mathbf{m_Q\neq 0}$}

The purpose of this appendix is to show that the gluino condensate $\langle S\rangle\neq 0$ at $m_Q\neq 0$ in all vacua of the $SU(N_c)$ theory with the broken flavor symmetry, $U(N_F)\ra U(n_1)\times U(n_2)$, in both the direct and dual theories. \\

\hspace*{1cm} {\bf 1\,.\,\, Direct theory}\\

We assume that there is at $N_c<N_F<2N_c$ a large number of {\it additional} vacua with either $1\leq n_1\leq N_c-1$ components $\langle\Qo\rangle=0$, or $n_2\geq n_1$ components $\langle\Qt\rangle=0$. Even in this case the relations at $\mu=\la$
\bq
\langle \Qo+\Qt\rangle-\frac{1}{N_c}{\rm Tr}\,\langle\qq\rangle=m_Q\mph,\quad\langle S\rangle=\frac{1}{\mph}\langle\Qo\rangle\langle\Qt\rangle,\quad
\langle\Qo\rangle\neq\langle\Qt\rangle,\,\, \label{(B.1)}
\eq
\bbq
\langle m^{\rm tot}_{Q,1}\rangle=\langle m_Q-\Phi_1\rangle=\frac{\langle\Qt\rangle}{\mph}\,,\quad
\langle m^{\rm tot}_{Q,2}\rangle=\langle m_Q-\Phi_2\rangle=\frac{\langle\Qo\rangle}{\mph}\,,
\eeq
following from the Konishi anomalies \eqref{(2.2)},\eqref{(2.4)} remain valid. Therefore, one obtains from \eqref{(B.1)} that either
\bq
\langle \Qt\rangle=0\,, \quad \langle\Qo\rangle=\frac{N_c}{N_c-n_1}\, m_Q\mph\,, \quad \langle S\rangle=0\,,\quad 1\leq n_1\leq N_c-1\,,\label{(B.2)}
\eq
\bbq
\langle m^{\rm tot}_{Q,1}\rangle=\frac{\langle \Qt\rangle}{\mph}=0\,,\quad \langle m^{\rm tot}_{Q,2}\rangle=\frac{\langle \Qo\rangle}{\mph}=\frac{N_c}{N_c-n_1}\, m_Q\,,
\eeq
or
\bq
\langle \Qo\rangle=0\,, \quad \langle\Qt\rangle=\frac{N_c}{N_c-n_2}\, m_Q\mph\,,
\quad \langle S\rangle=0\,, \quad n_2\neq N_c\,,\label{(B.3)}
\eq
\bbq
\quad \langle m^{\rm tot}_{Q,2}\rangle=\frac{\langle \Qo\rangle}{\mph}=0\,,\quad \langle m^{\rm tot}_{Q,1}\rangle=\frac{\langle \Qt\rangle}{\mph}=\frac{N_c}{N_c-n_2}\, m_Q\,
\eeq
in these vacua. We will show below that this assumption is not self-consistent. I.e., we will start with \eqref{(B.2)} or \eqref{(B.3)} and calculate then explicitly $\langle S\rangle\neq 0$ in these vacua. For this, using a holomorphic dependence of $\langle S\rangle$ on $\mph$, it will be sufficient to calculate $\langle S\rangle\neq 0$ in some range of most convenient values of $\mph$. Hence, we take $\mph\sim \la^2/m_Q$.\\

In vacua \eqref{(B.2)} with $\langle\Qt\rangle=0,\, \langle\Qo\rangle\sim m_Q\mph\sim \la^2$ the quarks ${\ov Q}_1,\, Q_1$ are higgsed with $\langle{\ov Q}_1\rangle=\langle Q^1\rangle\sim \la$. At $n_1<N_c-1$ the lower energy theory at $\mu<\la$ contains $SU(N_c-n_1)$ unbroken gauge symmetry with the scale factor of the gauge coupling $(\Lambda^\prime)^{\bo^\prime}\sim\la^{\bo}/\det \Pi_{11},\, \langle\Lambda^\prime\rangle\sim\la$, $n^2_1$ pions $\Pi_{11}$ and ${\ov Q}_2,\, Q^2$ quarks with zero condensate and the running mass $\langle m^{\rm tot}_{Q,2}\rangle=\langle m_Q-\Phi_2\rangle=\langle\Qo\rangle/\mph\sim m_Q$ at $\mu=\la$. For this reason, the variant with the $Higgs_2$ phase of these quarks is excluded, they will be always in the heavy quark $HQ_2$ - phase. At all $n_1<N_c-1$, proceeding as in \cite{ch1,ch2,ch3}, i.e. lowering the scale down to $\mu<m^{\rm pole}_{Q,2}\sim m_Q/z_Q(\la,m^{\rm pole}_{Q,2})$ and integrating out ${\ov Q}_2, Q_2$ quarks as heavy particles, there remains the pure $SU(N_c-n_1)$ Yang-Mills theory (and $n^2_1$ pions $\Pi_{11}$) with the scale factor of its gauge coupling
\bq
\lym^3=\Biggl (\frac{\la^{\bo}\det (m_Q-\Phi_{22})}{\det\Pi_{11}}\Biggr )^{1/(N_c-n_1)},\label{(B.4)}
\eq
and, finally, with the Lagrangian of the form \eqref{(3.22)} at $\mu<\langle\lym\rangle$. From \eqref{(B.4)}
\bq
\langle S\rangle=\langle\lym^3\rangle\sim \la^3\Bigl (\frac{\la}{\mph}\Bigr )^{\frac{n_1}{N_c-n_1}}\Bigl (\frac{m_Q}
{\la}\Bigr )^{\frac{n_2-n_1}{N_c-n_1}}\neq 0\,.\label{(B.5)}
\eq

At $n_1=N_c-1$ the gauge group will be broken completely and \eqref{(B.4)} originates from the instanton contribution.

The vacua \eqref{(B.3)} with $\langle\Qo\rangle=0$ are considered the same way and one obtains \eqref{(B.4)},\eqref{(B.5)} with the replacement $n_1\leftrightarrow n_2$. (In vacua \eqref{(B.3)} the cases with $n_2>N_c$ are excluded from the beginning as the rank of $\langle Q^2\rangle$ is $\leq N_c$ and the unbroken $U(n_2)$ flavor symmetry cannot be maintained; the case $n_2=N_c$ is also excluded as $\langle\Pi_1\rangle\neq 0$ in this case, see \eqref{(B.1)}\,). Hence, only the cases with $n_2\leq N_c-1$ remain).

On the whole, the assumption about the existence of additional vacua \eqref{(B.2)} or \eqref{(B.3)} with $\langle\Qo\rangle=0$ or $\langle\Qt\rangle=0$ at $N_c<N_F<2N_c$ is not self-consistent.\\

\hspace*{1cm} {\bf 2\,.\,\, Dual theory}\\

The dual analog of \eqref{(B.1)}-\eqref{(B.3)} looks as, see \eqref{(2.8)},
\bq
\langle M_1+M_2\rangle-\frac{1}{N_c}{\rm Tr}\,\langle M\rangle=m_Q\mph,\quad\langle S\rangle=\frac{1}{\mph}\langle M_1\rangle\langle M_2\rangle,\quad
\langle M_1\rangle\neq\langle M_2\rangle\,, \label{(B.6)}
\eq
By assumption, there is a large number of {\it additional} vacua with either
\bq
\langle M_2\rangle=0\,, \quad \langle M_1\rangle=\frac{N_c}{N_c-n_1}\, m_Q\mph\,, \quad \langle S\rangle=0\,,\quad 1\leq n_1\leq N_c-1\,,\label{(B.7)}
\eq
\bbq
\langle N_1\rangle=\langle m^{\rm tot}_{Q,1}\rangle\la=\frac{\langle M_2\rangle\la}{\mph}=0\,,\quad \langle N_2\rangle=\langle m^{\rm tot}_{Q,2}\rangle\la=\frac{\langle M_1\rangle\la}{\mph}=\frac{N_c}{N_c-n_1}\, m_Q\la\,,
\eeq
or
\bq
\langle M_1\rangle=0\,, \quad \langle M_2\rangle=\frac{N_c}{N_c-n_2}\, m_Q\mph\,, \quad \langle S\rangle=0\,,\quad n_2\neq N_c\,,\label{(B.8)}
\eq
\bbq
\langle N_2\rangle=\langle m^{\rm tot}_{Q,2}\rangle\la=\frac{\langle M_1\rangle\la}{\mph}=0\,,\quad \langle N_1\rangle=\langle m^{\rm tot}_{Q,1}\rangle\la=\frac{\langle M_2\rangle\la}{\mph}=\frac{N_c}{N_c-n_2}\, m_Q\la\,.
\eeq
In this case, it is more convenient for our purposes to choose the regions $\la\ll\mph\ll\mo$ at $3N_c/2<N_F<2N_c$ and $\la\ll\mph\ll\la(\la/m_Q)^{1/2}$ at $N_c<N_F<3N_c/2$.

We start from \eqref{(B.7)}. It is not difficult to check that in these ranges of $\mph$ and at all $N_c<N_F<2N_c$ the largest mass is ${\ov\mu}_{\rm gl,2}\gg \mu^{\rm pole}_{q,1}$ due to higgsing of ${\ov q}^2, q_2$ quarks. Hence, in these \eqref{(B.7)} vacua, the cases with $n_2>\nd$ are excluded from the beginning as the rank of $\langle q_2\rangle$ is $\leq \nd$ and the unbroken $U(n_2)$ flavor symmetry cannot be maintained. But this excludes all such vacua as $n_1\leq N_c-1$ and $n_2=N_F-n_1\geq\nd+1$.

Therefore, there remain only \eqref{(B.8)} vacua. In these, in the above ranges of $\mph$, the largest mass is ${\ov\mu}_{\rm gl,1}\gg \mu^{\rm pole}_{q,2}$ due to higgsing of ${\ov q}^1, q_1$ quarks. Hence, one obtains from similar considerations that $n_1\leq\nd-1$ (the case $n_1=\nd$ is also excluded from \eqref{(B.6)},\eqref{(B.8)}\,). And similarly, because their condensate $\langle{\ov q}^2 q_2 \rangle=0$, the quarks ${\ov q}^2, q_2$ will be always in the heavy quark $HQ_2$ - phase only. Hence, at all $n_1<N_c-1$, proceeding as in \cite{ch1,ch2,ch3}, i.e. integrating out first higgsed gluons and ${\ov q}^1, q_1$ quarks at $\mu<{\ov\mu}_{\rm gl,1}$, then ${\ov q}^2, q_2$ quarks with unhiggsed colors at $\mu<\mu^{\rm pole}_{q,2}$ and, finally, unhiggsed
gluons at $\mu<\langle\lym\rangle$, one obtains the low energy Lagrangian of the form \eqref{(9.6)} with
\bq
\lym^3=\Biggl (\frac{\la^{\bd}\det\Bigl (M_{22}/\la\Bigr )}{\det N_{11}}\Biggr )^{1/(\nd-n_1)}\,,\quad\langle S\rangle=\langle\lym^3\rangle\sim \la^3
\Bigl (\frac{\mph}{\la}\Bigr )^{\frac{n_2}{n_2-N_c}}\Bigl (\frac{m_Q}{\la}\Bigr )^{\frac{n_2-n_1}{n_2-N_c}}\neq 0\,.\label{(B.9)}
\eq

At $n_1=\nd-1$ the dual gauge group will be broken completely and \eqref{(B.9)} originates from the instanton contribution.

On the whole, the assumption about the existence of additional vacua \eqref{(B.7)} or \eqref{(B.8)} with $\langle M_1\rangle=0$ or $\langle M_2\rangle=0$ at $N_c<N_F<2N_c$ is also not self-consistent.\\

The {\it additional} $N_{\rm add}=1\cdot C_{N_F}^{\nd}$ special GK-vacua with $n_1=\nd,\, n_2=N_c$ and (in our notations) $\langle\Qo\rangle=\langle M_1\rangle=0,\,\langle\Qt\rangle=\langle M_2\rangle=m_Q\mph,\, \langle S\rangle=0$ have been found in the $SU(N_c)$ theory with $N_F$ quark flavors considered
in \cite{GK}, were  instead of \eqref{(2.3)} the superpotential looks as
\bq
\cw_{GK}=m_Q{\rm Tr}(\qq)-\frac{1}{2\mph}{\rm Tr}\, (\qq)^2\,. \label{(B.10)}
\eq
Due to this, the Konishi anomalies for \eqref{(B.10)} in vacua with the spontaneously broken flavor symmetry look as
\bq
\langle\Qo+\Qt\rangle_{\rm br}=m_Q\mph\,,\quad \langle S\rangle_{\rm br}=\frac{\langle\Qo\rangle
_{\rm br}\langle\Qt\rangle_{\rm br}}{\mph}\,.\label{(B.11)}
\eq

At the same time, for the superpotential \eqref{(2.3)}
\bq
\cw_Q=m_Q{\rm Tr}(\qq)-\frac{1}{2\mph}\Biggl ({\rm Tr}\,(\qq)^2-\frac{1}{N_c}\Bigl({\rm Tr}\,\qq \Bigr)^2 \Biggr )\label{(B.12)}
\eq
in this paper the Konishi anomalies look as
\bq
\langle\Qo+\Qt-\frac{1}{N_c}{\rm Tr}\,\qq\rangle_{\rm br}=m_Q\mph\,,\quad\quad\langle S\rangle_{\rm br}=\frac{\langle\Qo\rangle_{\rm br}\langle\Qt\rangle_{\rm br}}{\mph}\,.\label{(B.13)}
\eq

We note here only that the difference between \eqref{(B.11)} and \eqref{(B.13)} is crucial for these special GK-vacua. \eqref{(B.13)} {\it does not allow for such additional GK-vacua} with $n_2=N_c$ and $\langle\Qo\rangle=\langle S\rangle=0$.

Besides, e.g. at $\la\ll\mph\ll\mo$, in $(2N_c-N_F)\cdot C_{N_F}^{\nd}$ vacua with $n_1=\nd\,,\,n_2=N_c$ and $\langle\Qo\rangle\neq 0,\,\langle\Qt\rangle\neq 0,\,\langle S\rangle\neq 0$ the parametric behavior of condensates  following from the superpotential \eqref{(B.12)} used in this paper is
\bq
\langle\Qo\rangle=m_Q\mph\ll \langle\Qt\rangle\sim \la^2\Bigl (\frac{\la}{\mph}\Bigr )^{\frac{\nd}
{2N_c-N_F}},\quad \langle S\rangle\sim m_Q\la^2\Bigl (\frac{\la}{\mph}\Bigr )^{\frac{\nd}{2N_c-N_F}}\,, \label{(B.14)}
\eq
while \eqref{(B.10)} allows only for the L -type behavior
\bq
\langle\Pi_1\rangle\approx -\langle\Pi_2\rangle\sim \la^2\Bigl (\frac{\la}{\mph}\Bigr )^{\frac{\nd}
{2N_c-N_F}},\quad \langle S\rangle\sim\la^3\Bigl (\frac{\la}{\mph}\Bigr )^{\frac{N_F}{2N_c-N_F}}\,.\label{(B.15)}
\eq

\end{document}